\documentclass{jfm}

\usepackage[]{amsmath}
\usepackage[dvipsnames]{xcolor}
\usepackage[]{subfigure}
\usepackage{upgreek}
\usepackage{overpic}
\usepackage[]{multirow}
\usepackage[]{url}
\usepackage[]{nicefrac}
\usepackage{bm}
\usepackage{enumerate}
\usepackage[normalem]{ulem} 
\usepackage{subfigure}

\begin{document}

\newtheorem{lemma}{Lemma}
\newtheorem{corollary}{Corollary}

\shorttitle{Galerkin force model for transient and post-transient dynamics} 
\shortauthor{N.~Deng, B.~R.~Noack, M.~Morzy\'nski and L.~R.~Pastur} 

\title{Galerkin force model for transient and post-transient dynamics of the fluidic pinball}

\author{Nan Deng\aff{1,2}, 
	Bernd R.~Noack\aff{3,4}\corresp{\email{Bernd.Noack@hit.edu.cn}},
        Marek Morzy\'nski\aff{5}
        \and Luc~R.~Pastur\aff{1}\corresp{\email{Luc.Pastur@ensta-paris.fr}}
}
\affiliation
{
\aff{1}
Institute of Mechanical Sciences and Industrial Applications, 
ENSTA-Paris, Institut Polytechnique de Paris, 
828 Bd des Mar\'echaux, 
F-91120 Palaiseau, France. 
\aff{2}
LIMSI,
CNRS, Universit\'e Paris-Saclay, B\^at 507, rue du Belv\'ed\`ere, Campus Universitaire,
F-91403 Orsay, France
\aff{3}
Center for Turbulence Control,
Harbin Institute of Technology,
Shenzhen, Room 312, Building C,
University Town, Xili, Shenzhen 518058, 
People's Republic of China
\aff{4}
Institut f\"ur Str\"omungsmechanik und Technische Akustik (ISTA),
Technische Universit\"at Berlin,
M\"uller-Breslau-Stra{\ss}e 8,
D-10623 Berlin, Germany
\aff{5}
Chair of Virtual Engineering, 
Pozna\'n University of Technology,
Jana Pawla II 24, PL 60-965 Pozna\'n, Poland
}

\maketitle

\begin{abstract} 
We propose an aerodynamic force model 
associated with a Galerkin model
for the unforced fluidic pinball,
the two-dimensional flow around three equal cylinders with one radius distance to each other.
The starting point is a Galerkin model of a bluff-body flow.
The force on this body is derived as a constant-linear-quadratic function
of the mode amplitudes
from first principles following the pioneering work of Noca (1997, 1999) and \citet{liang2014virtual}.
The force model is simplified for the mean-field model
of the unforced fluidic pinball \citep{deng2020jfm}
using symmetry properties and sparse calibration.
The model is successfully applied to transient and post-transient dynamics 
in different Reynolds number regimes:
the periodic vortex shedding after the Hopf-bifurcation and
the asymmetric vortex shedding after the pitchfork bifurcation
comprising six different Navier-Stokes solutions.
We foresee many applications of the Galerkin force model 
for other bluff bodies and flow control.
\end{abstract}

\section{Introduction}
\label{Sec:Introduction}
The literature on aerodynamic forces on bodies 
associated with POD or any other Galerkin model is surprisingly sparse.
On the one hand, force computations are at the heart of engineering fluid mechanics.
On the other hand, systematic investigations and interpretations 
of the aerodynamic force in the Galerkin framework are mostly missing. Considering POD as a linear decomposition of the flow field realizations, \citet{brunton2009modeling} observed that 
\begin{quote}
``While POD modes and the low order model allow for accurate reconstruction 
of the flow field and preserve Lagrangian coherent structures, 
it is not clear that this model is directly useful for reconstructing body forces quickly 
and accurately, since lift and drag forces depend nonlinearly on the flow field, meaning that contributions from different POD modes cannot be added independently.'' 
\end{quote}

The pioneering early work of \citet{Noca1997phd,noca1999comparison} reveals
that the instantaneous fluid dynamic forces on the body can be expressed with only the velocity fields and their derivatives.
\citet{liang2014virtual} applied it to the velocity based POD modes, and
derived a force expression in terms of the force of each POD mode and the force from the interaction between the POD modes.
The Galerkin force model proposed in this work reveals that any force component is a constant-linear-quadratic function of the mode amplitudes.

The starting point of our investigation is a working Galerkin model
based on a low-dimensional modal expansion of an incompressible viscous fluid flow
around a stationary body.
Intriguingly, 
mean-field theory \citep{Stuart1958jfm,Stuart1971arfm} 
was  the first foundation of many Galerkin models,
building on weakly nonlinear generalizations of stability analyses.
Mean-field theory delivered the first derivation 
of the Landau model \citep[see, e.g.,][]{Landau1987book} for super- and subcritical Hopf bifurcations.
The Landau model is experimentally supported for the onset of vortex shedding 
behind the cylinder wake  \citep{Schumm1994jfm,Zielinska1995pf}.
Generalizations explain the cross-talk between 
different frequencies over the base flow \citep{Luchtenburg2009jfm,shaabani2020jfm},
special cases of `quasi-laminar' interactions foreshadowed by \citet{Reynolds1972jfm}. 

A few decades later, 
the pioneering wall turbulence POD model by \citet{Aubry1988jfm} 
allows employing snapshot data far a low-dimensional encapsulation of the Navier-Stokes dynamics.
Since then, numerous empirical reduced-order models have been proposed
\citep{taira_AIAA2017,kunisch2002galerkin,bergmann2009enablers,ilak2006bpod,rempfer2000cfd,rowley2004model}.
Control-oriented versions have been developed by \citet{rowley2017arfm, barbagallo2009jfm, bagheri2009jfm,hinze2005proper,gerhard2003model}.

A working Galerkin model can predict the flow and thus the force.
Theories for aerodynamic forces have a rich history documented in virtually every 
fluid mechanics textbook \citep[see, e.g.,][]{Panton1984book}.
There are several force formulae for different cases.
Potential flow theory for finite bodies
can only explain the force due to accelerations of the body
and predicts vanishing drag (d'Alambert paradox).
The Zhukovsky formula derives the lift for the potential flow around streamlined cylinders,
while the drag computation is still excluded by the d'Alambert paradox.
The lifting line theory by \citet{prandtl1921applications} 
extends Zukovsky's formula for finite wings
and adds a drag estimate from the created trailing edge vortices.
\citet{kirchhoff1869theorie}  laid the first practical foundation
for bluff-body drag by allowing for a separation with infinitely thin shear layer.
Until today, the drag and lift forces of a body are inferred from the downstream velocity profile 
\citep{schlichting2016book}.
These are arguably the most common force theories.

In the Galerkin modeling literature,
unsteady forces have been formulated as functions of mode coefficients, 
like in \citet{bergmann2008jcp} and in \citet{Luchtenburg2009jfm}. 
The force formulae are generally calibrated from the reconstructed flow field.
\citet{noca1999comparison} offered an expression of the unsteady forces on an immersed body in an incompressible flow, 
which only requires the knowledge of the velocity field and its time derivative. 
Based on this idea, \citet{liang2014virtual} presented a velocity POD mode force survey method 
to measure the forces from POD modes on a flat plate. 
It has shown that the force superposition of each mode of a full POD model can accurately predict the instantaneous forces, 
and the leading six POD modes are enough to predict the drag force with $5\%$ error.

In this study, we focus on the unforced ``fluidic pinball'',
the flow around three equidistantly placed cylinders in crossflow
\citep{bansal2017experimental}. 
Following \citet{ChenAlam2020jfm}, 
the gap distance between the cylinders is chosen one radius 
and the triangle formed by the centers of three cylinders points upstream.
This distance allows for an interesting `flip-flopping' dynamics.
The advantage of the fluidic pinball is that already the two-dimensional laminar flow 
exhibits a surprisingly rich dynamics which has recently been accurately modeled \citep{deng2020jfm}.
As the Reynolds number increases, the flow behaviour 
changes from a globally stable fixed dynamics
to a periodic symmetric vortex shedding after a Hopf bifurcation,
to asymmetric vortex shedding after a subsequent pitchfork bifurcation,
followed by quasi-periodic and chaotic behavior.
Intriguingly, the post-pitchfork regime with three unstable steady solutions
as well as two stable asymmetric limit cycles and one unstable symmetric limit cycle 
is adequately described by a single five-dimensional Galerkin model.
Apparently, the force model for multiple transients 
of this pitchfork regime is already a challenge.

In the present work, we propose a Galerkin force model for the transient dynamics of the unforced fluidic pinball at different Reynolds numbers. We derive the unsteady forces 
from the Navier-Stokes equations yielding  a constant-linear-quadratic expression of the mode amplitudes of the Galerkin expansion. 
The consistent form with \citet{liang2014virtual} strengthens the theoretical basis of the force expression. 
Any known symmetric property of the modes is usually considered 
in the relative modal analysis \citep{rigas2014JFM, Podvin2020prf}, 
particularly advised for symmetry-breaking instabilities
of flows around a symmetric configuration \citep{Fabre2008pof, boronska2010pre}.
Since the fluidic pinball exhibits a mirror-symmetry, we further investigate the force expression under the $Z2$-symmetry. The drag and lift contributions must come from the specific subsets of the constant-linear-quadratic polynomial functions, which is consistent with the drag- and lift-producing modes identified in \citet{liang2015pof}.

The manuscript is organized as follows.
\S~\ref{Sec:ForceModel} derives the aerodynamic force from a Galerkin model.
\S~\ref{Sec:configuration} describes the simulation and Galerkin model of the fluidic pinball.
In \S~\ref{Sec:HopfPitchfork}, the force model for the transition 
of a simple Hopf bifurcation and for the transition of a simple pitchfork bifurcation are discussed.
Next, the force model with the elementary modes of two successive bifurcations for the multi-attractor case is investigated in \S~\ref{Sec:Results}, together with a optimization based on the correction of mean-field distortion.
We summarize the results and outline future directions of research in \S~\ref{Sec:Conclusions}.

\section{Galerkin force model}
\label{Sec:ForceModel}

In this section, the derivation of a Galerkin force model is described and discussed. 
Based on the framework of a Galerkin expansion (\S~\ref{Sec:Galerkin}), 
the drag and lift forces are expressed as constant-linear-quadratic functions of the mode amplitudes in \S~\ref{Sec:ForceSurface}.
Alternatively, the forces can consistently be derived from the momentum balance as elaborated in Appendix~\ref{Sec:ForceVolume}. 
The force model can be further simplified under symmetry considerations 
in \S~\ref{Sec:Symmetries}. 

\subsection{The Galerkin framework}
\label{Sec:Galerkin}

The fluid flow satisfies the non-dimensionalized incompressible Navier-Stokes equations 
\begin{equation}
\label{Eqn:NSE}
\partial_t \bm{u} + \nabla \cdot \bm{u} \otimes \bm{u} = \nu \triangle \bm{u} - \nabla p,
\end{equation}
where $p$ and $\bm{u}$ are respectively the pressure and velocity flow fields, $\nu = 1/Re$, with the Reynolds number $Re$. 
Here,  $\partial_t$, $\nabla$, $\triangle$, $\otimes$ and $\cdot$ respectively 
denote the partial derivative in time, the Nabla and Laplace operator as well as the outer and inner tensor product.
All the variables have been  non-dimensionalized, with the cylinder diameter $D$, the oncoming velocity $U$, the time scale $D/U$, and the density $\rho$ of the fluid.

It is assumed that there exists at least one steady solution $(\bm{u}_s,p_s)$, satisfying the steady Navier-Stokes equations
\begin{equation}
\label{Eqn:NSE:Steady}
 \nabla \cdot \bm{u}_s \otimes \bm{u}_s = \nu \triangle \bm{u}_s - \nabla p_s.
\end{equation}

For the Galerkin framework,
the space of the square-integrable vector fields $\mathcal{L}^2 ( \Omega ) $ is introduced
in the observation domain $\Omega$. 
The associated inner product for two velocity fields $\bm{u} (\bm{x})$ and $\bm{v} (\bm{x})$ reads
\begin{equation}
\label{Eqn:InnerProduct}
\left( \bm{u} , \bm{v} \right)_{\Omega} := \int\limits_{\Omega} \!\! d\bm{x} \> \bm{u}(\bm{x}) \cdot \bm{v}(\bm{x}).
\end{equation}
The velocity field is decomposed in a basic mode $\bm{u}_0$ and a fluctuating contribution.
The basic mode may be the steady Navier-Stokes solution $\bm{u}_s$ or the time-averaged flow $\overline{\bm{u}}$.
The fluctuation is represented by 
a Galerkin approximation of $N$ orthonormal space-dependent modes $\bm{u}_i ( \bm{x} )$, $i=1,\ldots,N$, with time-dependent amplitudes $a_i(t)$:
\begin{eqnarray} 
\label{Eqn:GalerkinExpansion}
\bm{u} (\bm{x},t ) & = & \sum\limits_{i=0}^N a_i(t) \bm{u}_i ( \bm{x} ),
\end{eqnarray}
where the basic mode $\bm{u}_0$ is associated with $a_0 \equiv 1$ following \citet{Rempfer1994jfm2}.
The orthonormality condition reads $\left( \bm{u}_i , \bm{u}_j \right)_{\Omega} = \delta_{ij}, \quad i,j \in \{1,\ldots,N\}$.

The Galerkin expansion \eqref{Eqn:GalerkinExpansion} satisfies the incompressibility condition and the boundary conditions by construction. 
The evolution equation for the mode amplitudes $a_i$ is derived by a Galerkin projection of the Navier-Stokes equation \eqref{Eqn:NSE} onto the modes $\bm{u}_i$:
\begin{equation}
\label{Eqn:GalerkinSystem_pressure}
\frac{d}{dt} a_i = \nu \sum\limits_{j=0}^N l_{ij}^{\nu} a_j + \sum\limits_{j,k=0}^N q_{ijk}^{c} a_j a_k + \sum\limits_{j,k=0}^N q_{ijk}^{p} a_j a_k,
\end{equation}
with the coefficients $l_{ij}^{\nu} = \left( \bm{u}_i, \triangle \bm{u}_j \right)_{\Omega}$,
$q_{ijk}^{c} = \left( \bm{u}_i, \nabla \cdot \bm{u}_j \otimes \bm{u}_k \right)_{\Omega}$ 
and  $q_{ijk}^{p} = \left( \bm{u}_i, -\nabla p_{jk} \right)_{\Omega}$ 
for the viscous, convective and pressure terms in the Navier-Stokes equations \eqref{Eqn:NSE}, respectively. 
Details are provided by \citet{Noack2005jfm}.
Thus, a  linear-quadratic Galerkin system \citep{Fletcher1984book} can be derived,
\begin{equation}
\label{Eqn:GalerkinSystem}
\frac{d}{dt} a_i = \nu \sum\limits_{j=0}^N l_{ij}^{\nu} \> a_j + \sum\limits_{j,k=0}^N \left[ q_{ijk}^{c} + q_{ijk}^{p}\right] \>  a_j \> a_k.
\end{equation}

\subsection{Drag and lift forces on a body}
\label{Sec:ForceSurface}

Let ${\Gamma}$ be the boundary of the body in the flow domain $\Omega$ and $\bm{n}$ the unit normal pointing outward the surface element $dS$. 
The $\alpha $-component $F_\alpha^\nu$  ($\alpha=x, y,z$) 
of the viscous force vector $\bm{F}^\nu$ 
on the boundary is expressed by
\begin{equation}
\label{Eqn:ViscousForce}
F_{\alpha}^{\nu} = \bm{F}^\nu\cdot \bm{e}_\alpha 
= 2 \nu \oint\limits_{\Gamma} \! 
\sum_{\beta=x,y,z} {S}_{\alpha,\beta}\,n_{\beta}\,dS \> ,
\end{equation}
where $\bm{e}_\alpha$ is the unit vector in $\alpha$-direction and
$S_{\alpha,\beta} = \left( \partial_{\alpha} u_{\beta} + \partial_{\beta}  u_{\alpha} \right)/2$ the strain rate tensor with  indices $\alpha, \beta = x, y,z$. 

Similarly, the $\alpha $-component of the global pressure force, exerted on an immersed body, is defined as
\begin{equation}
\label{Eqn:PressureForce}
F_{\alpha}^{p} = \bm{F}^p\cdot \bm{e}_\alpha 
= - \oint\limits_{\Gamma} dS \> n_{\alpha}p .
\end{equation}
Without external forces, the viscous and pressure forces in $\Omega$ counter-balance the inertial terms provided by the left-hand side of Eq.~\eqref{Eqn:NSE}. 
The drag force is defined as the projection on $\bm{e}_x$ of the pressure and viscous forces exerted on the body 
\begin{equation}
\label{Eqn:DragForce}
F_D(t) = F_{x}^{p}(t) + F_{x}^{\nu}(t).
\end{equation}
The lift force is similarly defined as the projection on $\bm{e}_y$ of the resulting pressure and viscous forces exerted on the body 
\begin{equation}
\label{Eqn:LiftForce}
F_L(t) = F_{y}^{p}(t) + F_{y}^{\nu}(t).
\end{equation}
The drag and lift coefficients read
\begin{equation}
\label{Eqn:ForceCoef}
C_D (t)= \frac{2F_D (t)}{\rho\, U^2},\quad\quad
C_L (t)= \frac{2F_L (t)}{\rho\, U^2}.
\end{equation}

Employing the Galerkin approximation \eqref{Eqn:GalerkinExpansion}, the viscous force \eqref{Eqn:ViscousForce} can be re-written as 
\begin{equation}
\label{Eqn:ViscousExpansion}
F_{\alpha}^{\nu} =  \sum\limits_{j=0}^N q^{\nu}_{\alpha; j} a_j,
\end{equation}
where $q^{\nu}_{\alpha; j}$ can easily be derived from \eqref{Eqn:ViscousForce} with the corresponding $S_{\alpha,\beta}$ of the velocity mode $\bm{u}_j$, with the form 
\begin{equation}
\label{Eqn:ViscousForce-real}
q^{\nu}_{\alpha; j} = 2 \nu \oint\limits_{\Gamma} \! \sum_{\beta=x,y,z} {S}_{\alpha,\beta}(\bm{u}_j)\> n_{\beta}\, dS .
\end{equation}
Note that the contribution of the viscous force is  linear with respect to the mode amplitudes $a_j$.

Similarly, from the pressure Poisson equation
\begin{equation}
\label{Eqn:PPE}
\nabla^2 p = \nabla \cdot \left ( - \nabla \cdot \bm{u} \otimes \bm{u}  \right ) = -\sum_{\alpha=x,y,z}\sum_{\beta=x,y,z}\partial_{\alpha} u_{\beta} \partial_{\beta} u_{\alpha},
\end{equation}
the expression of the pressure field is derived as
\begin{equation}
\label{Eqn:PressureExpansion}
p(\bm{x},t) = \sum\limits_{j,k=0}^N p_{jk} (\bm{x}) \> a_j(t) \> a_k(t),
\end{equation}
with 
\begin{equation}
\label{Eqn:PressureForce-real}
\nabla^2 p_{jk}= \nabla \cdot  \left ( - \nabla \cdot \bm{u}_j \otimes \bm{u}_k  \right )= -\sum_{\alpha=x,y,z}\sum_{\beta=x,y,z}\partial_{\alpha} u_{\beta} (\bm{u}_j) \> \partial_{\beta} u_{\alpha}(\bm{u}_k). 
\end{equation}
The boundary conditions for partial pressures $p_{jk}$ are discussed by \citet{Noack2005jfm}.
Integrating \eqref{Eqn:PressureForce} with \eqref{Eqn:PressureExpansion} 
shows that the pressure force is a quadratic polynomial of the $a_j$'s
\begin{equation}
\label{Eqn:PressureForceExpansion}
F_{\alpha}^{p} =  \sum\limits_{j,k=0}^N  q^{p}_{\alpha; j k} a_j a_k,
\quad \hbox{where} \quad 
q^{p}_{\alpha; j k} = - \oint\limits_{\Gamma} dS \> n_{\alpha} p_{jk}.
\end{equation}

Taking the steady solution as the basic mode $\bm{u}_0 = \bm{u}_s$ with $a_0 \equiv 1$ implies that $\bm{a}$ with $a_i = \delta_{0i}$ is a fixed point of Eq.~\eqref{Eqn:GalerkinSystem} and the total force can be expressed as a constant-linear-quadratic expression in terms of the mode coefficients
\begin{equation}
\label{Eqn:ForceExpansion}
F_{\alpha} = F_{\alpha}^{\nu} + F_{\alpha}^{p}
= c_{\alpha} 
+ \sum\limits_{j=1}^N l_{\alpha; j} a_j
+ \sum\limits_{j,k=1}^N q_{\alpha; j k} a_j a_k,
\end{equation}
where 
\begin{equation}
\label{Coef_ForceExpansion}
c_{\alpha} =  q^{\nu}_{\alpha; 0} + q^{p}_{\alpha; 0 0}, \quad\quad
l_{\alpha; j}  =  q^{\nu}_{\alpha; j} 
                      + q^{p}_{\alpha; j 0} 
                      + q^{p}_{\alpha; 0 j}, \quad \quad
q_{\alpha; j k}  = q^{p}_{\alpha; j k}.
\end{equation}
The force expression in Eq.~\eqref{Eqn:ForceExpansion} can be alternatively derived from the residual of the Navier-Stokes equations in the flow domain $\Omega$, as demonstrated in Appendix~\ref{Sec:ForceVolume}.

With constant $\rho$ and $U$, the drag and lift coefficients in \eqref{Eqn:ForceCoef} can be rewritten in the form
\begin{subequations}
\label{Eqn:cdcl}
\begin{eqnarray}
C_D = c_{x} 
+ \sum\limits_{j=1}^N l_{x; j} a_j
+ \sum\limits_{j,k=1}^N q_{x; j k} a_j a_k, \label{Eqn:cd}\\
C_L = c_{y} 
+ \sum\limits_{j=1}^N l_{y; j} a_j
+ \sum\limits_{j,k=1}^N q_{y; j k} a_j a_k. \label{Eqn:cl}
\end{eqnarray}
\end{subequations}

A crucial step relies on the choice of the $\bm{u}_i$ modes for the decomposition of Eq.~\eqref{Eqn:GalerkinExpansion}. These could be the POD modes, as usually considered in fluid flows. However, a better choice could be to decompose the flow field on a basis of modes that are becoming active when the system is undergoing a bifurcation. This choice of the so-called \textit{bifurcation modes} will be investigated in \S~\ref{Sec:PinballBifurcationModes}.

\subsection{The Navier-Stokes equations under the $Z_2$-symmetry}
\label{Sec:Symmetries}
When the fluid flow configuration exhibits a mirror-symmetry, the Navier-Stokes equations \eqref{Eqn:NSE} possess at least one symmetric steady solution $(\bm{u}_s,p_s)$, satisfying
Eq.~\eqref{Eqn:NSE:Steady}. The $Z_2$-symmetry of the velocity and pressure fields, with respect to the ($x,z$)-plane defined by $y=0$, implies
\begin{subequations}
\label{Eqn:Symmetry}
\begin{eqnarray}
\notag u^s(x,-y,z) &=& u^s(x,y,z), \quad 
v^s(x,-y,z) = -v^s(x,y,z), \\ 
p^s(x,-y,z) &=& p^s(x,y,z),\\
\notag u^a(x,-y,z) &= &-u^a(x,y,z), \quad 
v^a(x,-y,z) =  v^a(x,y,z), \\ 
p^a(x,-y,z) &=& -p^a(x,y,z),
\end{eqnarray}
\end{subequations}
where the symmetric components $(u^s, v^s, p^s) \in \mathcal{U}^s$ and the antisymmetric components $(u^a, v^a, p^a) \in \mathcal{U}^a$, $\mathcal{U}^s$ and $\mathcal{U}^a$ being respectively the symmetric and antisymmetric subspaces of the system. 
Other steady solutions can exist, which break the symmetry of the system. We will consider the symmetric steady solution $(\bm{u}_s,p_s)$ as the reference point of Eq.~\eqref{Eqn:NSE} in the Reynolds decomposition of the flow field as Eq.~\eqref{Eqn:ReynoldsDecomposition}.

The dynamics under consideration can include transient and post-transient regimes.
Here, we introduce the $T$-averaged flow fields $\bar{\bm{u}}_T(\bm{x},t)$ as
\begin{equation}
\label{Eqn:MeanField1}
\bar{\bm{u}}_T(\bm{x},t) = \frac{1}{T}\int_{t-T/2}^{t+T/2}\>\bm{u}(\bm{x},\tau)\>d\tau,
\end{equation}
where $T$ is a time-scale to be chosen. When the flow field is oscillating in time, an appropriate choice for $T$ is the period of the local oscillation. 
The mean flow field is further defined as
\begin{equation}
\label{Eqn:Taveragedflow}
\bar{\bm{u}}(\bm{x}) = \lim _{T\rightarrow \infty}\bar{\bm{u}}_T(\bm{x},t) 
\end{equation}
and only focuses on the post-transient limit.

When two mirror-conjugated attractors co-exist, it is convenient to introduce the ensemble-averaged flow field $\bar{\bm{u}}_T^\bullet (\bm{x},t)$ as
\begin{equation}
\label{Eqn:MeanField2}
\bar{\bm{u}}_T^\bullet (\bm{x},t) = \frac{1}{2}(\bar{\bm{u}}_T^+(\bm{x},t)+\bar{\bm{u}}_T^-(\bm{x},t)). 
\end{equation}
where $\bar{\bm{u}}_T^\pm(\bm{x},t)$ are the $T$-averaged  flow field on the way to each individual attractor. This definition could be readily extended to more than two conjugated attractors. As an ensemble average on mirror-conjugated attracting sets, the ensemble-averaged flow field $\bar{\bm{u}}_T^\bullet (\bm{x},t)$ belongs to the symmetric subspace $\mathcal{U}^s$.

At this point, it is most convenient to introduce the Reynolds decomposition of the flow field, in the form  
\begin{equation}
\label{Eqn:ReynoldsDecomposition}
 \bm{u}(\bm{x},t) = \bar{\bm{u}}_T^\bullet (\bm{x},t) + \bm{u}'(\bm{x},t) = \bm{u}_s(\bm{x}) + \bm{u}_{\Delta}(\bm{x},t) + \bm{u}'(\bm{x},t).
\end{equation}
where the mean-field deformation $\bm{u}_{\Delta}(\bm{x},t)$ accounts for the distortion of the flow field from the symmetric steady solution $\bm{u}_s(\bm{x})$ to the ensemble-averaged flow field $\bar{\bm{u}}_T^\bullet (\bm{x},t)$ as
\begin{equation}
\label{Eqn:ShiftField}
 \bm{u}_{\Delta}(\bm{x},t)= \bar{\bm{u}}_T^\bullet (\bm{x},t)  -  \bm{u}_s(\bm{x}).
\end{equation}
The fluctuation flow field $\bm{u}'(\bm{x},t)$ has a vanishing time average, meaning that $\bm{u}(\bm{x},t)$ is centered on $\bar{\bm{u}}_T^\bullet (\bm{x},t)$. 
By construction, $\bar{\bm{u}}_T^\bullet (\bm{x},t), \bm{u}_{\Delta}(\bm{x},t), \bm{u}_s(\bm{x})$ belongs to the symmetric subspace $\mathcal{U}^s$ and $\bm{u}'(\bm{x},t)$  to the anti-symmetric subspace $\mathcal{U}^a$.
Thus, a symmetry-based decomposition of Eq.~\eqref{Eqn:NSE} results into a symmetric and an anti-symmetric part, yielding
\begin{subequations}
\label{Eqn:NSE-Antisym-Sym}
\begin{eqnarray}
 \partial_t \bm{u}_{\Delta} + \nabla \cdot  \left[ \bm{u}_s        \otimes \bm{u}_{\Delta}
 + \bm{u}_{\Delta} \otimes \bm{u}_s 
 + \bm{u}_{\Delta} \otimes \bm{u}_{\Delta}
 + \bm{u}^{\prime} \otimes \bm{u}^{\prime}  \right] 
& = & \nu \triangle \bm{u}_{\Delta} - \nabla p_{\Delta},\label{Eqn:NSE-Sym} \\
 \partial_t \bm{u}^{\prime} + \nabla \cdot  \left[ \bar{\bm{u}}_T^\bullet  \otimes \bm{u}^{\prime} 
 + \bm{u}^{\prime} \otimes \bar{\bm{u}}_T^\bullet   \right] 
& = & \nu \triangle \bm{u}^{\prime} - \nabla p^{\prime}.\label{Eqn:NSE-Antisym}
\end{eqnarray}
\end{subequations}

Integrating \eqref{Eqn:NSE-Sym} on the spatial domain $\Omega$, both the left and right hand sides yield a time-evolving force vector aligned on $\bm{e}_y$, while integrating \eqref{Eqn:NSE-Antisym} yields a time-evolving force vector aligned on $\bm{e}_x$. The former is the resulting lift force applying to the boundaries of the fluid domain, while the latter is the drag force. 
Thus, the $Z_2$-symmetry applied to equations \eqref{Eqn:cd} and
\eqref{Eqn:cl} yields
\begin{subequations}
\label{Eqn:cdcl2}
\begin{eqnarray}
C_D = C_D^\circ
+ \sum\limits_{j=1}^N \underbrace{ [l_{x; j} a_j]}_{\in \mathcal{U}^s}
+ \sum\limits_{j,k=1}^N \underbrace{ [q_{x; j k} a_j a_k]}_{\in \mathcal{U}^s}, \label{Eqn:cd2}\\
C_L = \sum\limits_{j=1}^N \underbrace{ [l_{y; j} a_j]}_{\in \mathcal{U}^a}
+ \sum\limits_{j,k=1}^N \underbrace{ [q_{y; j k} a_j a_k]}_{\in \mathcal{U}^a},\label{Eqn:cl2}
\end{eqnarray}
\end{subequations}
where $C_D^\circ$ is the drag coefficient of the symmetric steady solution.

The vanishing terms in \eqref{Eqn:cdcl2} can be easily derived from the definition of $q^{\nu}_{\alpha; j}$ and $q^{p}_{\alpha; j k}$ in \S~\ref{Sec:ForceVolume} as:
\begin{subequations}
\label{Eqn:ForceExpansion_SymConstraint}
\begin{eqnarray}
l_{x; j} &= q_{x;j}^{\nu} + q_{x; 0j}^{p} + q_{x; j0}^{p} &= 0, \quad\quad \bm{u}_j \in \mathcal{U}^a ,\\
l_{y; j} &= q_{y;j}^{\nu} + q_{y; 0j}^{p} + q_{y; j0}^{p} &= 0, \quad\quad \bm{u}_j \in \mathcal{U}^s ,\\
q_{x; j k} &= q_{x; j k}^{p} &= 0, \quad\quad \bm{u}_j \otimes \bm{u}_k  \in \mathcal{U}^a ,\\
q_{y; j k} &= q_{y; j k}^{p} &= 0, \quad\quad \bm{u}_j \otimes \bm{u}_k \in \mathcal{U}^s . 
\end{eqnarray}
\end{subequations}
As a result, the drag contribution must come from the symmetric subsets of the constant-linear-quadratic polynomial functions, and from the antisymmetric subsets for the lift contribution.

\section{Galerkin model of the fluidic pinball}
\label{Sec:configuration}

The force model derived in \S\,\ref{Sec:ForceModel} is applied to a configuration of three equidistantly placed cylinders in a cross-flow, known as the ``fluidic pinball'' configuration \citep{Noack2017put}. The flow configuration and the direct Navier-Stokes solver are described in \S~\ref{Sec:Pinball}.  
As the Reynolds number is increased, the flow undergoes two subsequent supercritical Hopf and pitchfork bifurcations. The corresponding force dynamics at different Reynolds numbers are reported in \S~\ref{Sec:ForceFeatures}. The \textit{bifurcation modes}, newly introduced by \cite{deng2020jfm}, are defined in \S\,\ref{Sec:PinballBifurcationModes}. 
They  provide the orthogonal basis for the Galerkin projection. 

\subsection{The fluidic pinball}
\label{Sec:Pinball}
\begin{figure}
 \centerline{
 \includegraphics[width = .55\linewidth]{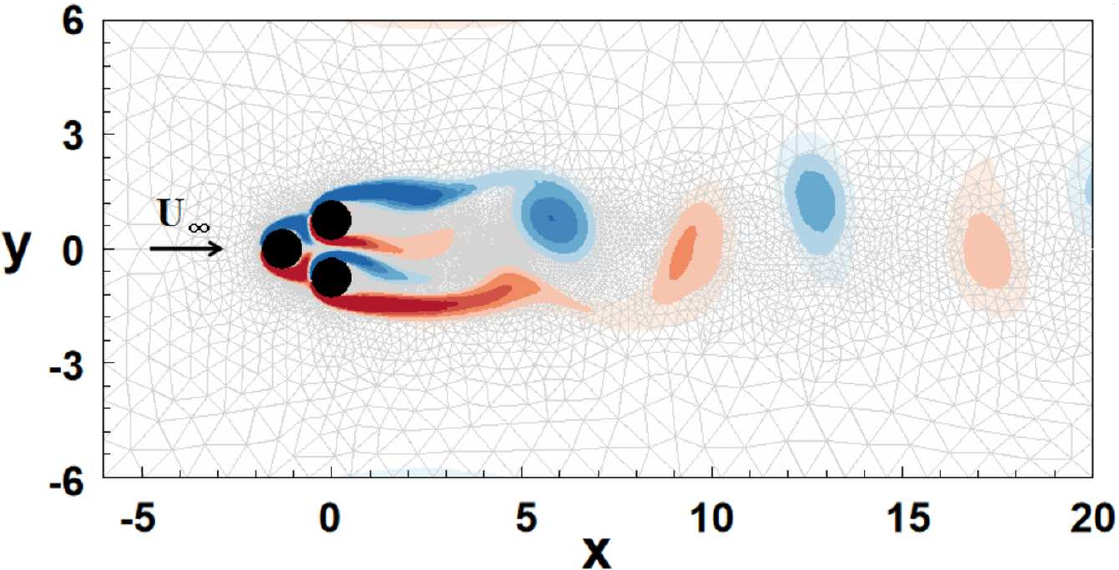}
 }
\caption{Configuration of the fluidic pinball and dimensions of the simulated domain. A typical field of vorticity is represented in color with $[-1.5, 1.5]$. The upstream velocity is denoted $ U_\infty $.}
\label{Fig:pinball}
\end{figure}
The geometric configuration, shown in figure~\ref{Fig:pinball}, consists of three fixed cylinders of unit diameter $ D $ mounted on the vertices of an equilateral triangle of side length $ 3D/2 $  in the $ (x, y) $ plane. The flow domain is bounded with a $[-6,+20]\times[-6,+6]$ box. 
The upstream flow, of uniform velocity $ U_\infty $ at the input of the domain, is transverse to the cylinder axis and aligned with the symmetry axis of the cylinder cluster.
All quantities will be non-dimensionalized with cylinder diameter $D$, the velocity $U_\infty$, and the unit fluid density $\rho$.
Considering the symmetry of this configuration, a Cartesian coordinate system will be used in the following discussion, with its origin in the middle of the rightmost two cylinders. 
In this study, no external force will be applied to these three cylinders.
A no-slip condition is applied on the cylinders and the velocity in the far wake is assumed to be $ U_\infty $. Here, the Reynolds number is defined as $ Re = U_\infty D/\nu $, where $ \nu $ is the kinematic viscosity of the fluid. A no-stress condition is applied at the output of the domain. 


The resolution of the Navier-Stokes equations \eqref{Eqn:NSE} is based on a second-order finite-element discretization method of the Taylor-Hood type \citep{taylor1973ef}, on an unstructured grid of 4\,225 triangles and 8\,633 vertices, and an implicit integration of the third-order in time. The instantaneous flow field is calculated with a Newton-Raphson iteration until the residual reaches a tiny tolerance prescribed. This approach is also used to calculate the steady solution, which is derived from the steady Navier-Stokes equations \eqref{Eqn:NSE:Steady}. The Direct Navier-Stokes solver used herein has been validated in \citet{Noack2003jfm} and \cite{deng2020jfm}, with a detailed technical report \citep{Noack2017put}. The grid used for the simulations was shown to provide a consistent flow dynamics, compared to a refined grid, see \citet{deng2020jfm}.

\subsection{Flow features and the corresponding force dynamics}
\label{Sec:ForceFeatures}

Different from \citet{deng2020jfm}, where the viscous contribution to the forces has been ignored,
the lift $ C_L$ and drag $ C_D $ coefficients are here calculated from the resulting force $ \bm{F} $ of pressure and viscous components exerted on the three cylinders. 

The flow characteristics depend on the Reynolds number $Re$.
Following the literature on clusters of cylinders \citep{ChenAlam2020jfm}, 
the characteristic length scale is chosen to be the cylinder diameter $D$ 
and not the transverse width $5 D/2$ of the configuration.
This width loses its dynamic significance for large distances considered in other studies.

\begin{figure}
\centering
\subfigure[Lift coefficients]{
    \begin{minipage}[b]{0.55\linewidth}
        \centering
        \includegraphics[width=0.9\linewidth]{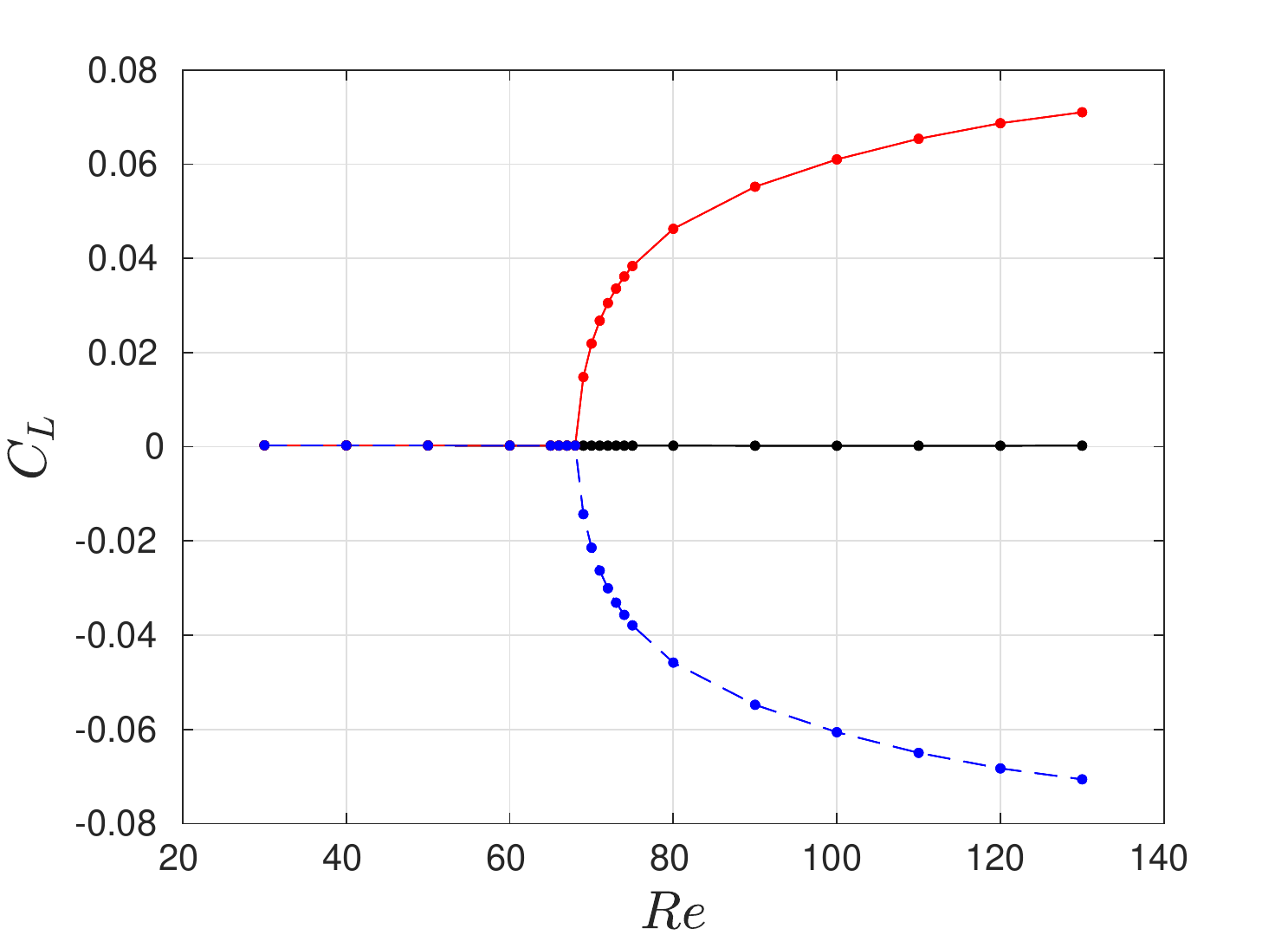} 
    \end{minipage}%
}
\subfigure[Three steady solutions]{
    \begin{minipage}[b]{0.4\linewidth}
        \centering
        \includegraphics[width=.8\linewidth]{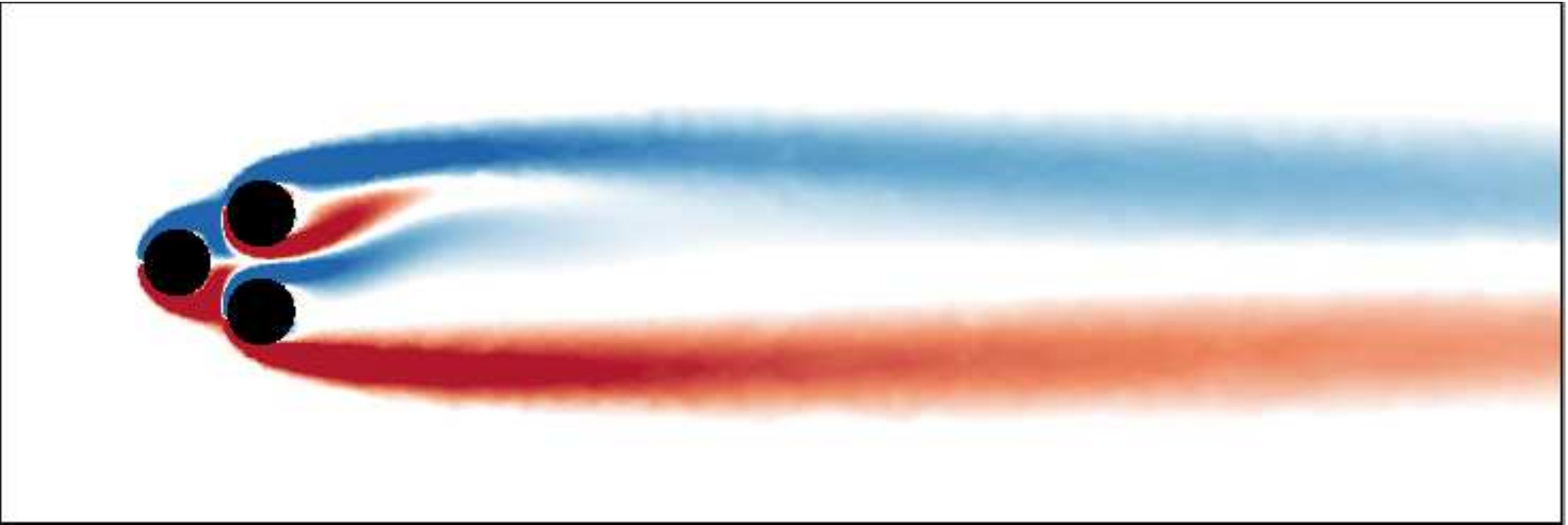}\vspace{0.03cm}
        \includegraphics[width=.8\linewidth]{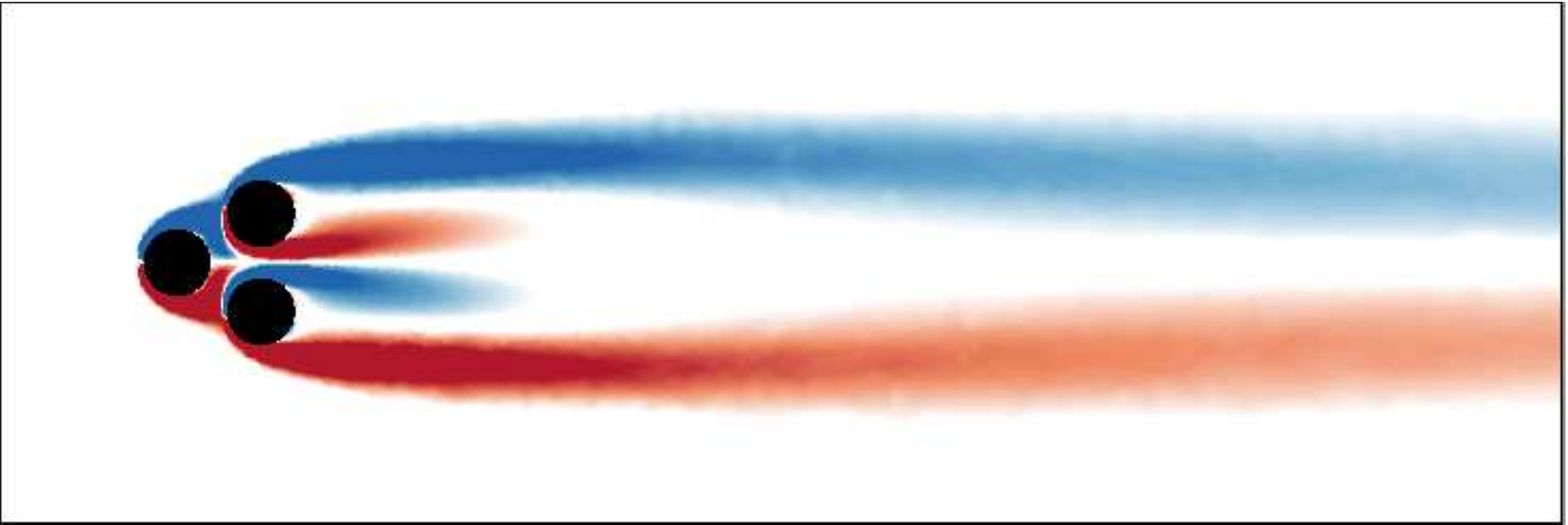}\vspace{0.03cm}
        \includegraphics[width=.8\linewidth]{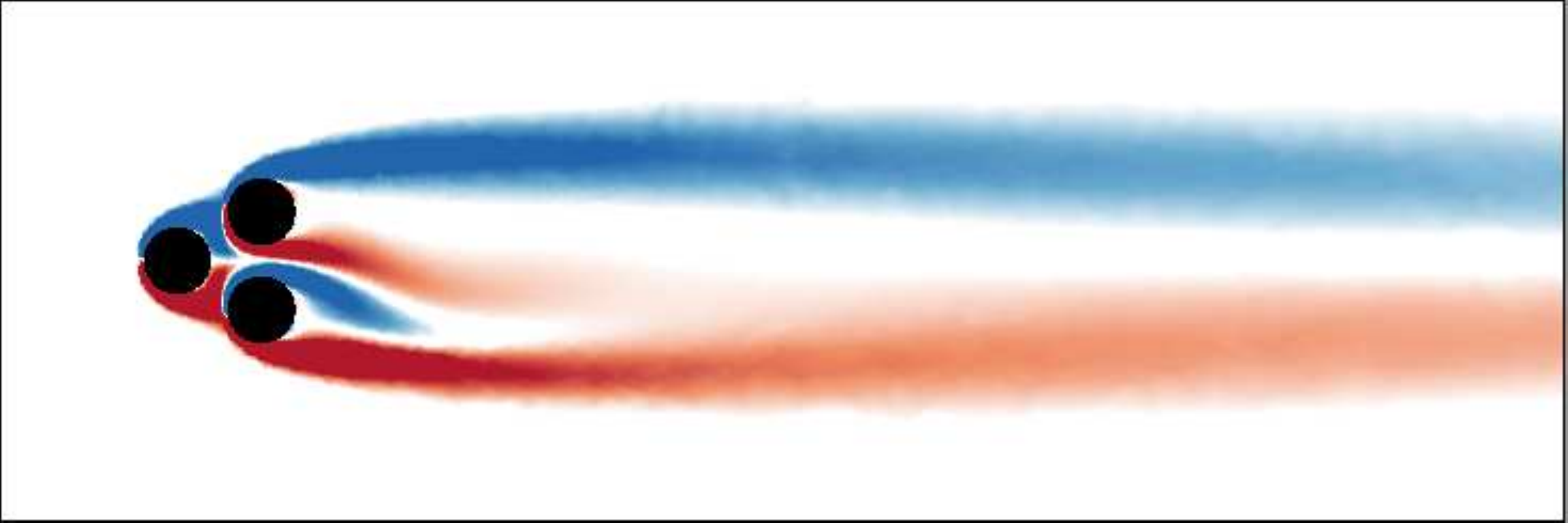}
    \end{minipage}%
}%
\caption{Lift coefficients at different Reynolds numbers (a) of the symmetric steady solutions $\bar{\bm{u}}_s$(black curve),the asymmetric steady solutions $\bar{\bm{u}}_s^-$(blue curve), the asymmetric steady solutions $\bar{\bm{u}}_s^+$(red curve), exemplified with the vorticity field of $\bar{\bm{u}}_s^+$, $\bar{\bm{u}}_s$, $\bar{\bm{u}}_s^-$ at $Re = 100$ from top to bottom (b). }
\label{Fig:Lift-SS}
\end{figure}

For Reynolds numbers $Re<Re_{\rm H}\approx 18$, 
the symmetric steady solution $\bm{u}_s(\bm{x})$ was found to be stable and is the only attractor of the system. 
A supercritical Hopf bifurcation occurs at $ Re = Re_{\rm H} $, associated with the cyclic release of counter-rotating vortices in the wake of the three cylinders from the shear-layers that delimit the configuration, forming a von K\'arm\'an street of vortices. 
The corresponding Reynolds number based on the transverse width of the fluidic pinball 
is $45$, i.e., is well-aligned with typical onsets of vortex shedding behind bluff bodies.
For the critical value $ Re = Re_{\rm PF} \approx 68$, the system undergoes a supercritical pitchfork bifurcation. As a result, two additional (unstable) steady solutions occur, namely $\bm{u}_s^+(\bm{x})$ and $\bm{u}_s^-(\bm{x})$, which break the reflectional symmetry of the configuration, as shown with the lift coefficients of the steady solutions in figure \ref{Fig:Lift-SS}.
The mean-field inherits the asymmetry of the steady solutions, with the jet between the two downstream cylinders being deflected upward or downward. As reported in \citet{deng2020jfm}, at $Re=Re_{\rm PF}$, the statistically symmetric limit cycle, associated with the statistically symmetric vortex shedding, becomes unstable with respect to two mirror-conjugated statistically asymmetric limit cycles, associated with statistically asymmetric von K\'arm\'an streets of vortices. 

\begin{figure}
 \centerline{
 \begin{tabular}{ccc}
 &$C_D$ &  $C_L$\\
 \hline \hline
(a) &  & \\
 & \includegraphics[width=.45\linewidth]{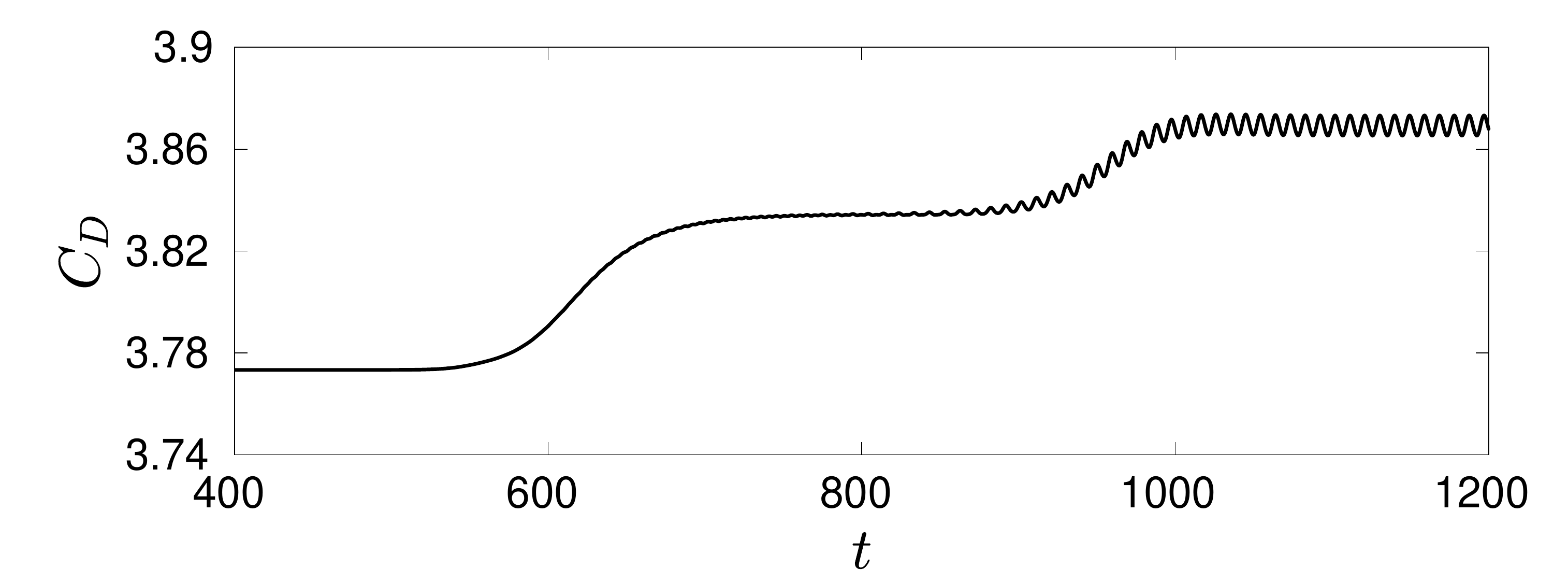} 
 & \includegraphics[width=.45\linewidth]{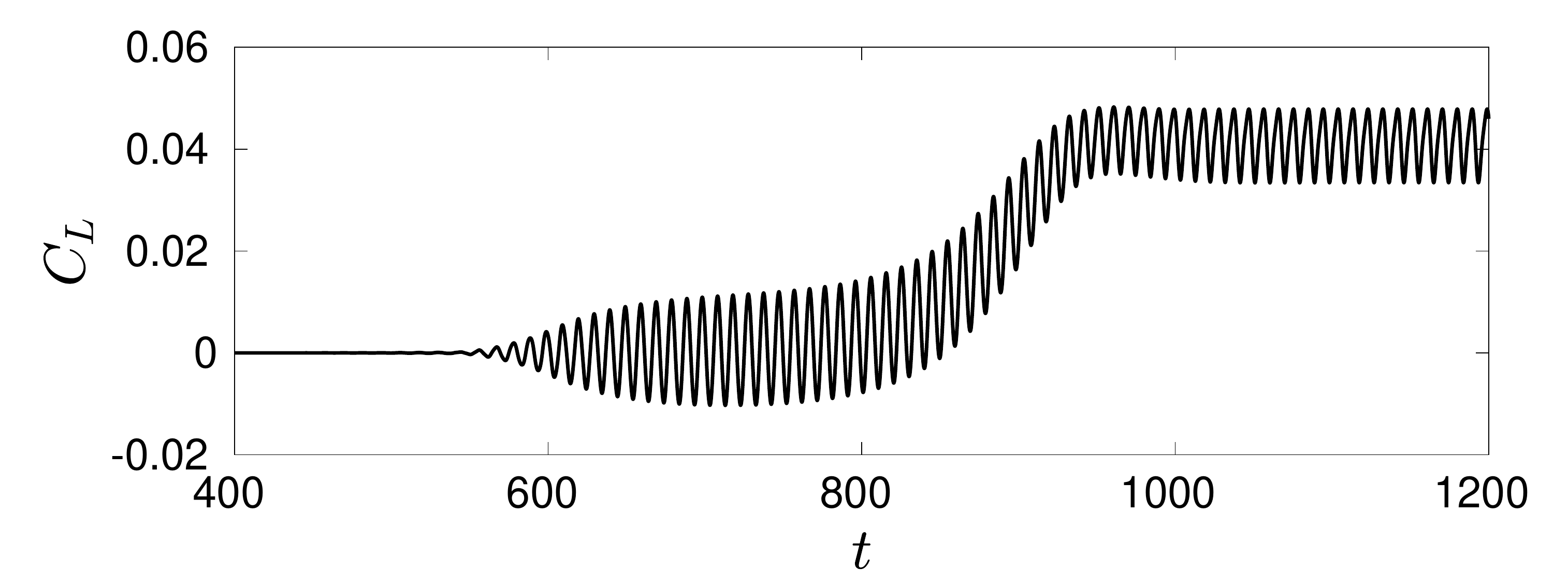} \\ 
(b) &  & \\
 & \includegraphics[width=.45\linewidth]{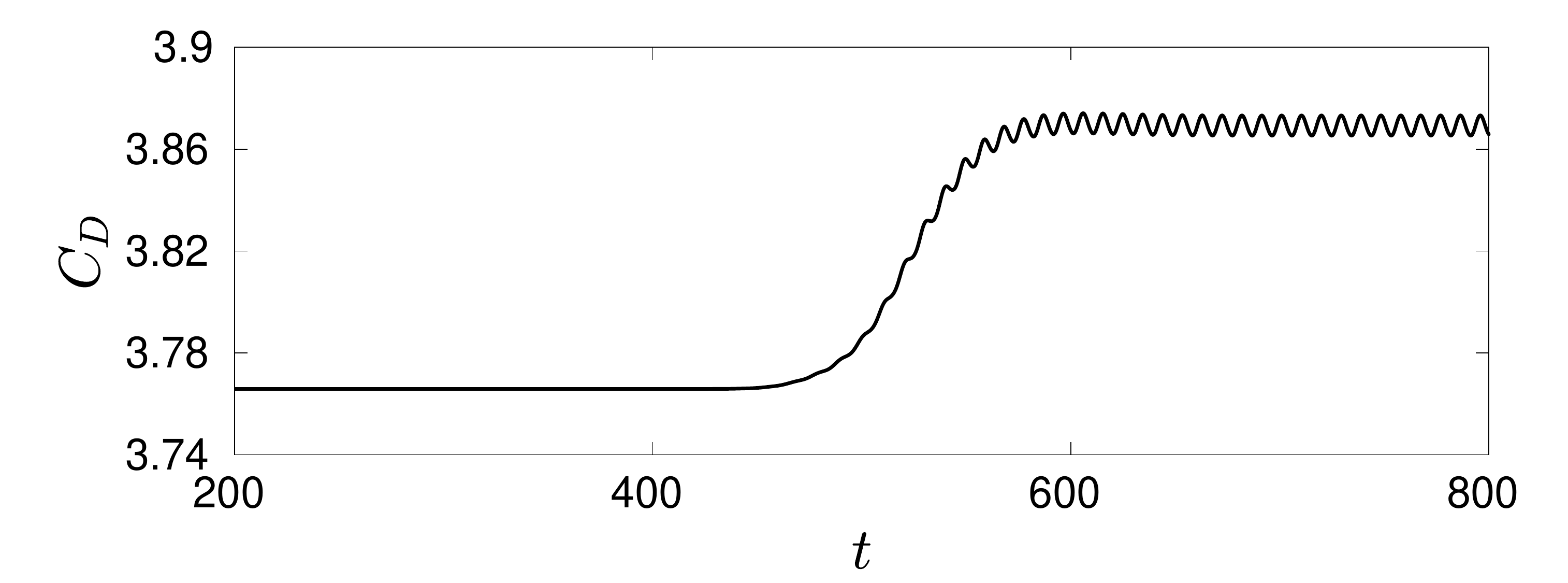} 
 & \includegraphics[width=.45\linewidth]{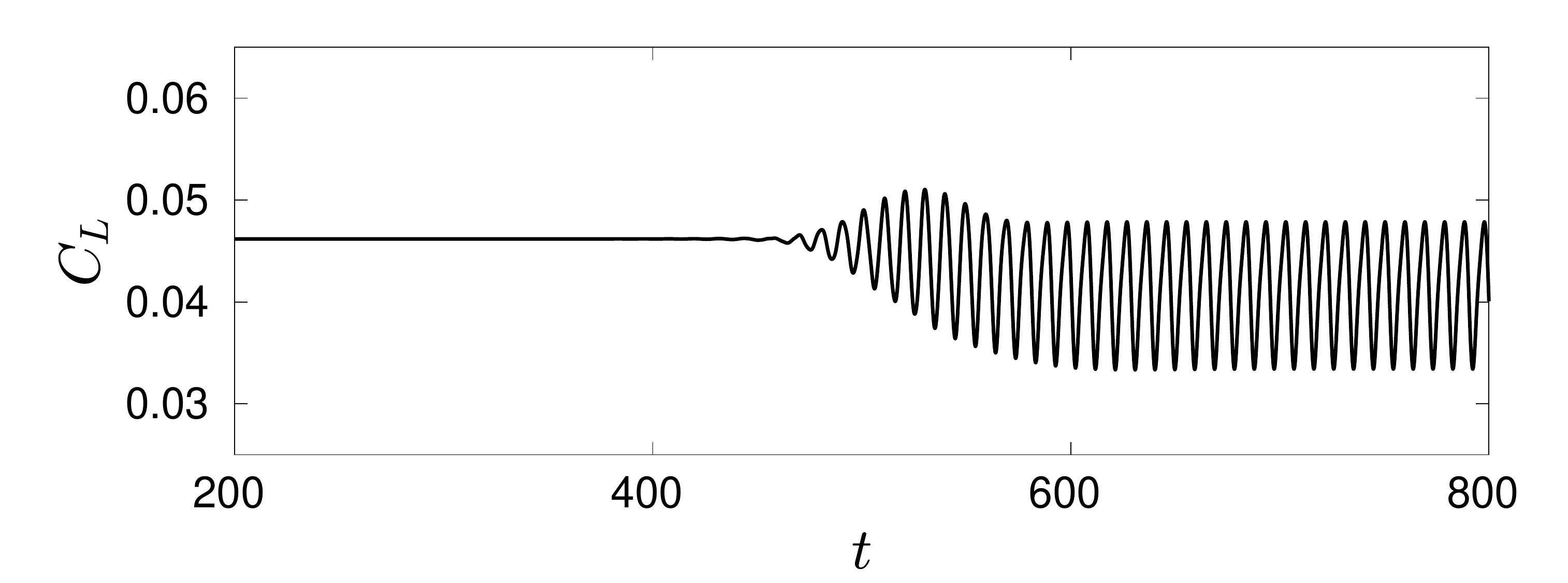} \\ 
 \end{tabular}
 }
\caption{Time evolution of the drag (left) and lift (right) coefficients, starting (a) from the symmetric steady solution $\bm{u}_s$, (b) from the asymmetric steady solution $\bm{u}_s^+$, at $Re=80$.}
\label{Fig:Re80}
\end{figure}

Figure \ref{Fig:Re80} shows the time evolution of the lift and drag coefficients at $Re=80$, when the initial condition is either the symmetric steady solution $\bm{u}_s$ (figure \ref{Fig:Re80}(a)) or the asymmetric steady solution $\bm{u}_s^+$ (figure \ref{Fig:Re80}(b)). In both cases, the asymptotic regime is the same. However, when starting from the symmetric steady solution in figure \ref{Fig:Re80}(a), a long-living plateau of the drag coefficient is reached around time $t\approx 775$, which corresponds to the transient exploration of the unstable limit cycle, centered on the symmetric T-averaged flow field $\bar{\bm{u}}_{98}(\bm{x},775)$. Note that during the transient dynamics from the steady solution to the unstable limit cycle, the drag coefficient is monotonically increasing, before reaching the transient plateau. The drag coefficient is further increasing when leaving the unstable limit cycle towards the asymptotically stable limit cycle, the latter being centered on the asymmetric mean flow field $\bar{\bm{u}}^+$. 

\begin{figure}
\centering
\includegraphics[width=.9\linewidth]{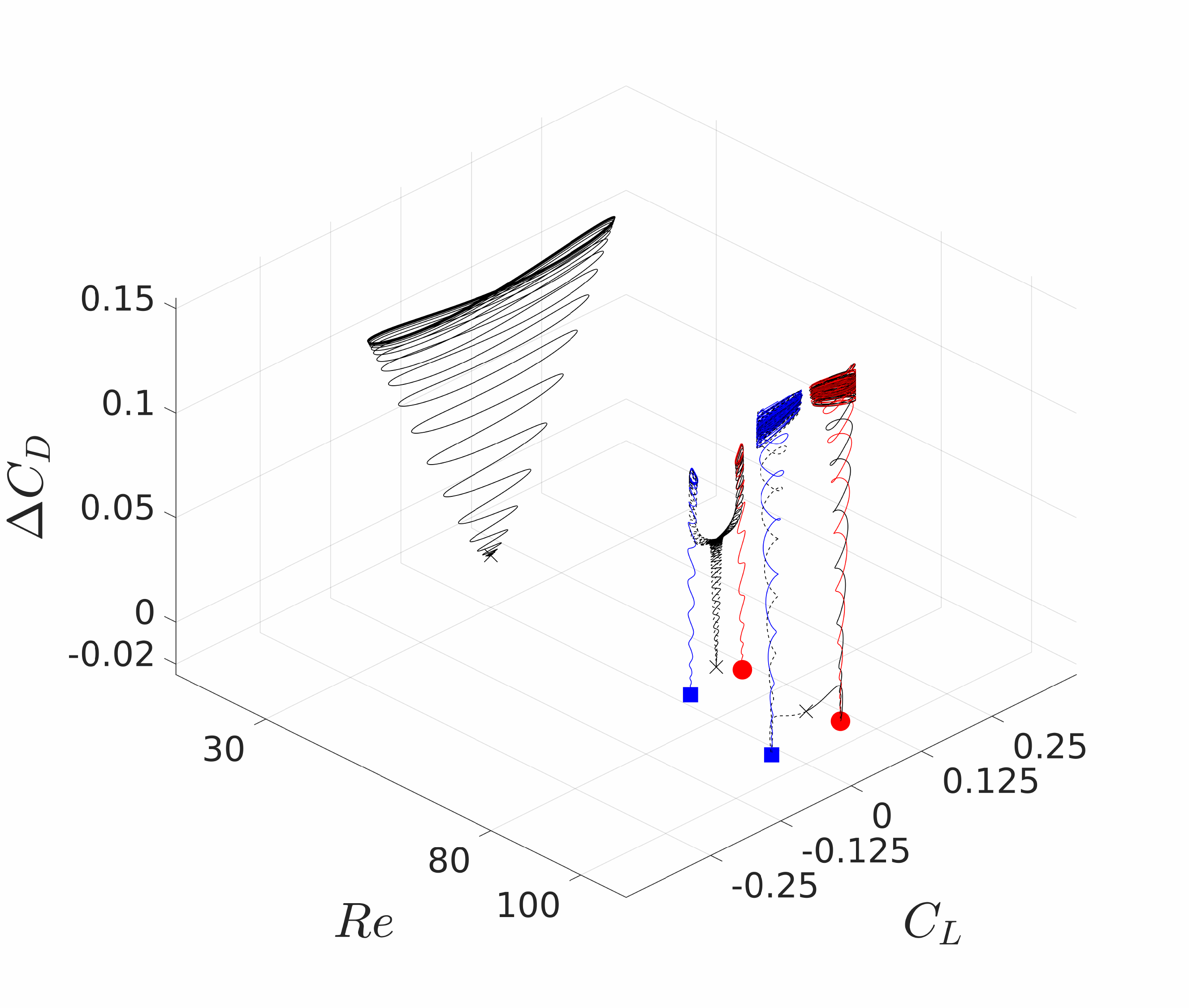}
\caption{Trajectories in the $(C_L,\Delta C_D)$ plane, for Reynolds numbers $Re=30$, 80 and 100, starting, for the black trajectories, close to the symmetric steady solution $\bm{u}_s$ ($\times$), for the red trajectories close to the asymmetric steady solution $\bm{u}_s^+$ ($\bullet$), and for the blue trajectories close to 
the asymmetric steady solution $\bm{u}_s^-$ ($\blacksquare$). 
$\Delta C_D = C_D - C_D^\circ$, where $C_D^\circ$ is the drag coefficient of the symmetric steady solution at the corresponding Reynolds number.}
\label{Fig:scheme_3cases}
\end{figure}

Figure \ref{Fig:scheme_3cases} shows another representation of the transient dynamics for $Re=30$, 80 and 100, starting from different initial conditions in the plane ($C_L,\Delta C_D$), where $\Delta C_D = C_D - C_D^\circ$, $C_D^\circ$ being the drag associated with the symmetric steady solution at the Reynolds number under consideration. The black cross ($\times$) stands for the symmetric steady solution $\bm{u}_s$ while the asymmetric $\bm{u}_s^+$ and $\bm{u}_s^-$ steady solutions are respectively represented by a red circle and a blue square, when they exist, at $Re=80$ and 100. As it can be observed in this figure, to the difference of what happens at $Re=80$,  
the transient dynamics from the symmetric steady solution at $Re=100$ first reaches one of the two asymmetric steady solutions, before evolving toward the stable attracting limit cycle.

\subsection{The bifurcation modes of the fluidic pinball}
\label{Sec:PinballBifurcationModes}
\begin{figure}
\centerline{
\begin{tabular}{cc}
(a) & (b) \\
 \includegraphics[width=.5\linewidth]{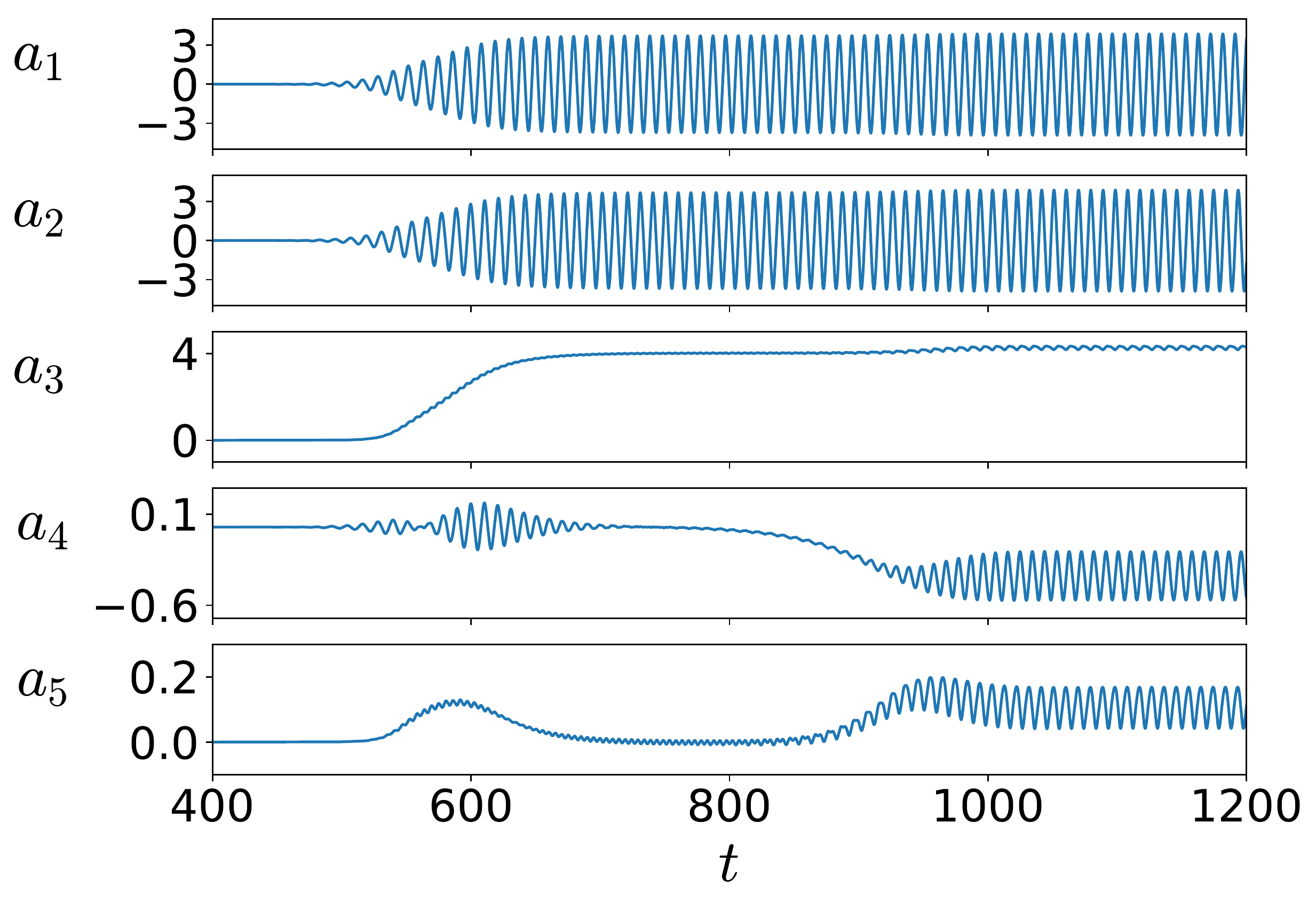} &
 \includegraphics[width=.5\linewidth]{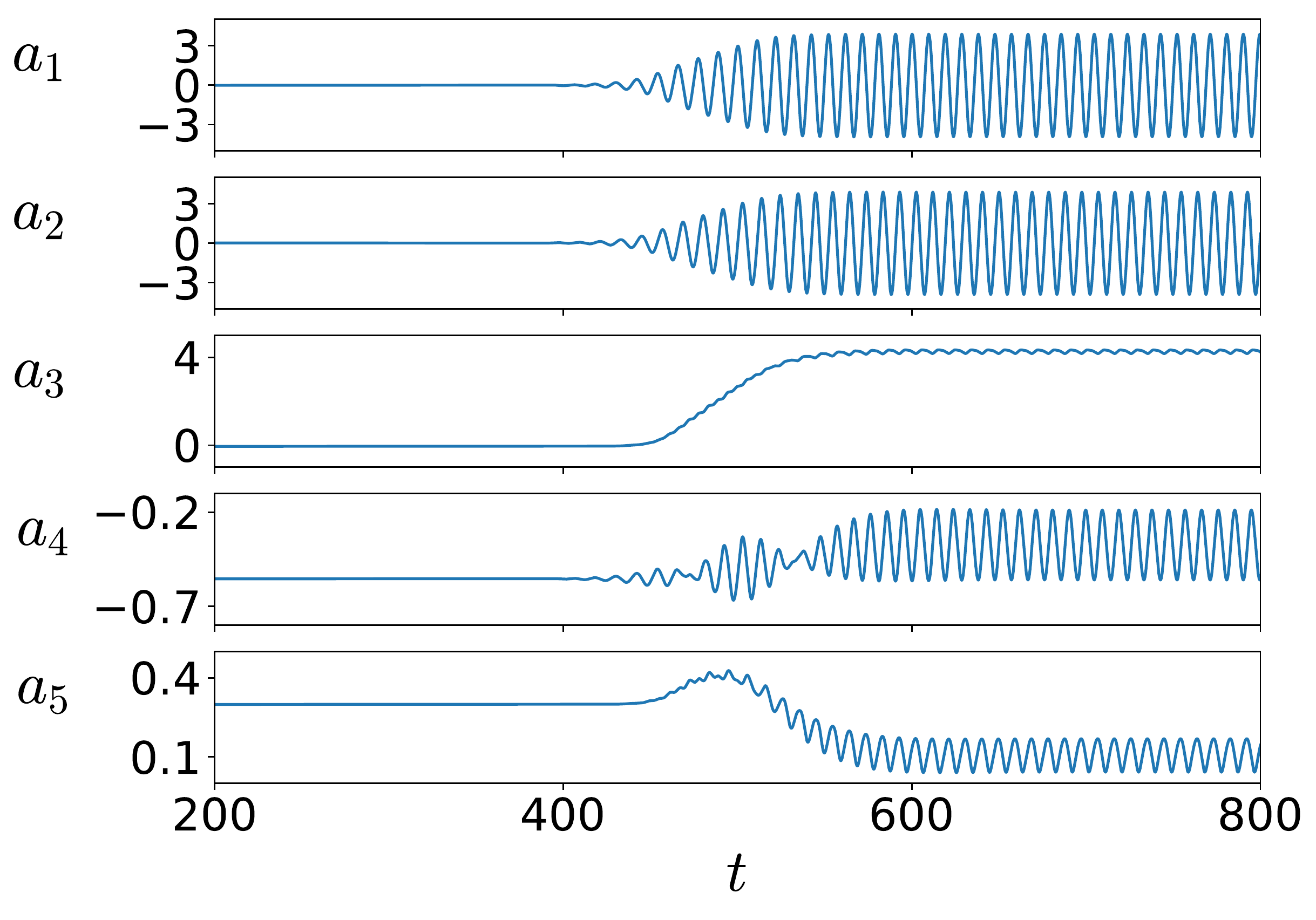} 
\end{tabular}
}
\caption{Mode amplitudes $a_i(t)$, $i=1,\dots,5$ in the full-flow dynamics starting (a) from the symmetric steady solution $\bm{u}_s$, (b) from the asymmetric steady solution $\bm{u}_s^+$, at $Re=80$.}
\label{Fig:Ampl_5dof}
\end{figure}

In the case of two subsequent supercritical Hopf and pitchfork bifurcations, 
\cite{deng2020jfm} have shown that the reduced-order model must comprise 5 modes: 
\begin{equation} 
\label{Eqn:BifurcationExpansion}
\bm{u} (\bm{x},t ) = \bm{u}_s ( \bm{x} ) + \sum\limits_{j=1}^5 a_j(t) \bm{u}_j ( \bm{x} ).
\end{equation}
Hence, in the decomposition of Eq.~\eqref{Eqn:GalerkinExpansion}, 
the number of modes is restricted to $N=5$.
For dynamic interpretability,  
the basic mode $\bm{u}_0(\bm{x})$ is chosen to be symmetric steady solution $\bm{u}_s(\bm{x})$. %
The first three modes $\bm{u}_{1,2,3}(\bm{x})$ are associated with the Hopf bifurcation, the last two modes $\bm{u}_{4,5}(\bm{x})$ with the pitchfork bifurcation. We will refer to these modes as the irreducible \textit{bifurcation modes} of the system. Modes $\bm{u}_3(\bm{x})$ and $\bm{u}_5(\bm{x})$ are symmetric. 
The instability-related modes $\bm{u}_{1,2}(\bm{x})$ and $\bm{u}_4(\bm{x})$ are anti-symmetric. Modes $\bm{u}_{1,2}(\bm{x})$ span the subspace associated with the limit cycle of the Hopf bifurcation, while $\bm{u}_4(\bm{x})$ accounts for the symmetry breaking of the pitchfork bifurcation. In \cite{deng2020jfm}, modes $\bm{u}_{1,2}(\bm{x})$ are provided by the first two dominant POD modes, while mode $\bm{u}_4(\bm{x})$ is defined as 
\begin{equation}
\label{Eqn:u4}
\bm{u}_4(\bm{x}) \propto \bm{u}_s^+(\bm{x})-\bm{u}_s^-(\bm{x}),    
\end{equation}
where $\bm{u}_s^\pm(\bm{x})$ are the two additional (asymmetric) steady solutions arising from the supercritical pitchfork bifurcation. 
Mode $\bm{u}_3(\bm{x})$ is slaved to $\bm{u}_{1,2}(\bm{x})$ while $\bm{u}_5(\bm{x})$ is slaved to $\bm{u}_4(\bm{x})$. The mode $\bm{u}_3(\bm{x})$ is usually defined as the shift mode from $\bm{u}_s(\bm{x})$ to the asymptotic mean flow field, $\bm{u}_3(\bm{x})\propto \bar{\bm{u}}(\bm{x})-\bm{u}_s(\bm{x})$, before being ortho-normalized to $\bm{u}_1(\bm{x})$ and $\bm{u}_2(\bm{x})$. Here, $\bar{\bm{u}}(\bm{x})$ will be restricted to the \textit{symmetric} mean flow field, associated with the statistically symmetric limit cycle, whether this limit cycle is stable or unstable. 
Similarly to $\bm{u}_4(\bm{x})$, mode $\bm{u}_5(\bm{x})$ is defined as
\begin{equation}
\label{Eqn:u5}
\bm{u}_5(\bm{x}) \propto (\bm{u}_s^+(\bm{x})+\bm{u}_s^-(\bm{x}))-2\bm{u}_s(\bm{x}),    
\end{equation}
\begin{figure}
\centerline{
 \includegraphics[width=.3\linewidth]{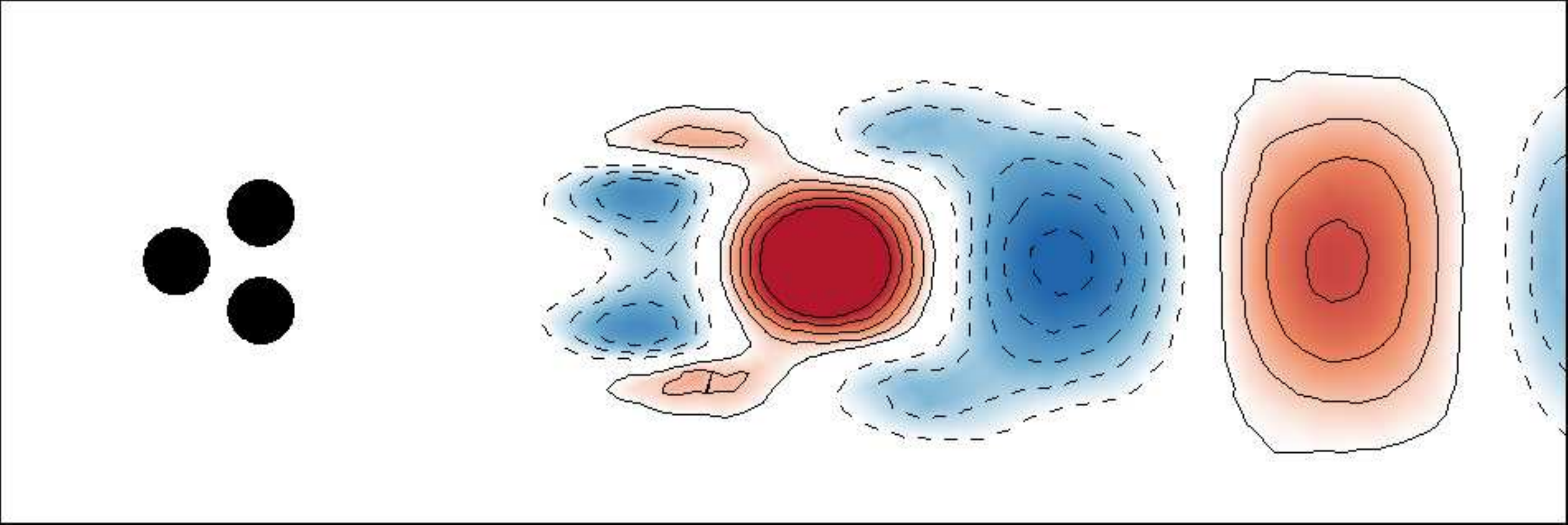} \hfill
 \includegraphics[width=.3\linewidth]{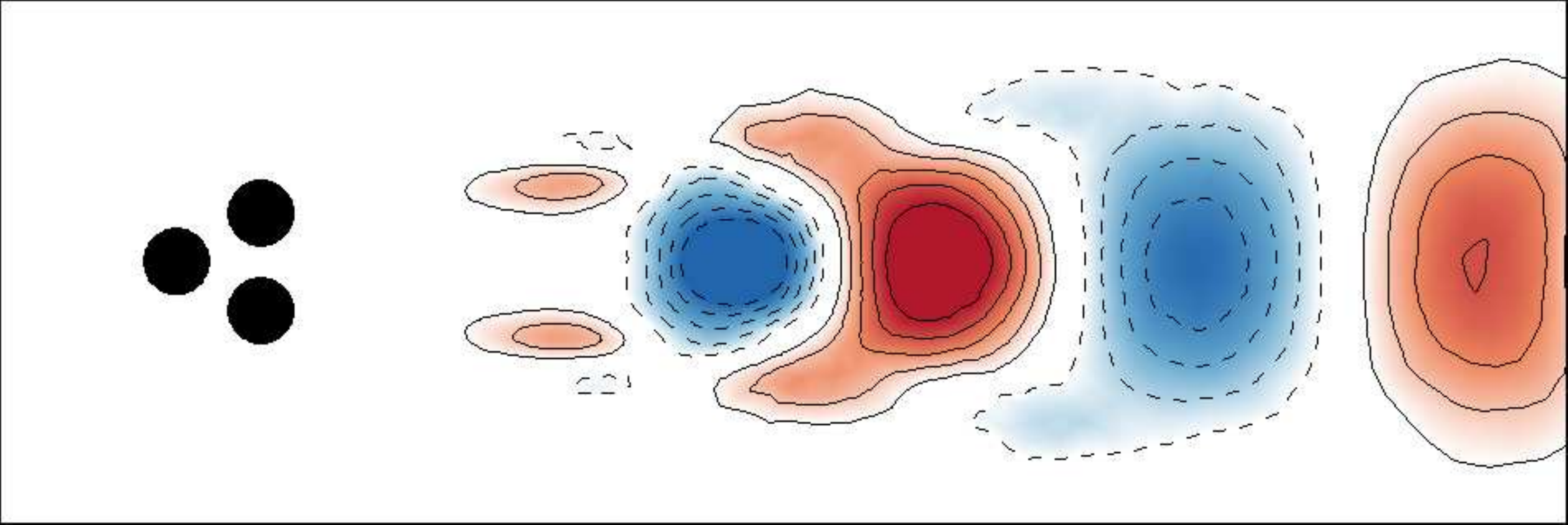} \hfill 
 \includegraphics[width=.3\linewidth]{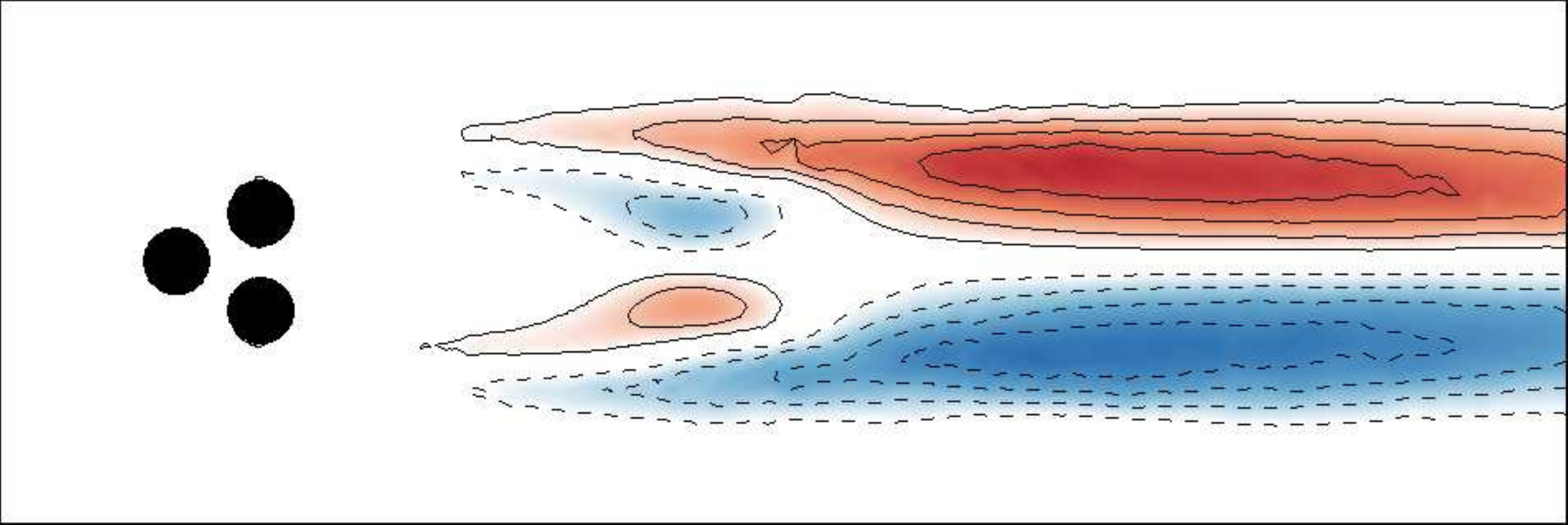} 
 }
\centerline{
 \includegraphics[width=.3\linewidth]{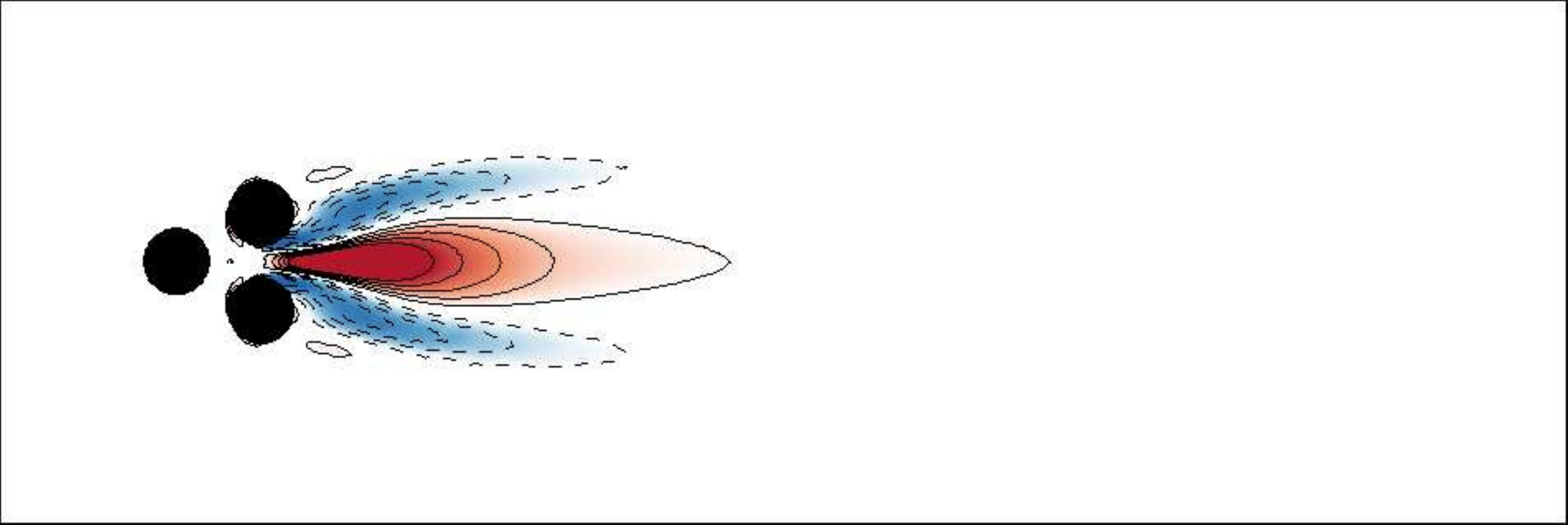} \quad 
 \includegraphics[width=.3\linewidth]{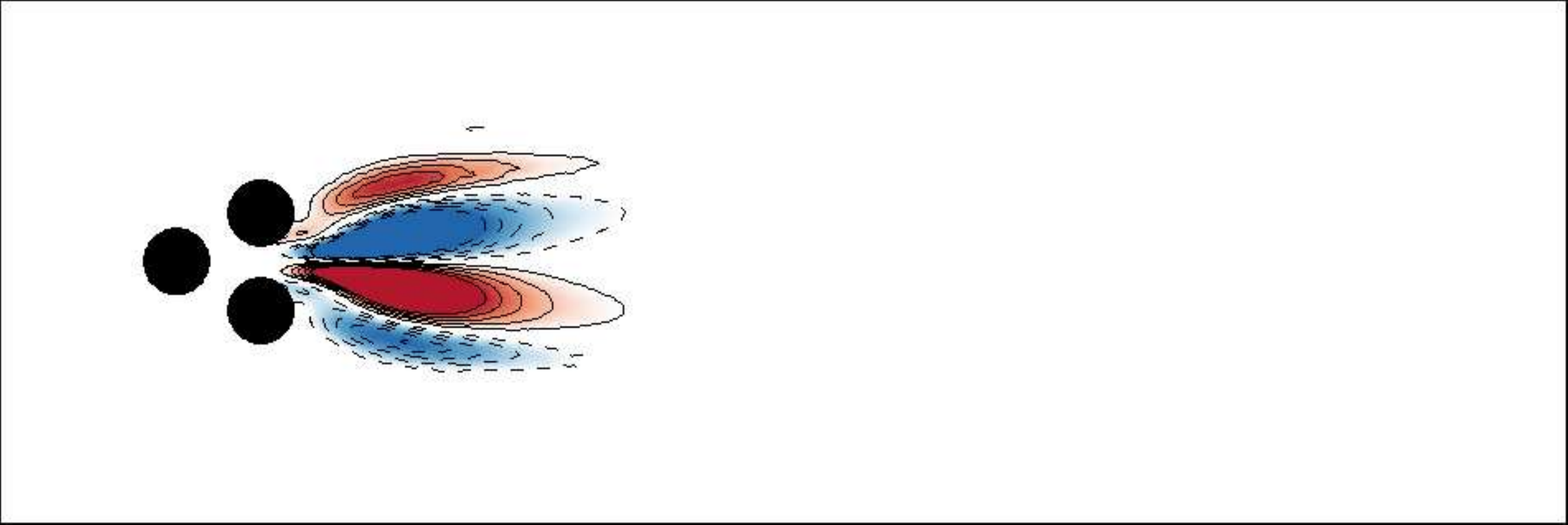} 
 }
\caption{Spatial structures of the modes $\bm{u}_1(\bm{x})$, $\bm{u}_2(\bm{x})$, $\bm{u}_3(\bm{x})$ (top), $\bm{u}_4(\bm{x})$, $\bm{u}_5(\bm{x})$ (bottom), of the velocity field associated with the five elementary degrees of freedom $\left\{a_1(t) - a_5(t)\right\}$, at $Re=80$. }
\label{Fig:ROM_5dof}
\end{figure}
These two modes, together with modes $\bm{u}_{1,2,3}$, are shown is figure \ref{Fig:ROM_5dof} after orthonormalization by a Gram-Schmidt procedure, and the corresponding time-dependent amplitudes $a_i(t)$, $i=1,\ldots,5$, in the full-flow dynamics are shown in figure \ref{Fig:Ampl_5dof} when starting from either the symmetric steady solution (figure \ref{Fig:Ampl_5dof}a) or the asymmetric steady solution (figure \ref{Fig:Ampl_5dof}b).

\section{Galerkin force model associated with the supercritical Hopf and pitchfork bifurcation}
\label{Sec:HopfPitchfork}

As already mentioned, the fluidic pinball undergoes a supercritical Hopf bifurcation at $Re=Re_{\rm HP}$ and a subsequent supercritical pitchfork bifurcation at $Re=Re_{\rm PF}>Re_{\rm HP}$. The Galerkin force models are derived for the supercritical Hopf bifurcation in \S~\ref{Sec:Hopf} and for the supercritical pitchfork bifurcation in \S~\ref{Sec:pitchfork}. 

\subsection{Force model associated with the supercritical Hopf bifurcation}
\label{Sec:Hopf}

The symmetric steady solution $\bm{u}_s \in \mathcal{U}^s$ is stable at low Reynolds numbers. At $Re \ge  Re_{\rm HP}$, it undergoes a supercritical Hopf bifurcation. 
The resulting Galerkin expansion reads
\begin{equation}
\label{Eqn:MFAnsatz:Hopf}
\bm{u} ( \bm{x},t ) = \bm{u}_s ( \bm{x} ) 
                      + \underbrace{a_1(t) \> \bm{u}_1(\bm{x}) + a_2(t) \> \bm{u}_2(\bm{x})}_{ \bm{u}^{\prime}}
                      + \underbrace{a_3(t) \> \bm{u}_3(\bm{x})}_{\bm{u}_{\Delta}},
\end{equation}
and the corresponding mean-field Galerkin system
\begin{subequations}
\label{Eqn:MFSystem:Hopf:Antisymmetric}
\begin{eqnarray}
   &da_1/dt &= \sigma a_1 - \omega a_2,
\\ &da_2/dt &= \sigma a_2 + \omega a_1, 
\\ &da_3/dt &= \sigma_3 a_3 + \beta_3 \left( a_1^2 + a_2^2 \right),
\end{eqnarray}
\end{subequations}
with $\sigma = \sigma_1 - \beta a_3$ and $\omega = \omega_1 + \gamma a_3$, where $\sigma_1$ and $\omega_1$ are the initial growth rate and frequency depending on the Reynolds number. For a direct supercritical Hopf bifurcation, $\sigma_1, \omega_1, \beta > 0$, $\sigma_3 < 0$ and $\beta_3>0$. We refer to \citet{deng2020jfm} for details.

Introducing \eqref{Eqn:MFAnsatz:Hopf} in equations \eqref{Eqn:ViscousForce} and \eqref{Eqn:PressureForce}, the total force can be written as \eqref{Eqn:ForceExpansion} with $N=3$ degrees of freedom. From symmetry considerations, as $\bm{u}_{1,2} \in \mathcal{U}^a$ and $\bm{u}_{0,3} \in \mathcal{U}^s$, the coefficients  $l_{x; 1}$, $l_{x; 2}$, $q_{x; 13}$, $q_{x; 23}$, $l_{y; 0}$, $l_{y; 3}$, $q_{y; 11}$, $q_{y; 12}$, $q_{y; 22}$, $q_{y; 33}$ are vanishing. Finally, the drag formulae \eqref{Eqn:cdcl2} simplify to
\begin{subequations}
\label{Eqn:cdcl-Hopf}
\begin{eqnarray}
&C_D &= C_D^\circ + l_{x; 3}\ a_3 + q_{x; 1 1}\ a_1^2 + q_{x; 1 2}\ a_1 a_2 + q_{x; 2 2}\ a_2^2 + q_{x; 3 3}\ a_3^2, \\
&C_L &= l_{y; 1}\ a_1 + l_{y; 2}\ a_2 + q_{y; 1 3}\ a_1 a_3 + q_{y; 2 3}\ a_2 a_3.
\end{eqnarray}
\end{subequations}

Here again, $C_D^\circ$ is the drag coefficient associated with the symmetric steady solution 
The unknown parameters in the force model can be identified by a least-squares approach, according to the known force dynamics and the relevant mode amplitudes. However, for the mean-field Galerkin system 
\eqref{Eqn:MFSystem:Hopf:Antisymmetric}, the slaving relation between the degree of freedom $a_3$ to the oscillating degrees of freedom $a_{1}$, $a_{2}$ imposes an additional sparsity in the force model. We employ the SINDy (Sparse Identification of Nonlinear Dynamics) algorithm \citep{Brunton2016pnas} to arrive at simpler and more interpretable models. A $L1$-regularization can be introduced in the LASSO (least absolute shrinkage and selection operator) regression process. Another option in the SINDy algorithm is the sequential thresholded least squares regression, which iteratively applies the least squares regression and eliminates terms with weight smaller than a given threshold.   
Both regression algorithms benefit from simplicity, only requiring one sparsity parameter $\lambda$. 
The optimal $\lambda$ balances the accuracy and complexity of the identified model.
To evaluate the performance of the identified model, the complexity is presented with the number of non-zero coefficients and the accuracy by the coefficient of determination,denoted as the $r^2\, score$ \citep{draper1998book}.
A detailed review of this sparsity parameter can be found in \citet{Loiseau2018Galerkin}.
A recent extension of the SINDy algorithm with physical constraints of energy-preserving quadratic nonlinearities successfully identifies the sparse model, benefiting from the Galerkin projection of the Navier-Stokes equations \citep{Loiseau2018jfm}.

\begin{figure}
\centering
\includegraphics[width=.45\linewidth]{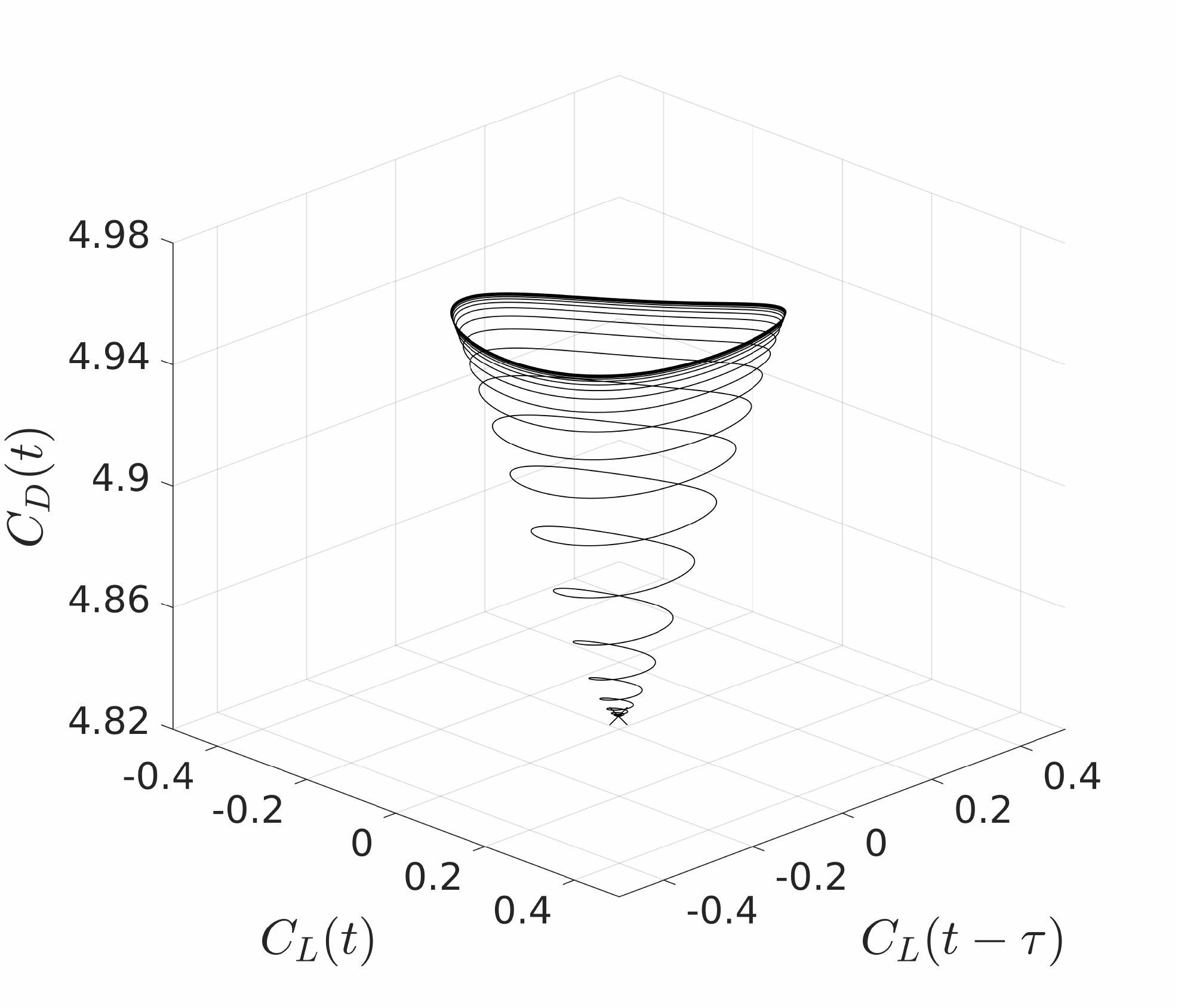} 
\caption{Transient dynamics from the unstable symmetric steady solution $\bm{u}_s$ ($\times$) to the asymptotic limit cycle (statistically symmetric vortex shedding), at $Re=30$, in the time-delayed embedding space of the lift $C_L$ and drag $C_D$ coefficients, with $\tau =2$. }
\label{Fig:scheme30}
\end{figure}

The LASSO algorithm is applied to a scenario starting with the unstable symmetric steady solution at $Re=30$. The training data used for the sparse regression is provided by the force coefficients and the mode amplitudes from the DNS starting with the symmetric steady solution to the final asymptotic regime.
The resulting transient dynamics and the asymptotically attracting limit cycle are shown in the three-dimensional space of the time-delayed coordinates of $C_L$ and $C_D$ in figure \ref{Fig:scheme30}.

The possible over-fitting terms, such as the slaving relation between $a_3$ and $a_1^2, a_2^2$, can be suppressed with a larger $L1$-penalty parameter for the LASSO algorithm. 
The choice of the $L1$-penalty parameter drives the sparsity of the identified model. 
A too small $L1$ will lead to a complex model with few eliminated terms; on the contrary, a too-large $L1$ can jeopardize accuracy. 
Both cases weaken the robustness of the identified model,and the same is observed for the sequential thresholded least squares regression. The influence of the sparsity parameter $\lambda$ and the comparison of these two regression methods are presented in Appendix~\ref{Sec:TwoRegressionMethods}. 

\begin{figure}
\centerline{
 \begin{tabular}{cc}
(a) & (b)\\
\includegraphics[width=.45\linewidth]{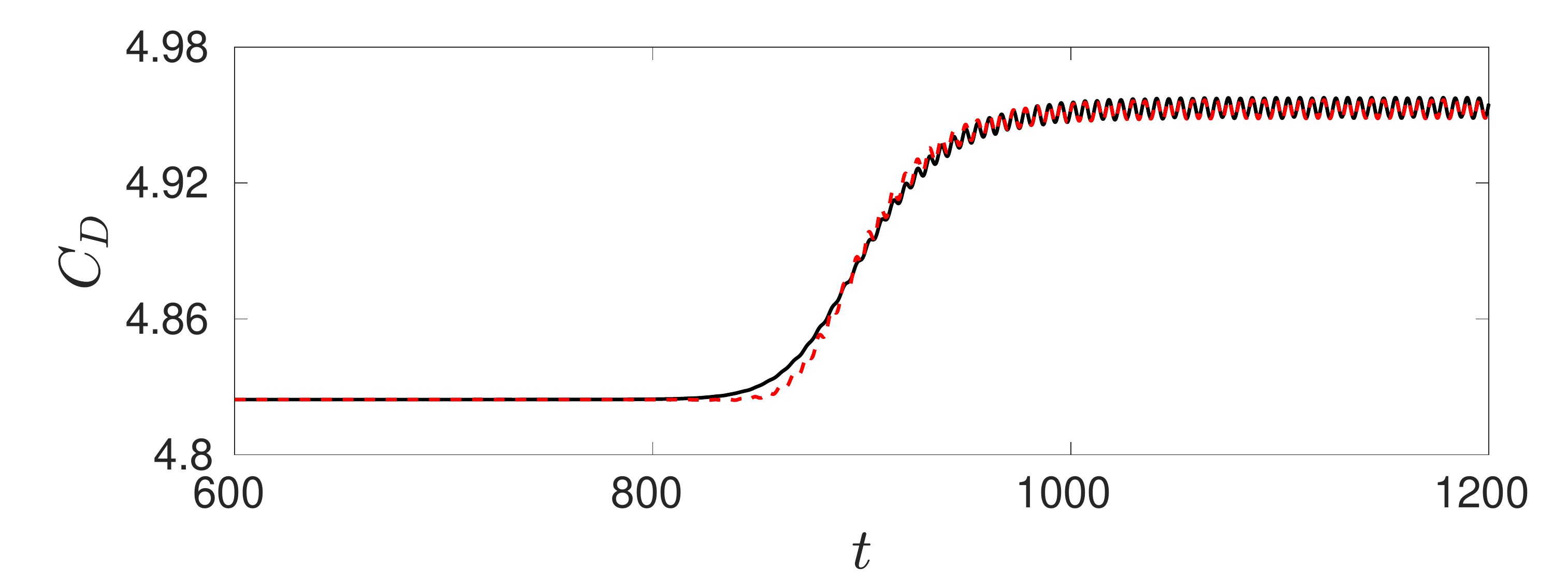} 
& \includegraphics[width=.45\linewidth]{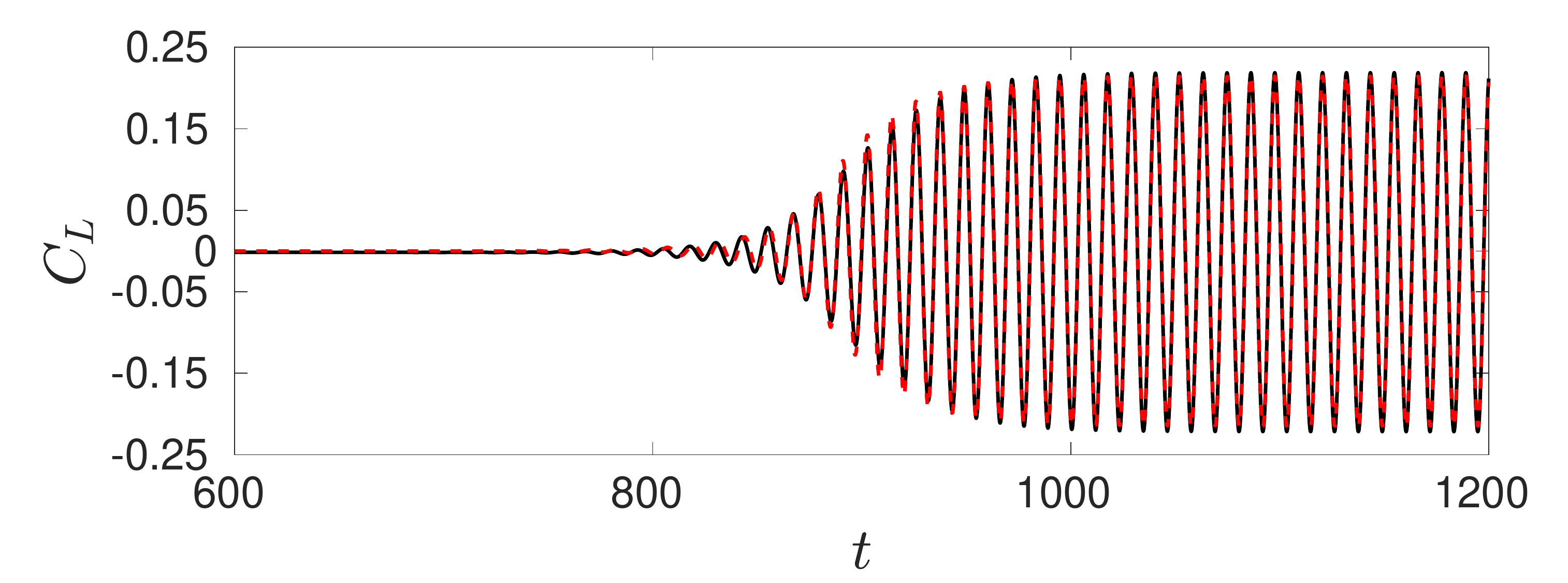} 
 \end{tabular}
}
\caption{Performance of the force model with the three elementary modes of the Hopf bifurcation. Time evolution of the drag $C_D$ (a) and lift $C_L$ (b) coefficients, in the full flow dynamics (solid black line) and for the force model (red dashed line), at $Re=30$. Initial condition: symmetric steady solution.  }
\label{Fig:ForceModel-Hopf}
\end{figure}

Gradually increasing the $L1$-penalty from $0$ to nearly $1$, the terms $a_1 a_2$, $a_3$, $a_2^2$, $a_1^2$ are eliminated subsequently in the drag model, while $a_3^2$ is always retained. 
The sparsity parameter $\lambda$, here the $L1$-penalty, is chosen as the largest value without any known over-fitting term. 
Hence, according to the order of elimination, $a_3$ is the over-fitting term in the drag model due to the slaving relation between $a_3$ and $a_1^2, a_2^2$. 
The details of this choice can be found in Appendix~\ref{Sec:TwoRegressionMethods}.
Finally, the identified force model reads
\begin{subequations}
\begin{eqnarray}
&C_D &= 4.82440448 \quad\> - 0.00037484 \> a_1^2 - 0.00098337 \> a_2^2 \>\>+ 0.01777408 \> a_3^2, \label{Eqn:forcemodel_3dof_a} \\
&C_L &= 0.00867623 \> a_1 + 0.01397362 \> a_2 + 0.0166239 \> a_1 a_3 - 0.01302317 \> a_2 a_3. \label{Eqn:forcemodel_3dof_b}
\end{eqnarray}
\label{Eqn:forcemodel_3dof}
\end{subequations}
The force model is highly accurate 
as corrobororated  by the $r^2$ scores of $0.9991$ and $0.9942$ for the drag and lift formulae, respectively.
As shown in figure \ref{Fig:ForceModel-Hopf}, the dynamics of the force model compares well with the real force transient dynamics, starting from the symmetric steady solution at $Re=30$.

In the drag model \eqref{Eqn:forcemodel_3dof_a}, the coefficient of $a_3$ is vanishing. Mode $\bm{u}_3$ actually contributes to the increase of the drag through $a_3^2$, as evidenced by the positive coefficient of the $a_3^2$ term.
This is an interesting result, since the effect of the bifurcation mode $\bm{u}_3$ is to decrease the length of the recirculation bubble in the $T$-averaged flow field $\bar{\bm{u}}_T(\bm{x},t)\approx \bm{u}_s(\bm{x})+a_3(t)\bm{u}_3(\bm{x})$, resulting in an increase of the drag through the quadratic term $a_3^2$. This quadratic dependency is also reported in \cite{Loiseau2018jfm}. 

It is also worth noticing that $a_3^2$ contributes to the mean value of $C_D$ while $a_1^2, a_2^2$ accounts for the instantaneous oscillations of $C_D$, as $C_D$ oscillates at twice the vortex shedding frequency. 
For $C_L$, the oscillatory pair $(a_1 , a_2)$ fits well with the phase of the initial transient part, while the pair $(a_1 a_3 , a_2 a_3)$ resolves the phase dependency of the post-transient part of the dynamics.

\subsection{Force model associated with the supercritical pitchfork bifurcation}
\label{Sec:pitchfork}

Next, we consider the supercritical pitchfork bifurcation, which breaks the symmetry of the symmetric steady solution $\bf{u}_s$ at $Re \ge  Re_{\rm PF}$. In this case the antisymmetric mode $\bm{u}_{4}$ describes the antisymmetric instability, which corresponds to an unstable eigenmode with a real eigenvalue. 
The resulting Galerkin expansion reads
\begin{equation}
\label{Eqn:MFAnsatz:Pitchfork}
\bm{u} ( \bm{x},t ) = \bm{u}_s ( \bm{x} ) 
                      + \underbrace{a_4(t) \bm{u}_4(\bm{x})}_{ \bm{u}^{\prime}}
                      + \underbrace{a_5(t) \bm{u}_5(\bm{x})}_{\bm{u}_{\Delta}},
\end{equation}
and the corresponding mean-field Galerkin system
\begin{subequations}
\label{Eqn:MFSystem:Pitchfork}
\begin{eqnarray}
   da_4/dt &=& \sigma_4 a_4 - \beta_4 a_4 a_5,
\label{Eqn:MFSystem:Pitchfork:Antisymmetric}
\\ da_5/dt &=& \sigma_5 a_5 + \beta_5 a_4^2,
\label{Eqn:MFSystem:Pitchfork:Symmetric}
\end{eqnarray}
\end{subequations}
where $\sigma_4$ is the positive initial growth rate, which depends on the Reynolds number. For a direct supercritical pitchfork bifurcation, $\sigma_4, \beta_4 > 0$, $\sigma_5 < 0$ and $\beta_5>0$, see \citet{deng2020jfm} for details.

Substituting \eqref{Eqn:MFAnsatz:Pitchfork} in equations \eqref{Eqn:ViscousForce} and \eqref{Eqn:PressureForce}, with $N=2$ in \eqref{Eqn:ForceExpansion}, and with $\bm{u}_{4} \in \mathcal{U}^a$ and $\bm{u}_s, \bm{u}_5 \in \mathcal{U}^s$, the force model becomes
\begin{subequations}
\label{Eqn:cdcl-pitchfork}
\begin{eqnarray}
&C_D &= C_D^\circ + l_{x; 5}\ a_5 + q_{x; 4 4}\ a_4^2 + q_{x; 5 5}\ a_5^2, \label{Eqn:cdcl-pitchfork_a} \\
&C_L &= l_{y; 4}\ a_4 + q_{y; 4 5}\ a_4 a_5. \label{Eqn:cdcl-pitchfork_b}
\end{eqnarray}
\end{subequations}
Five parameters, namely $l_{x; 0}$, $l_{x; 5}$, $q_{x; 44}$, $q_{x; 55}$, $l_{y; 4}$, $q_{y; 45}$ need to be identified. 

\begin{figure}
\centering
\includegraphics[width=.45\linewidth]{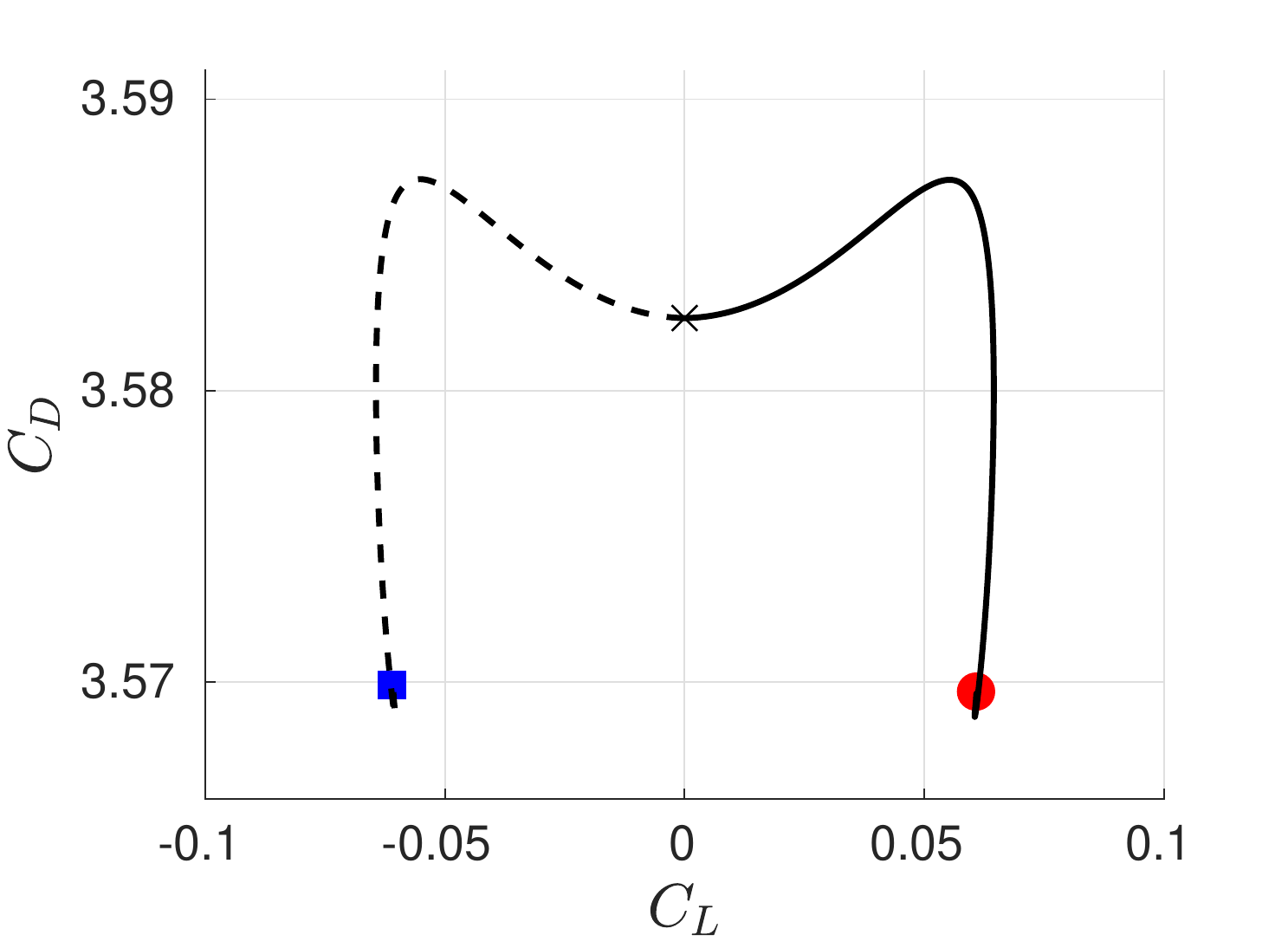}
\caption{Transient trajectories (solid and dashed lines) starting from two initial conditions close to the symmetric steady solution, at $Re=100$. Asymmetric steady solution $\bm{u}_s^+$ ($\bullet$), asymmetric steady solution $\bm{u}_s^-$ ($\blacksquare$).}
\label{Fig:scheme100t700}
\end{figure}

\begin{figure}
\centerline{
 \begin{tabular}{cc}
(a) & (b)\\
\includegraphics[width=.45\linewidth]{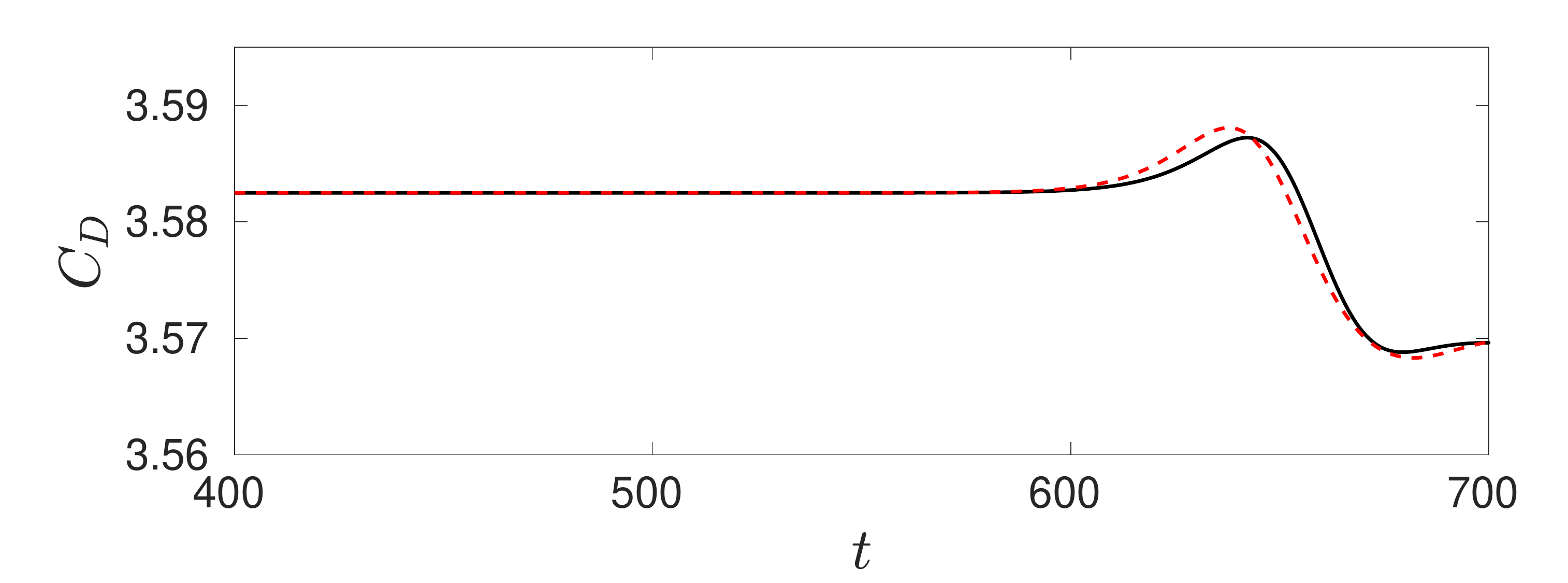} & 
\includegraphics[width=.45\linewidth]{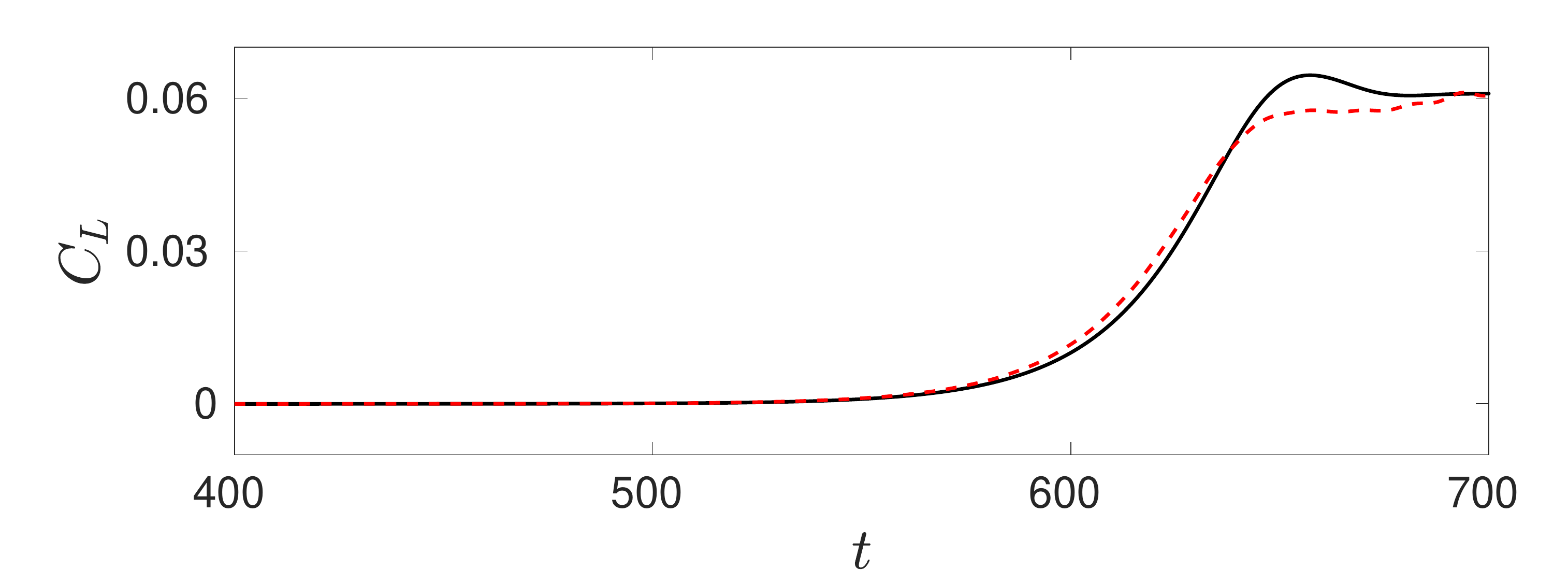} 
 \end{tabular}
}
\caption{Performance of the force model with the two elementary modes of the pitchfork bifurcation. Time evolution of the drag $C_D$ (a) and lift $C_L$ (b)  coefficients in the full flow dynamics (solid black line) and for the force model (red dashed line), at $Re=100$. Initial condition: symmetric steady solution.}
\label{Fig:ForceModel-PF}
\end{figure}

In the fluidic pinball, the pitchfork bifurcation occurs after the primary Hopf bifurcation as the Reynolds number is increased. However, the transient dynamics observed at $Re=100$, when starting close to the symmetric steady solution, first exhibits the static symmetry breaking, which is typical of the pitchfork bifurcation, before developing the cyclic release of vortices, which is characteristic of the Hopf bifurcation. The early stage of the transient dynamics, starting from the symmetric steady solution and evolving toward one of the asymmetric steady solutions, is shown in figure~\ref{Fig:scheme100t700}. The time evolutions of the lift $C_L(t)$ and drag $C_D(t)$ coefficients are shown in figure~\ref{Fig:ForceModel-PF}.  

Only the degrees of freedom associated with the pitchfork bifurcation are active in this early stage of the transient dynamics, as also shown in figure~\ref{Fig:Ampl_5dof}(a). The degrees of freedom associated with the Hopf bifurcation will only become active further in time during the transient dynamics, which will be further discussed in \S~\ref{Sec:Validation}.
Accordingly,  a force model is derived 
for the transition after a simple pitchfork bifurcation.
The training data are the lift $C_L(t)$ and drag $C_D(t)$ coefficients 
and the relevant mode amplitudes in Eq.~\eqref{Eqn:cdcl-pitchfork} 
from the early to final stage of the transient dynamics.
The observed slaving  of $a_5$ in $a_4^2$ may reduce the robustness of the identified model. 
Gradually increasing the $L1$-penalty parameter in the LASSO regression, the optimized force model reads
\begin{subequations}
\begin{eqnarray}
&C_D &= 3.58248992 + 0.04367604 \>a_5 - 0.08525184 \> a_5^2, \label{Eqn:forcemodel_2dof_a} \\
&C_L &= -0.13611053 \> a_4 + 0.09194312 \> a_4 a_5, \label{Eqn:forcemodel_2dof_b}
\end{eqnarray}
\label{Eqn:forcemodel_2dof}
\end{subequations}
with $r^2 =0.9949$ for the drag model and $r^2 =0.9992$ for the lift model. 
The over-fitting term $a_4^2$ has been eliminated in the sparse formula of the drag force. 
Note that the mode $\bm{u}_5$ contributes to the drag through $a_5$, while $a_5^2$ acts in decreasing the drag, as indicated by the sign of their associated coefficients in Eq.~\eqref{Eqn:forcemodel_2dof_a}.

Figure~\ref{Fig:ForceModel-PF} compares the evolution of the drag and lift coefficients in the full flow dynamics (solid black line) to their prediction by the force model \eqref{Eqn:forcemodel_2dof} (red dashed curve), during the early stage of the transient dynamics at $Re=100$. The derived force model is well aligned with the real force dynamics using only two active degrees of freedom of the pitchfork bifurcation in the dynamics of the system.

\section{Galerkin force model for multiple invariant sets}
\label{Sec:Results}

We focus on the regime after the pitchfork bifurcation $Re \ge Re_{\rm PF}=68$
and before the quasi-periodic behaviour $Re \le Re_{\rm QP}=104$.
This flow has 6 invariant sets:
3 unstable fixed points, 2 stable asymmetric mirror-conjugated periodic orbits, and one meta-stable symmetric limit cycle.
 \S\,\ref{Sec:ForceModel5DOF} investigates the dynamics of the fluidic pinball at 
$Re=80$, when the degrees of freedom associated with the Hopf bifurcation are first activated before the degrees of freedom associated with the pitchfork bifurcation. The predictive power of the force model is assessed in \S\,\ref{Sec:PredictivePower}. \S\,\ref{Sec:ForceModel7DOF} introduces two additional degrees of freedom in the force model, in order to take into account the distortion of the shift mode when the attractor is reached. The robustness of the force model is emphasized in \S\,\ref{Sec:Validation} by considering the flow dynamics at $Re=100$, where the pitchfork degrees of freedom are activated before the Hopf degrees of freedom during the transient dynamics. 

\subsection{Force model at $Re=80$}
\label{Sec:ForceModel5DOF}
\begin{figure}
\centering
\includegraphics[width=.45\linewidth]{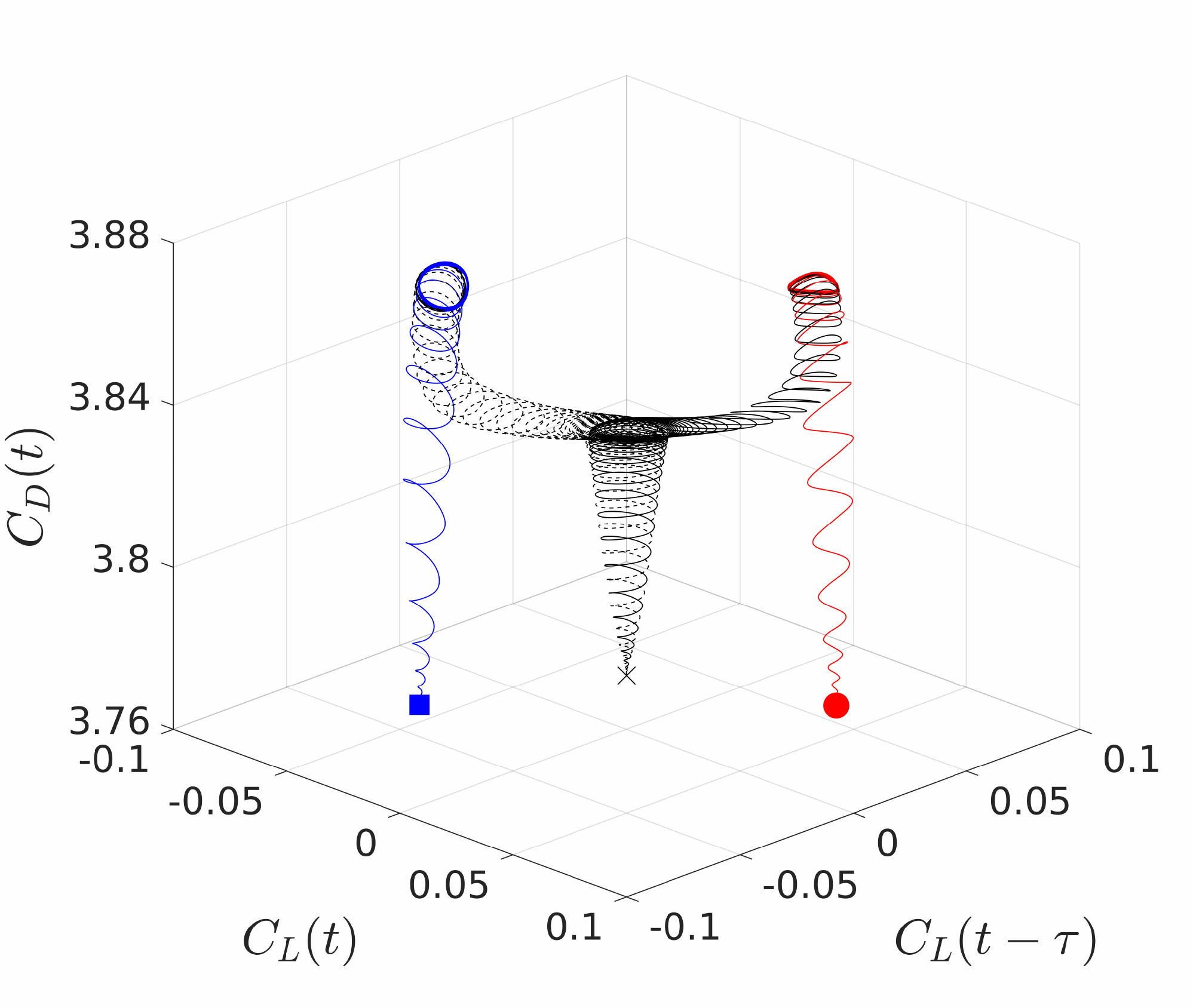}\\
\caption{Trajectories in the time-delayed embedding space of the lift $C_L$ and drag $C_D$ coefficients, with $\tau =2$, at $Re=80$. Black trajectories starting close to the symmetric steady solution $\bm{u}_s$ ($\times$); red trajectory starting close to the asymmetric steady solution $\bm{u}_s^+$ ($\bullet$), blue trajectory starting close to 
the asymmetric steady solution $\bm{u}_s^-$ ($\blacksquare$).}
\label{Fig:scheme80}
\end{figure}

\begin{figure}
\centerline{
 \begin{tabular}{ccc}
 &$C_D$ &  $C_L$\\
 \hline \hline
(a) &  & \\
 & \includegraphics[width=.45\linewidth]{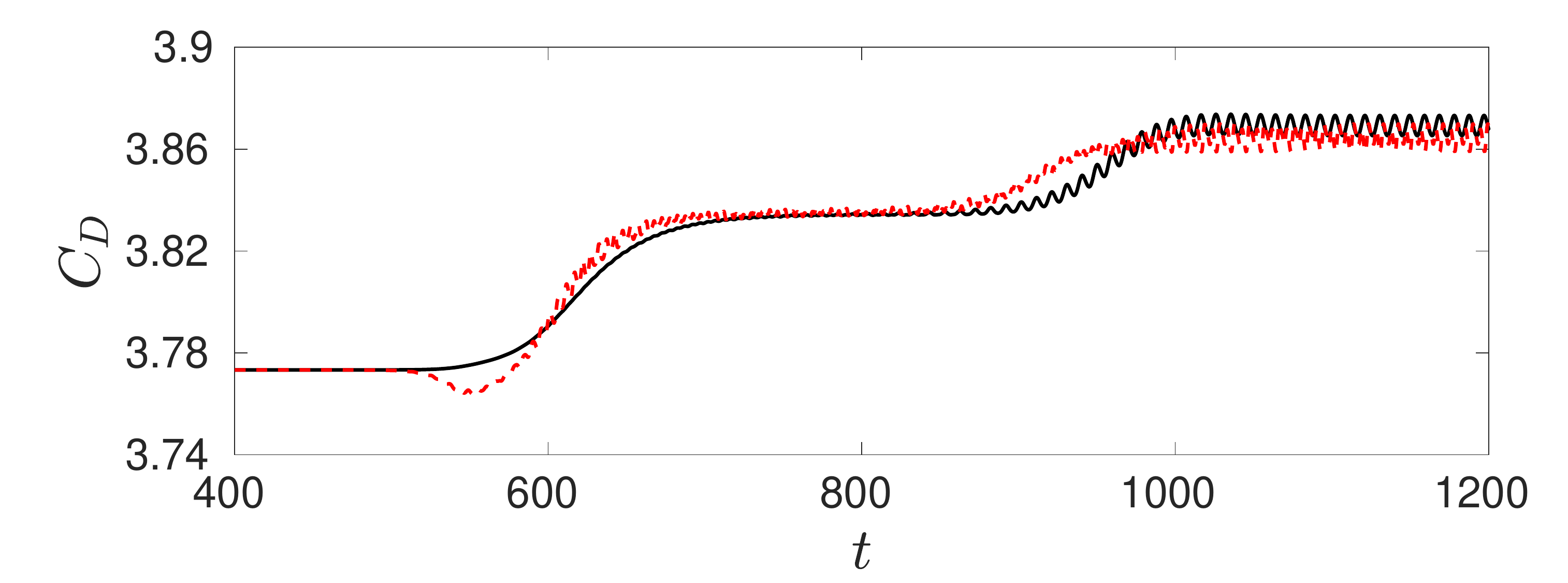} 
 & \includegraphics[width=.45\linewidth]{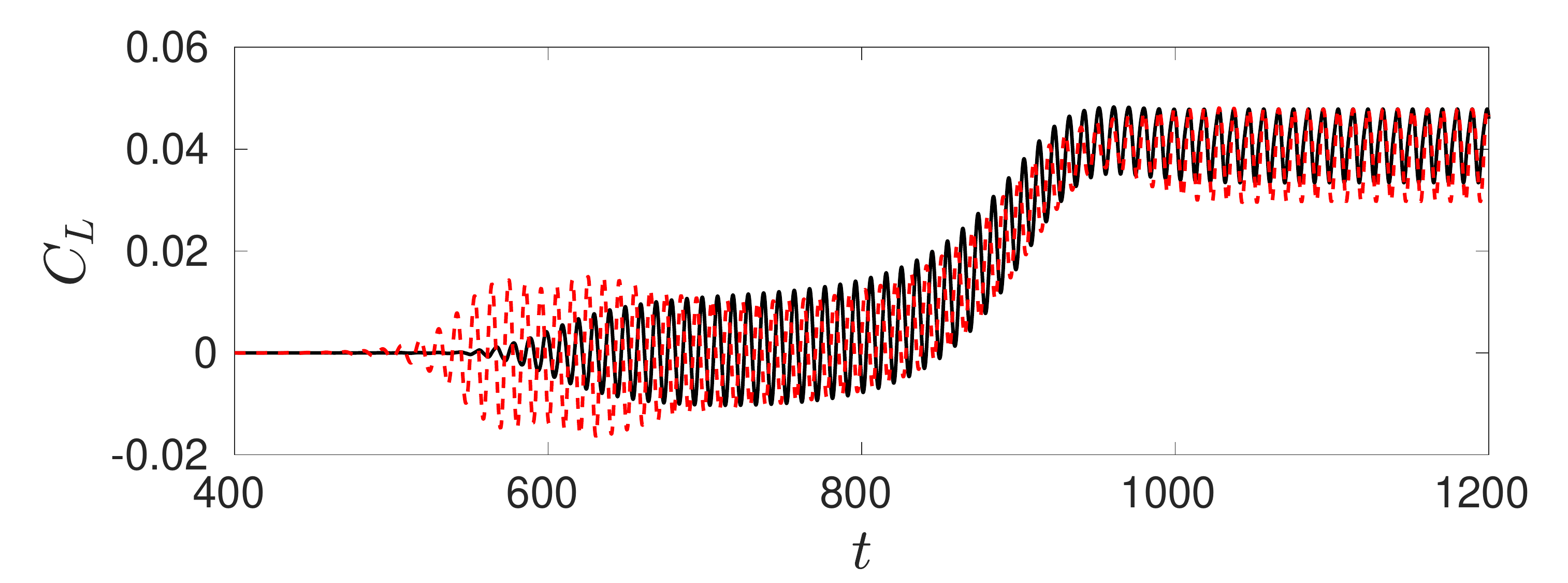} \\ 
(b) &  & \\
 & \includegraphics[width=.45\linewidth]{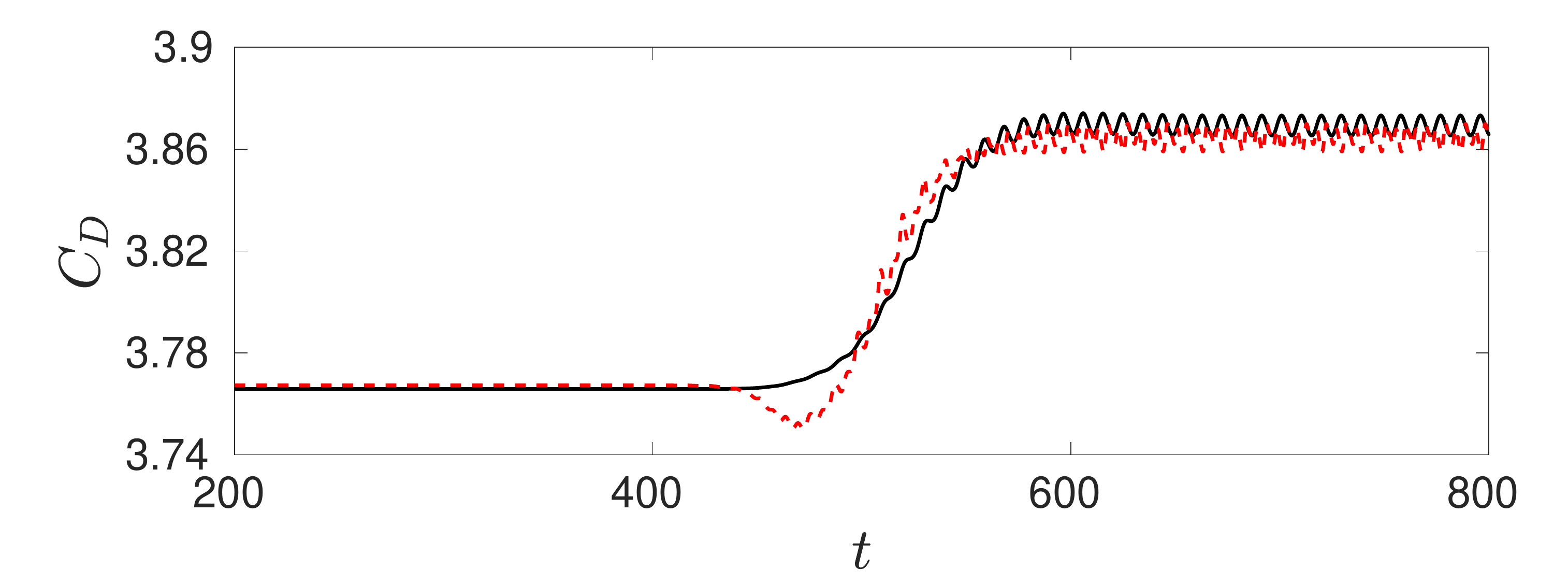} 
 & \includegraphics[width=.45\linewidth]{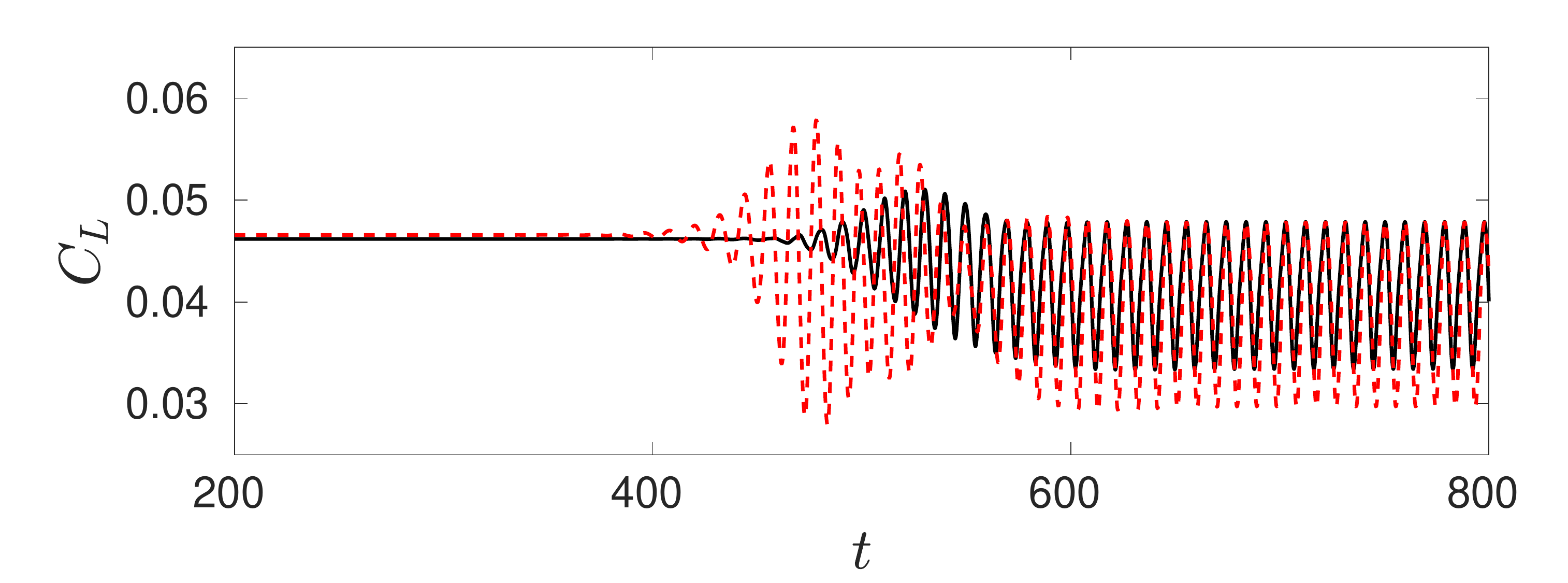} \\ 
 \end{tabular}
}
\caption{Performance of the force model with the five elementary modes. Time evolution of the drag $C_D$ (left) and the lift $C_L$ (right) coefficients in the full flow dynamics (solid black line) and for the force model (red dashed line), at $Re=80$. Initial condition: (a) symmetric steady solution, (b) asymmetric steady solution.}
\label{Fig:ForceTrans_Re80}
\end{figure}

At $Re=80$, the system has already undergone a supercritical Hopf bifurcation and a supercritical pitchfork bifurcation.  
The trajectories issued from $\bm{u}_s$ and $\bm{u}_s^\pm $ are shown in the time-delayed embedding state space $(C_L(t),C_L(t-\tau),C_D(t))$ of figure \ref{Fig:scheme80}. 
The force model will rely on five degrees of freedom at minimum, namely the three degrees of freedom associated with the Hopf bifurcation $a_i$, $i=1,2,3$ 
and the two degrees of freedom $a_i$, $i=4,5$, associated with the pitchfork bifurcation. As a generalization of \eqref{Eqn:cdcl-Hopf} and \eqref{Eqn:cdcl-pitchfork}, the force model reads
\begin{subequations}
\label{Eqn:cdcl-Hopf-pitchfork}
\begin{eqnarray}
C_D & =  C_D^\circ& + l_{x; 3}\ a_3 + q_{x; 1 1}\ a_1^2 + q_{x; 1 2}\ a_1 a_2 + q_{x; 2 2}\ a_2^2 + q_{x; 3 3}\ a_3^2  \\
\notag & & + l_{x; 5}\ a_5 + q_{x; 4 4}\ a_4^2 + q_{x; 5 5}\ a_5^2 \\
\notag & & + q_{x; 1 4}\ a_1 a_4 + q_{x; 2 4}\ a_2 a_4 + q_{x; 3 5}\ a_3 a_5 ,\label{Eqn:cdcl-Hopf-pitchfork_a}\\
C_L & = \quad\quad & \>\>\>l_{y; 1}\ a_1 + l_{y; 2}\ a_2 + q_{y; 1 3}\ a_1 a_3 + q_{y; 2 3}\ a_2 a_3 \\ \notag & & + l_{y; 4}\ a_4 + q_{y; 4 5}\ a_4 a_5 \\
\notag & & + q_{y; 1 5}\ a_1 a_5 + q_{y; 2 5}\ a_2 a_5 + q_{y; 3 4}\ a_3 a_4. \label{Eqn:cdcl-Hopf-pitchfork_b}
\end{eqnarray}
\end{subequations}
Due to symmetry reasons, only 2 linear terms ($a_{3},a_{5}$) and 9 quadratic terms ($a_1^2$, $a_1a_2$, $a_1a_4$, $a_2^2$, $a_2a_4$, $a_3^2$, $a_3a_5$, $a_4^2$, $a_5^2$) are left in Eq.~\eqref{Eqn:cdcl-Hopf-pitchfork_a} for the drag coefficient. For the lift coefficient, only 3 linear terms, $a_{1}$, $a_{2}$, $a_{4}$, and 6 quadratic terms, $a_1a_3$, $a_1a_5$, $a_2a_3$, $a_2a_5$, $a_3a_4$, $a_4a_5$, are left in Eq.~\eqref{Eqn:cdcl-Hopf-pitchfork_b}. 
The training data is taken from the DNS starting from the three steady solutions, with the real force dynamics, see the black curves in figure~\ref{Fig:ForceTrans_Re80}, and the relevant mode amplitudes, see figure~\ref{Fig:Ampl_5dof}. 
The coefficients of the force models are identified by the sequential thresholded least-squares regression with the optimal sparsity parameter $\lambda$. 
We note that the LASSO regression can also be used here. See Appendix~\ref{Sec:TwoRegressionMethods} for the comparison of these two methods. 
The resulting force model reads
\begin{subequations}
\label{Eqn:forcemodel_5dof}
\begin{eqnarray}
\notag C_D &=& 3.77331204 + 0.05888312 \>a_5 - 0.01115552\> a_1^2 - 0.01088109 \>a_2^2\\
&& +  0.01323449 \>a_3^2 + 0.02949701\> a_3a_5  - 0.25910470\> a_5^2 ,\label{Eqn:forcemodel_5dof_a}\\
\notag C_L &=& \>\>\> 0.00953160 \>a_1 \>\>\>\>\> + 0.00720164 \>a_2  \>\>\>\>- 0.10179203 \> a_4 \\
\notag && - 0.00303677 \> a_1a_3 - 0.00197075 \> a_2 a_3 - 0.00200840 \> a_3a_4 \\
&& + 0.05914386 \> a_4a_5.\label{Eqn:forcemodel_5dof_b}
\end{eqnarray}
\end{subequations}
The good accuracy of the identified drag model can be determined from the high $r^2$ score of $0.9816$. The drag model of Eq.~\eqref{Eqn:forcemodel_5dof_a} preserves both the basic forms of the drag model for the Hopf and pitchfork bifurcations and the signs of the coefficients.
This indicates that the identified model is robust.
The only remaining cross-term $a_3a_5$ provides the coupling relation between the degrees of freedom associated with both bifurcations.

A robust sparse formula for the lift model is more difficult to derive, due to the oscillating dynamics of the lift and the fact that $a_4$ and $a_5$ also oscillate at the fundamental frequency. 
With respect to the basic lift model of two bifurcations, a balanced method is used here to solve the difficulty of the identification. 
Starting with a large $L1$-penalty, the derived under-fitted system can figure out the most elementary features of the dynamics,  eliminating $a_1a_5$, $a_2a_5$, $a_4a_5$. 
This is reasonable as $a_3$ is about ten times larger than $a_5$, which means that most of the mean-field distortion comes from $\bm{u}_3$.
However, if the $L1$-penalty is too large, the term $a_4a_5$ can disappear from the lift model, making the resulting model non-consistent with Eq.~\eqref{Eqn:forcemodel_5dof_b}. 
In order to balance sparsity and robustness, $a_4a_5$ needs to be reintroduced into the library. The sparse formula of the lift model in Eq.~\eqref{Eqn:forcemodel_5dof_b} is determined by least-squares regression, constraining the parameters of $a_1a_5$, $a_2a_5$ to zero. The $r^2$ score of the identified lift model is $0.9673$.

The identified force dynamics in Eq.~\eqref{Eqn:forcemodel_5dof} (dashed red line) is compared to the real force dynamics (solid black line) at $Re=80$ in figure \ref{Fig:ForceTrans_Re80}. 
The force model based on the least-order model can reproduce the main features of the real force dynamics. The drag model of Eq.~\eqref{Eqn:forcemodel_5dof_a} shows how the degrees of freedom of the Hopf ($a_1^2, a_2^2, a_3^2$) and pitchfork ($a_5, a_5^2$) bifurcations contribute to the drag force, as well as the coupling between these degrees of freedom ($a_3a_5$). The lift model of Eq.~\eqref{Eqn:forcemodel_5dof_b} shows that the lift oscillations occur through the coupling of the oscillating degrees of freedoms $a_{1}$, $a_{2}$ to $a_3$, while the coupling between the degrees of freedoms $a_4$ and $a_5$ contribute to the mean value of $C_L$. Hence, the mean lift coefficient can be simplified with fewer terms, as $\overline{C_L} = l_{y; 4}\ a_4 + q_{y; 4 5}\ a_4 a_5  + q_{y; 3 4}\ a_3 a_4$, which meets well with the Krylov-Bogoliubov assumption \citep{jordan1999nonlinear}.

\subsection{Assessing the predictive power of the force model}
\label{Sec:PredictivePower}

The time-evolution of the drag and lift coefficients in the fluidic pinball are shown in figure \ref{Fig:ForceTrans_Re80} as solid black lines. The evolutions of the drag and lift coefficients in the model  \eqref{Eqn:forcemodel_5dof} are shown with dashed red lines. 
The model reproduces correctly the time scales of the force dynamics as well as the transient and asymptotic amplitudes of the forces. 
However, it is observed that the fine details of the transient dynamics, at the early stage of the linear instability, are not satisfactorily reproduced in the identification process (figure \ref{Fig:ForceTrans_Re80}(a,b) at $t\approx 590 $ and 475 respectively). 
The ranges of time concerned, in both cases, are also associated with oscillations in $a_4$, as observed during the initial stage at $t\approx 590$ in figure \ref{Fig:Ampl_5dof}(left) and  $t\approx 475$ in  figure \ref{Fig:Ampl_5dof}(right). This strongly suggests that the oscillations of $a_4$ be triggered by the degrees of freedom associated with the Hopf bifurcation. This means that the  degrees of freedom of the pitchfork bifurcation are affected by the degrees of freedom of the Hopf bifurcation, at least when the distance from the bifurcation point is large enough, which is the case at $Re=80$.  

In addition, as recalled in \S~\ref{Sec:PinballBifurcationModes}, at $Re\approx 68$, both the steady symmetric solution and the symmetric-based limit cycle undergo a supercritical pitchfork bifurcation. We  emphasize that this coincidence of two local pitchfork bifurcations might not occur by chance, as mentioned in \citet{deng2020jfm}. As a result of these two simultaneous bifurcations, the degrees of freedom involved in the pitchfork bifurcation of the fixed point might not coincide with those involved in the pitchfork bifurcation of the limit cycle. For this reason, it is reasonable to introduce two distinct sets of degrees of freedom for each of them, namely $a_4$, $a_5$ at the fixed point and $a_6$, $a_7$ at the limit cycle. These two additional degrees of freedom will complete the mean-field model with more details and will take into account the mean-field distortion during the transition from the fixed point to the limit cycle. The new resulting mean-field Galerkin system, with seven degrees of freedom, is derived in appendix~\ref{Sec:MFM-7dof}, while the new resulting force model is discussed in the next subsection. 
%
\subsection{The need for additional modes}
\label{Sec:ForceModel7DOF}

All our attempts to smooth out the kicks observed at the beginning of the exponential growth, in both $C_D$ and $C_L$ in the frame of the force model \eqref{Eqn:forcemodel_5dof}, failed, even when over-fitting the model without any sparsity. This strongly indicates that five degrees of freedom might not be sufficient to account for the force evolution on the full-time range.  

Digging into this idea, it becomes manifest that the way $\bm{u}_{3}(\bm{x})$ is built, namely as the difference between the statistically symmetric mean flow field, associated with the unstable limit cycle, and the symmetric steady solution $\bm{u}_s(\bm{x})$, 
\begin{equation}
\label{Eqn:u3}
 \bm{u}_3(\bm{x}) = \bar{\bm{u}}_{T}(\bm{x},775) - \bm{u}_s(\bm{x}), 
\end{equation}
see figure \ref{Fig:ForceTrans_Re80}(a), does not allow to satisfactorily account for the complete dynamics of the lift and drag forces. 
This also indicates that $\bm{u}_3(\bm{x})$ gets \textit{distorted} when the system is evolving along the manifold, which connects the unstable limit cycle to one of the two conjugated stable limit cycles. In other words, the mean-field distortion on the attractors associated with the two asymmetric mean flow fields $\bar{\bm{u}}^\pm$, namely 
\begin{equation}
\label{Eqn:u3distorted}
\bm{u}_3^\pm(\bm{x}) = \bar{\bm{u}}^\pm(\bm{x}) - \bm{u}_s^\pm(\bm{x})
\end{equation}
do not coincide exactly with $\bm{u}_3(\bm{x})$. The asymmetric mean flow fields $\bar{\bm{u}}^\pm$ only focus on the post-transient dynamics, as shown in figure \ref{Fig:ForceTrans_Re80}(c), which can be expressed with $\bar{\bm{u}}_{T}^\pm(\bm{x},700)$.
The difference between $\bm{u}_3^\pm(\bm{x})$ and $\bm{u}_3(\bm{x})$ is asymmetric and can be decomposed into a symmetric and an anti-symmetric part, respectively $\bm{u}_6(\bm{x})$ and $\bm{u}_7 (\bm{x})$:
\begin{equation}
\label{Eqn:u3distortion}
 \bm{u}_3^\pm(\bm{x})-\bm{u}_3(\bm{x})=  \pm \bm{u}_6(\bm{x}) + \bm{u}_7(\bm{x}).
\end{equation}
As a result, the modes $\bm{u}_6(\bm{x})$ and $\bm{u}_7(\bm{x})$ can be defined as,
\begin{subequations}
\label{Eqn:u6u7}
\begin{eqnarray}
 \bm{u}_6 (\bm{x}) &\propto&  \bar{\bm{u}}^+ (\bm{x}) -  \bar{\bm{u}}^-(\bm{x}) ,\\
 \bm{u}_7 (\bm{x}) &\propto& (\bar{\bm{u}}^+ (\bm{x}) +  \bar{\bm{u}}^-(\bm{x})) - 2 \bar{\bm{u}}_T(\bm{x},775).
\end{eqnarray}
\end{subequations}

\begin{figure}
\centerline{
\begin{tabular}{cc}
(a) & (b) \\
 \includegraphics[width=.45\linewidth]{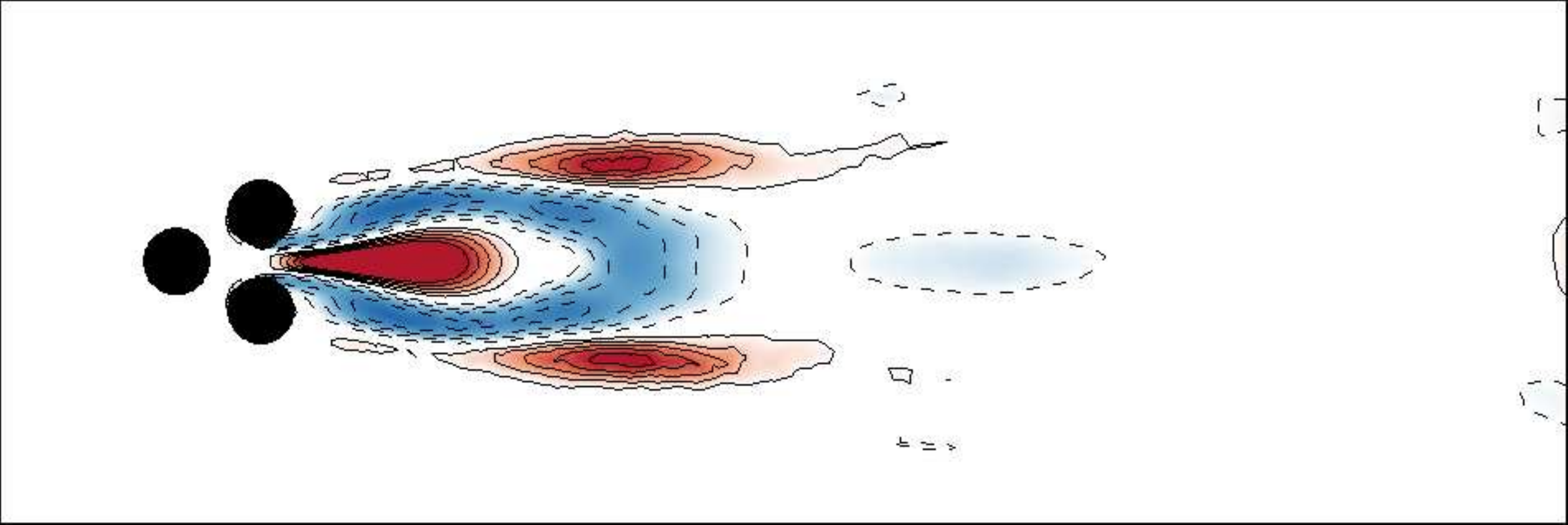} & 
 \includegraphics[width=.45\linewidth]{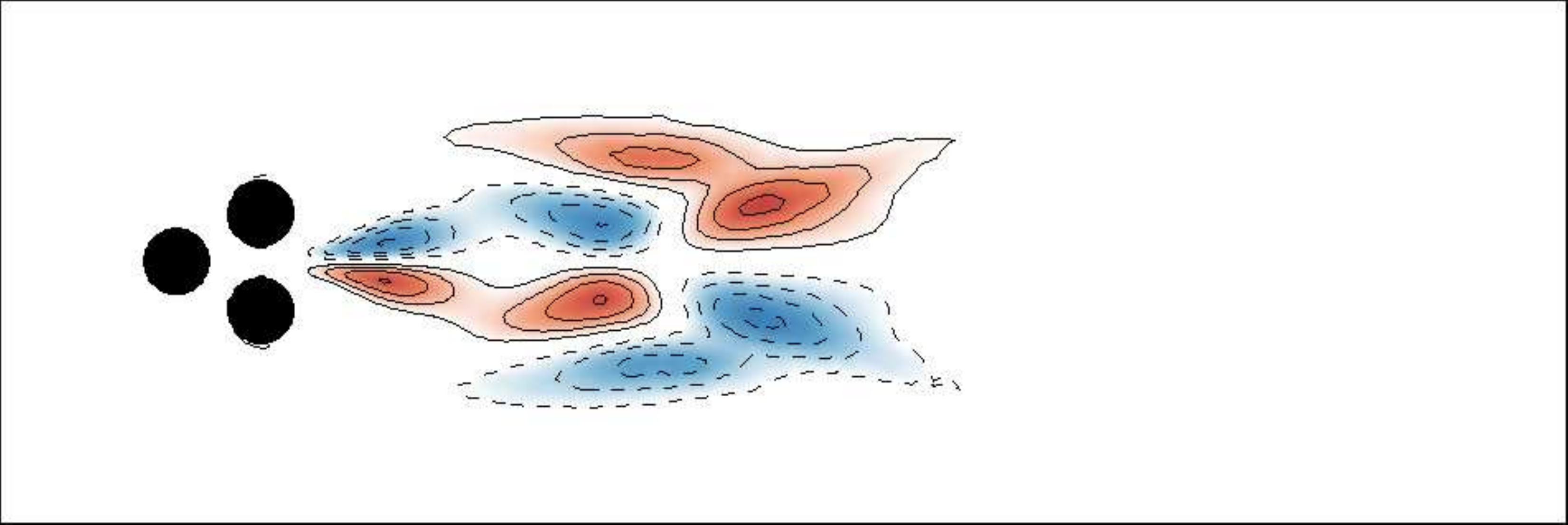}
\end{tabular}
}
\caption{Vortical structure (color) of the modes $\bm{u}_6(\bm{x})$ (a), $\bm{u}_7(\bm{x})$ (b), at $Re=80$. }
\label{Fig:Correctionmodes}
\end{figure}

\begin{figure}
\centerline{
\begin{tabular}{cc}
(a) & (b) \\
 \includegraphics[width=.5\linewidth]{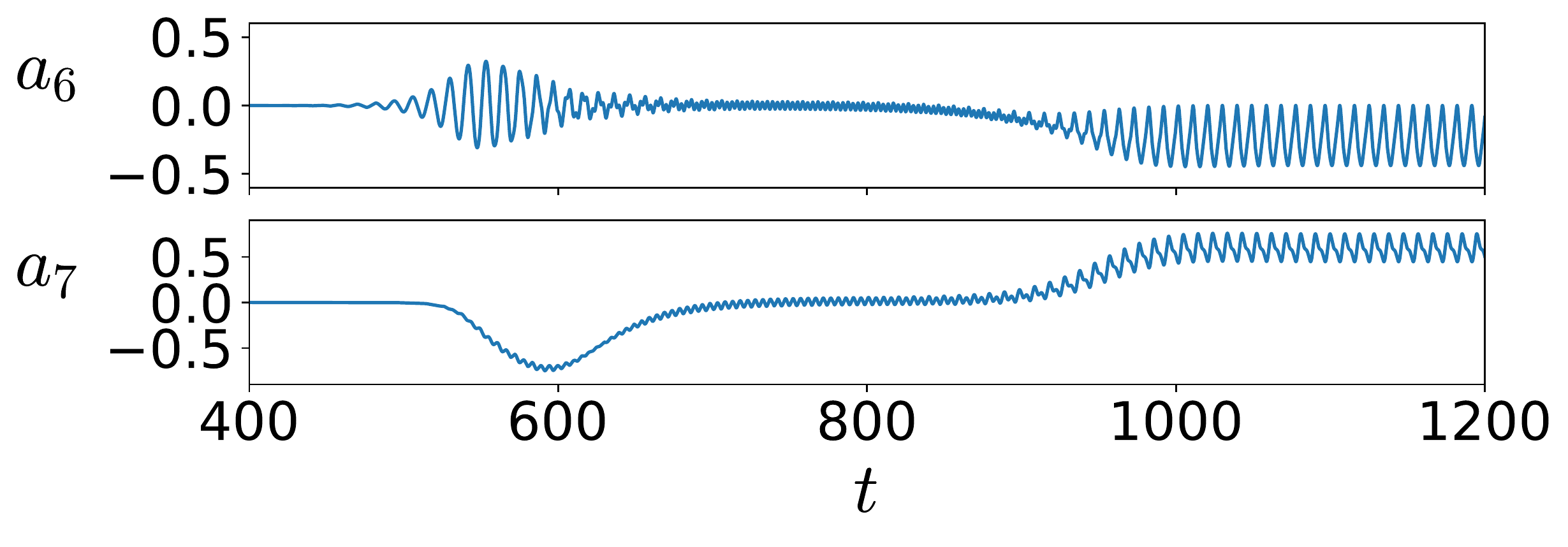} &
 \includegraphics[width=.5\linewidth]{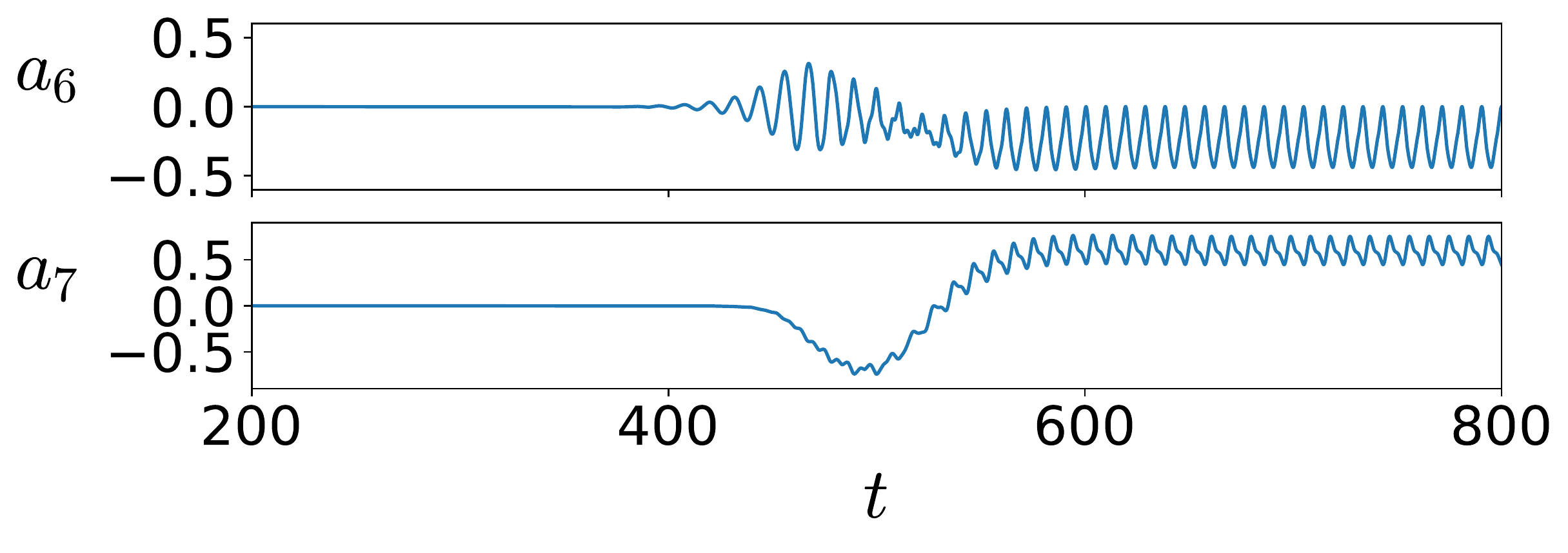} 
\end{tabular}
}
\caption{Mode amplitudes $a_{6,7}(t)$ in the full-flow dynamics starting (a) from the symmetric steady solution $\bm{u}_s$, (b) from the asymmetric steady solution $\bm{u}_s^+$, at $Re=80$.}
\label{Fig:Ampl_2dof}
\end{figure}

After orthogonal normalization by a Gram-Schmidt procedure, the resulting modes are shown in figure \ref{Fig:Correctionmodes}, with their mode amplitudes in figure \ref{Fig:Ampl_2dof}. When comparing the definitions of $\bm{u}_6$ and $\bm{u}_7$ in Eq.~\eqref{Eqn:u6u7} and of $\bm{u}_4$ and $\bm{u}_5$ in Eq.~\eqref{Eqn:u4}--\eqref{Eqn:u5}, it is not surprising that the spatial structure of $\bm{u}_6$, resp. $\bm{u}_7$ (see figure \ref{Fig:Correctionmodes}), be so similar to the spatial structure of $\bm{u}_4$, resp. $\bm{u}_5$ (see figure \ref{Fig:ROM_5dof}). To be mentioned, $\bm{u}_6$, $\bm{u}_7$ as defined in \eqref{Eqn:u6u7}, would be equivalent to the pitchfork modes $\bm{u}_4$, $\bm{u}_5$ built on the periodic solutions instead of being built on the steady solutions. However, after the Gram-Schmidt procedure, the $\bm{u}_6$, $\bm{u}_7$ modes of figure \ref{Fig:Correctionmodes} have been transformed into corrective modes of $\bm{u}_4$, $\bm{u}_5$ when departing from the steady solutions and approaching the asymptotic limit cycles. The corrective modes $\bm{u}_6$, $\bm{u}_7$  should be slaved to $\bm{u}_4$, $\bm{u}_5$ along the mean field distortion of $\bm{u}_{3}$. The corresponding slaving relation will not be discussed in this paper. Hence, the combination of $\bm{u}_i$, $i=4,\ldots,7$, works as a flexible pitchfork mode expansion, which adapts the whole phase space where all the invariant sets (steady/periodic) locate. 

In figure \ref{Fig:Ampl_2dof}, the transient dynamics of $a_6$, $a_7$ shows to be also similar to $a_4$, $a_5$ in figure \ref{Fig:Ampl_5dof}. Not surprisingly, the opposite initial bump of $a_6$, $a_7$ helps to better fit the dynamics on the manifold. Besides, $a_6$, $a_7$ show no contribution close to the steady solutions, as their role is to adapt the modes $\bm{u}_4$, $\bm{u}_5$ when approaching the stable limit cycle.   

The force model identification is more challenging with these two additional modes.
High robustness is required for our force model without losing the identified terms in \S~\ref{Sec:ForceModel5DOF}.
Compared to the force formula \eqref{Eqn:cdcl-Hopf-pitchfork} with five modes, 8 new terms are introduced in the drag formula, namely $a_7,a_1a_6,a_2a_6,a_4a_6,a_6^2,a_3a_7,a_5a_7,a_7^2$, and 7 additional terms are considered in the lift formula, namely $a_6,a_3a_6,a_5a_6,a_1a_7,a_2a_7,a_4a_7,a_6a_7$.
Due to the similar transient dynamics of $a_4$, $a_5$ and $a_6$, $a_7$, the corrective degrees of freedom $a_6$, $a_7$ can easily replace $a_4$, $a_5$ in the identified model. Hence, the original structure of the force model with five modes could be lost. 
To avoid  possible over-fitting, we need to free the active terms gradually and constraint the parameters of $a_4$, $a_5$ during the sparse regression to ensure the robustness of the result.
In addition, the newly introduced terms should work as a corrective function to the original force model with five degrees of freedom. In other words, the new force model with seven degrees of freedom should inherit the original structure of Eq.~\eqref{Eqn:forcemodel_5dof}.

\begin{figure}
\centerline{
 \begin{tabular}{ccc}
 &$C_D$ &  $C_L$\\
 \hline \hline
(a) &  & \\
 & \includegraphics[width=.45\linewidth]{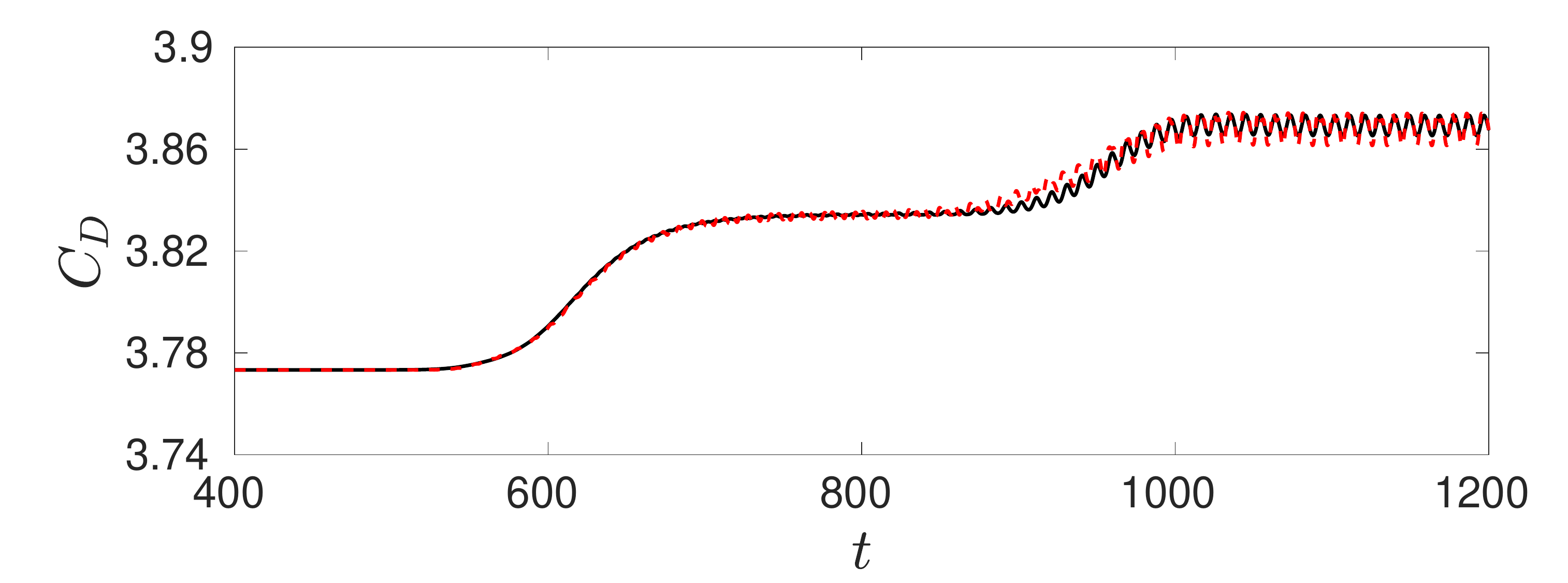} 
 & \includegraphics[width=.45\linewidth]{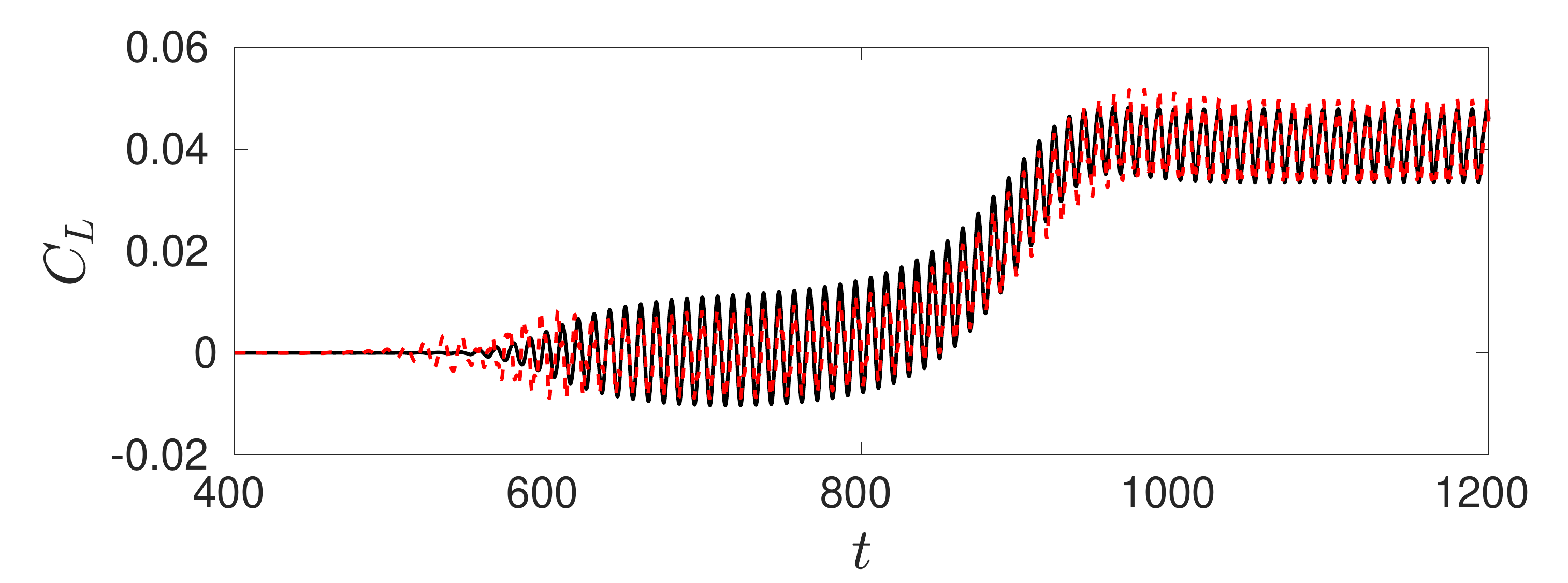} \\ 
(b) &  & \\
 & \includegraphics[width=.45\linewidth]{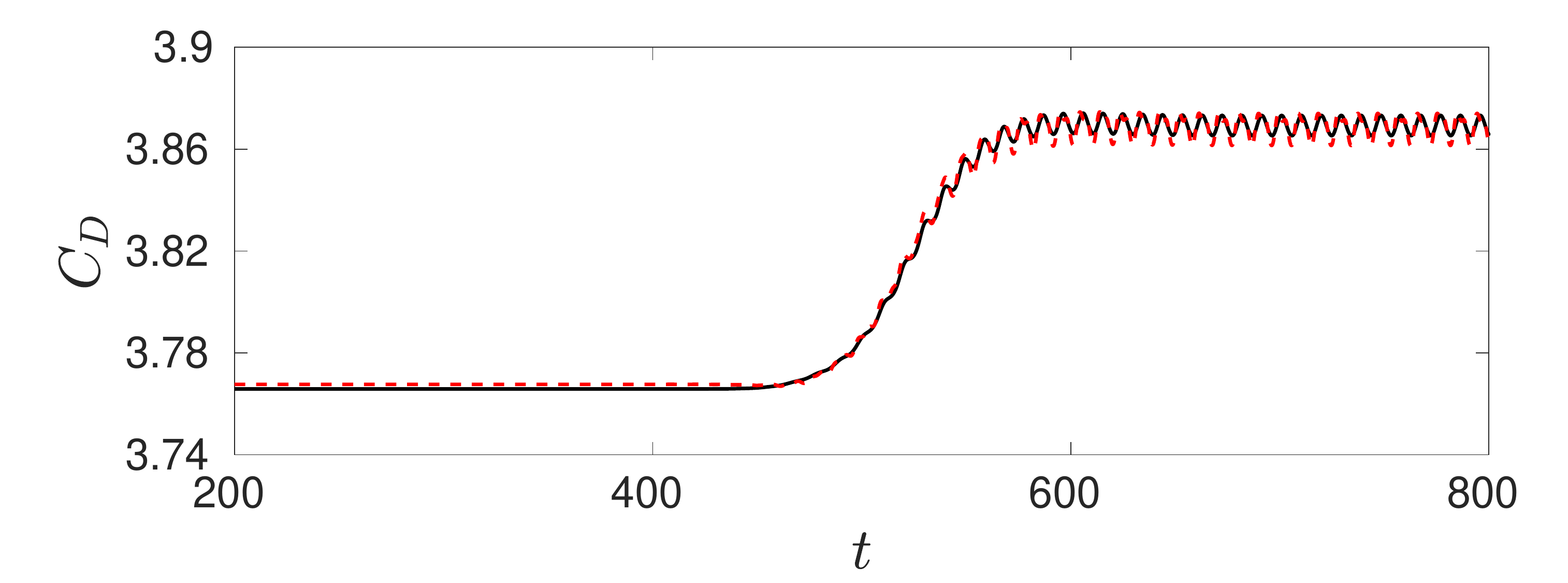} 
 & \includegraphics[width=.45\linewidth]{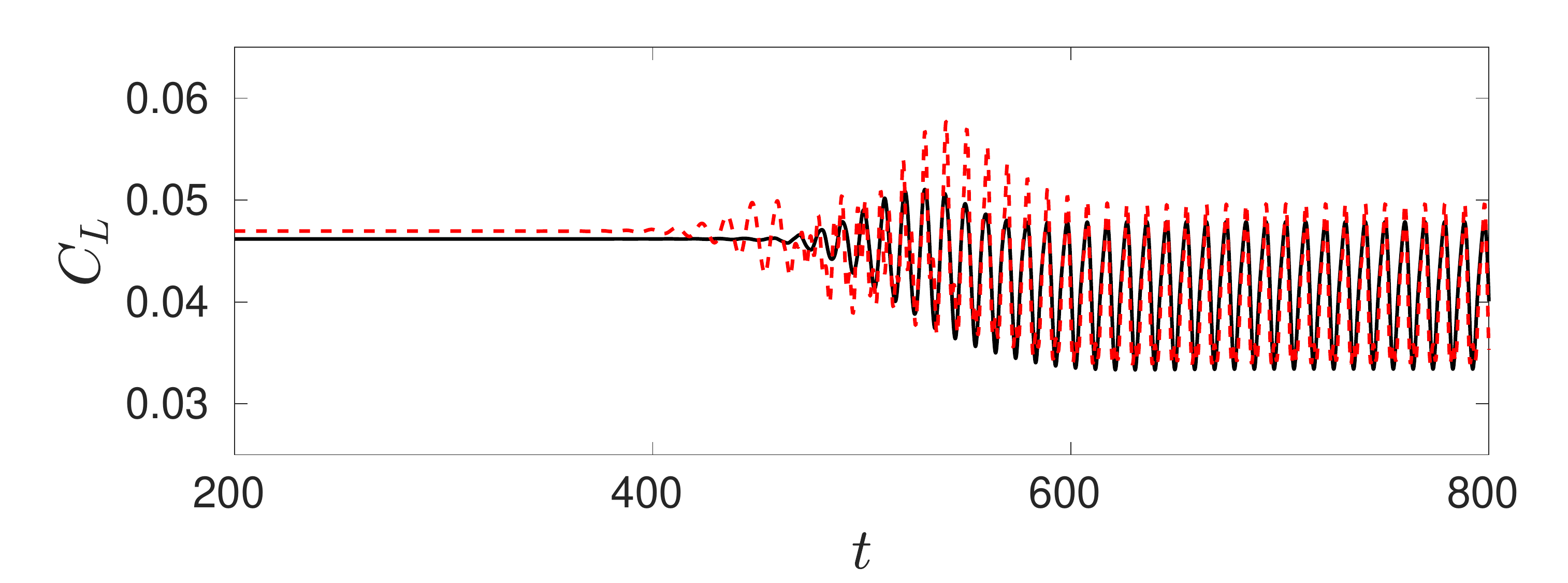} \\ 
 \end{tabular}
}
\caption{Performance of the force model with two additional slaved corrective modes. Time evolution of the drag $C_D$ (left) and lift $C_L$ (right) coefficients in the full flow dynamics (solid black line) and for the force model (red dashed line), at $Re=80$. Initial condition: (a) symmetric steady solution $\bm{u}_s$, (b) asymmetric steady solution $\bm{u}_s^+$. }
\label{Fig:ForceTrans_Re80-7dof}
\end{figure}

Based on the structure of the drag model \eqref{Eqn:forcemodel_5dof_a}, the terms $a_7$, $a_7a_7$, $a_3a_7$, $a_4a_6$ and $a_5a_7$ are introduced in the extended model. The terms $a_1a_6,a_2a_6,a_6^2$ are firstly set to zero because their corresponding terms  $a_1a_4,a_2a_4,a_4^2$ in Eq.~\eqref{Eqn:forcemodel_5dof_a} are vanishing. In order to improve the robustness of the regression results, the terms $a_5$ and $a_5^2$ are constrained with the values from Eq.~\eqref{Eqn:forcemodel_5dof_a}.
Increasing the $L1$-penalty of the LASSO regression, $l_{x; 7}$, $q_{x; 4 6}$ and $q_{x; 7 7}$ vanish successively, and an obvious under-fitting starts when losing $q_{x; 3 5}$. The introduced terms $q_{x; 3 7}$, $q_{x; 5 7}$ are robust with few possibility of over-fitting. Eventually, the drag model reads
\begin{eqnarray}
\label{Eqn:forcemodel_7dof_CD}
\notag C_D &=& 3.77331204  + 0.05888312 a_5 - 0.00169970 a_1^2  - 0.00156775 a_2^2 \\
\notag && + 0.00513885 a_3^2 + 0.00786294 a_3a_5 + 0.00950204 a_3a_7 \\
&& - 0.25910470 a_5^2 - 0.06264888 a_5a_7 .
\end{eqnarray}
Eq.~\eqref{Eqn:forcemodel_7dof_CD} preserves the original form of Eq.~\eqref{Eqn:forcemodel_5dof_a}, with tiny changes of the coefficients. This extended model fits well the dynamics of the drag coefficient, with the $r^2$ score increasing to $0.9981$, also can be seen with the red dashed curve of figure \ref{Fig:ForceTrans_Re80-7dof}(left). 

As already mentioned, the drag monotonously increases with the development of the vortex shedding.
This is obvious, for instance, from figure \ref{Fig:ForceTrans_Re80-7dof}, when the lift starts to oscillate and the drag to increase. The positive signs of $q_{y; 3 3}$, $q_{y; 3 5}$ and $q_{y; 3 7}$, in the drag model of Eq.~\eqref{Eqn:forcemodel_7dof_CD}, are responsible for this monotonous increase of the drag.  
Compared to the drag model with only $a_5$ in \S~\ref{Sec:ForceModel5DOF}, the contribution to the drag of $a_5$ and $a_7$ is more subtle. They contribute to an increase of the drag through $a_5$, $a_5a_3$ and $a_7a_3$, while they promote a decrease of the drag through $a_5^2$ and $a_5a_7$. As a non-trivial result, the statistically asymmetric (stable) limit cycles have a larger drag than the statistically symmetric (unstable) limit cycle, while the asymmetric steady solutions have a lower drag than the symmetric steady solution. This is obvious in figure~\ref{Fig:scheme80} when considering the relative positions of the three steady solutions and three limit cycles along the $C_D$ axis. 
Note that the parameters $q_{x; 1 1}$, $q_{x; 2 2}$ and $q_{x; 5 7}$ all own negative signs but are relatively small. The two parameters $q_{x; 1 1}$, $q_{x; 2 2}$ solely contribute to the oscillating dynamics, as discussed in \S \ref{Sec:Hopf}, while $q_{x; 5 7}$ optimizes the fitting result when evolving toward the attracting limit cycles. 

Analogously, for the lift model, the values of $l_{y; 4}$ and $q_{y; 4 5}$ are taken from the identified lift model in Eq.~\eqref{Eqn:forcemodel_5dof_b}, while $q_{y; 1 7}$, $q_{y; 2 7}$ are set to zero for consistency with the structure of Eq.~\eqref{Eqn:forcemodel_5dof_b}, in which $q_{y; 1 5}$, $q_{y; 2 5}$ are absent. Based on the structure of model \eqref{Eqn:forcemodel_5dof_b}, the terms $a_6$, $a_6a_7$, $a_3a_6$, $a_5a_6$ and $a_4a_7$ are introduced in the extended model. The final sparse form is identified by the LASSO regression with gradually increasing the $L1$-penalty.
A sparse lift model, compatible with the structure of Eq.~\eqref{Eqn:forcemodel_5dof_b}, is derived as 
\begin{eqnarray}
\label{Eqn:forcemodel_7dof_CL}
\notag C_L &=&  0.00762433 \>a_1 + 0.01102097 \> a_2  - 0.10179203 \> a_4 - 0.03129798 \> a_6 \\
\notag && - 0.00141416\> a_1a_3 - 0.00289952 \>a_2 a_3 + 0.00656293 \>a_3a_4 -  0.01082375 \> a_3a_6\\
&& + 0.05914386 \> a_4a_5 + 0.02365784 \>a_4a_7 - 0.03348935 \>a_5a_6. 
\end{eqnarray}
In addition to the lift model of Eq.~\eqref{Eqn:forcemodel_5dof_b}, the lift model of Eq.~\eqref{Eqn:forcemodel_7dof_CL} contains the terms $a_{6}$, $a_3a_6$ and $a_5a_6$, as well as the coupling between $a_4$ to $a_7$. The $r^2$ score has increased to $0.9952$.
Both the oscillating dynamics in the early stage and the symmetry-breaking stage are better reproduced for the lift coefficient, as the red dashed curve of figure~\ref{Fig:ForceTrans_Re80-7dof}(right) proves.

With the two additional degrees of freedom $a_6$, $a_7$, the time evolution of the drag and lift coefficients are well reproduced, as shown in figure~\ref{Fig:ForceTrans_Re80-7dof}(a,b). 
Without notable changes of the original lift structure, the phase of the lift dynamics is now correctly caught along with the complete transient dynamics. 

\subsection{Force model at $Re=100$}
\label{Sec:Validation}

\begin{figure}
\centering
\includegraphics[width=.45\linewidth]{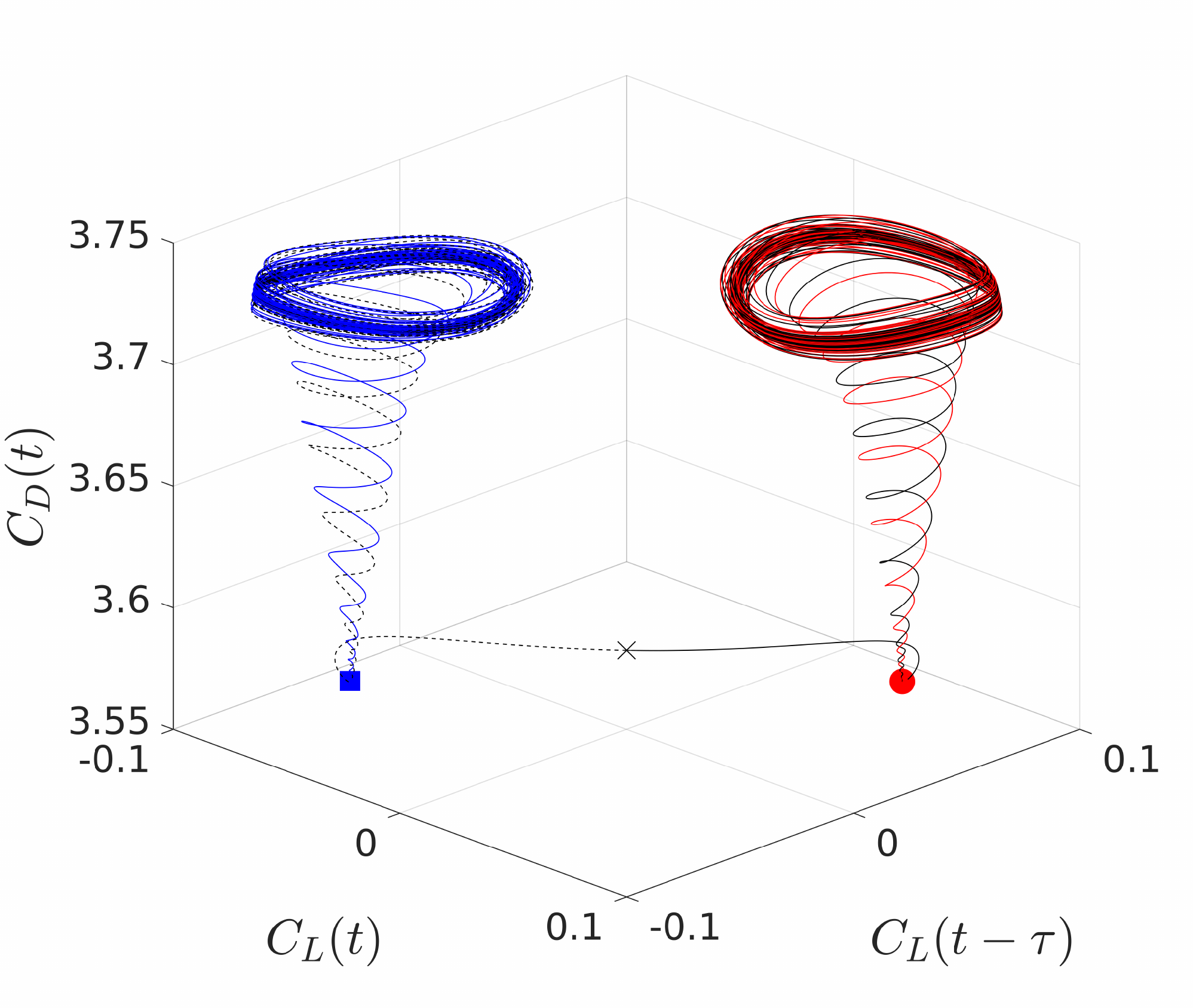}
\caption{Trajectories in the time-delayed embedding space of the lift $C_L$ and drag $C_D$ coefficients, with $\tau =2$, at $Re=100$. Black trajectories starting close to the symmetric steady solution $\bm{u}_s$ ($\times$); red trajectory starting close to the asymmetric steady solution $\bm{u}_s^+$ ($\bullet$), blue trajectory starting close to 
the asymmetric steady solution $\bm{u}_s^-$ ($\blacksquare$).
}
\label{Fig:scheme100}
\end{figure}

In \S~\ref{Sec:pitchfork}, we derived a basic force formula for the primary stage of the transient evolution at $Re=100$, when only the degrees of freedom of the pitchfork bifurcation were involved.  We now consider the complete force evolution at $Re=100$.
Figure \ref{Fig:scheme100} shows trajectories issued from the three different steady solutions in the three-dimensional time-delayed embedding space of $C_L$ and $C_D$. The black trajectory, issued from the symmetric steady solution $\bm{u}_s$ (black cross $\times$ in figure \ref{Fig:scheme100}) first approaches the asymmetric steady solution $\bm{u}_s^+$ (red point) before escaping out of it and eventually reaching the stable (statistically asymmetric) limit cycles around $\bar{\bm{u}}^+$. 

\begin{figure}
\centerline{
\begin{tabular}{cc}
(a) & (b) \\
 \includegraphics[width=.5\linewidth]{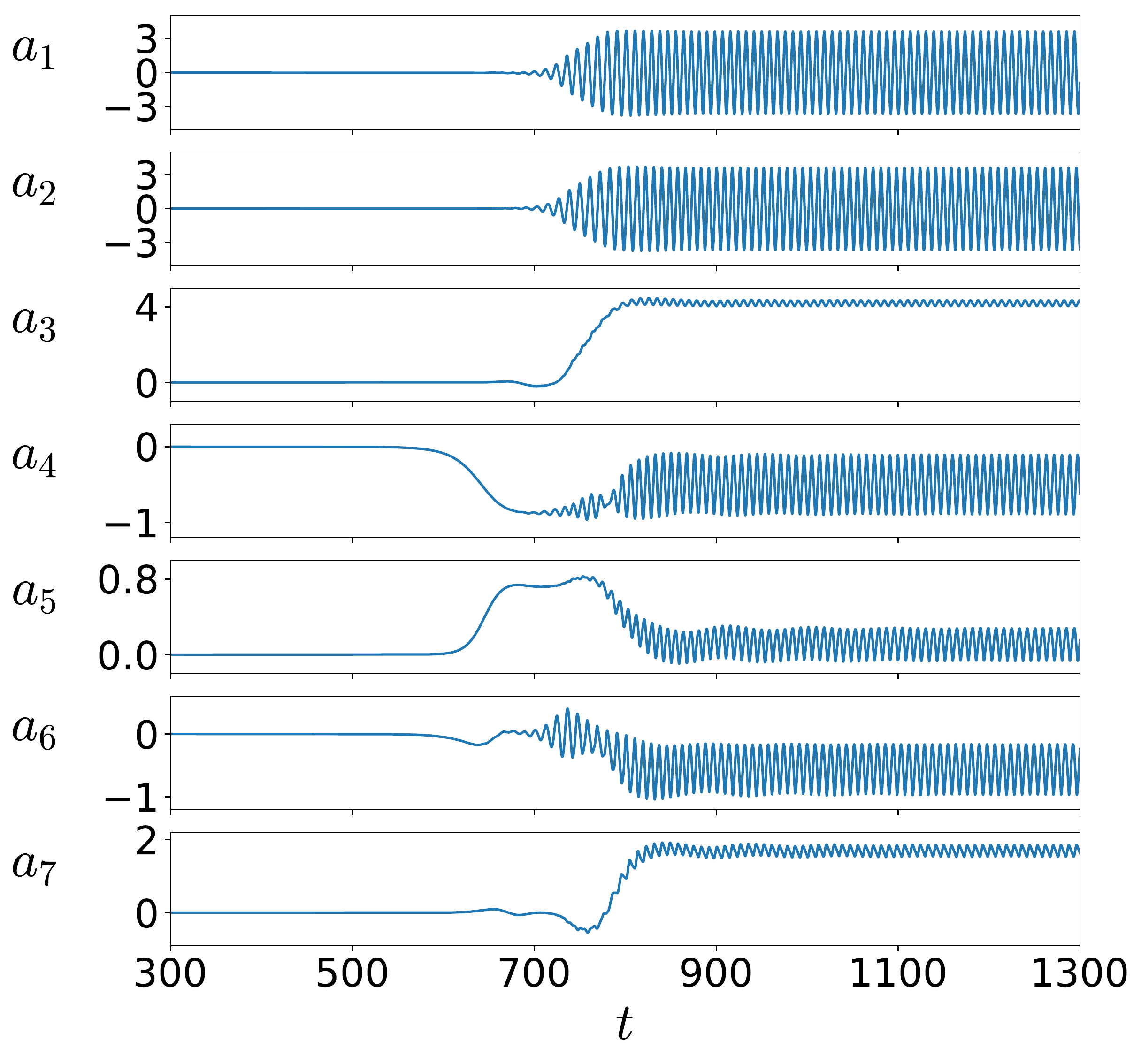} & 
 \includegraphics[width=.5\linewidth]{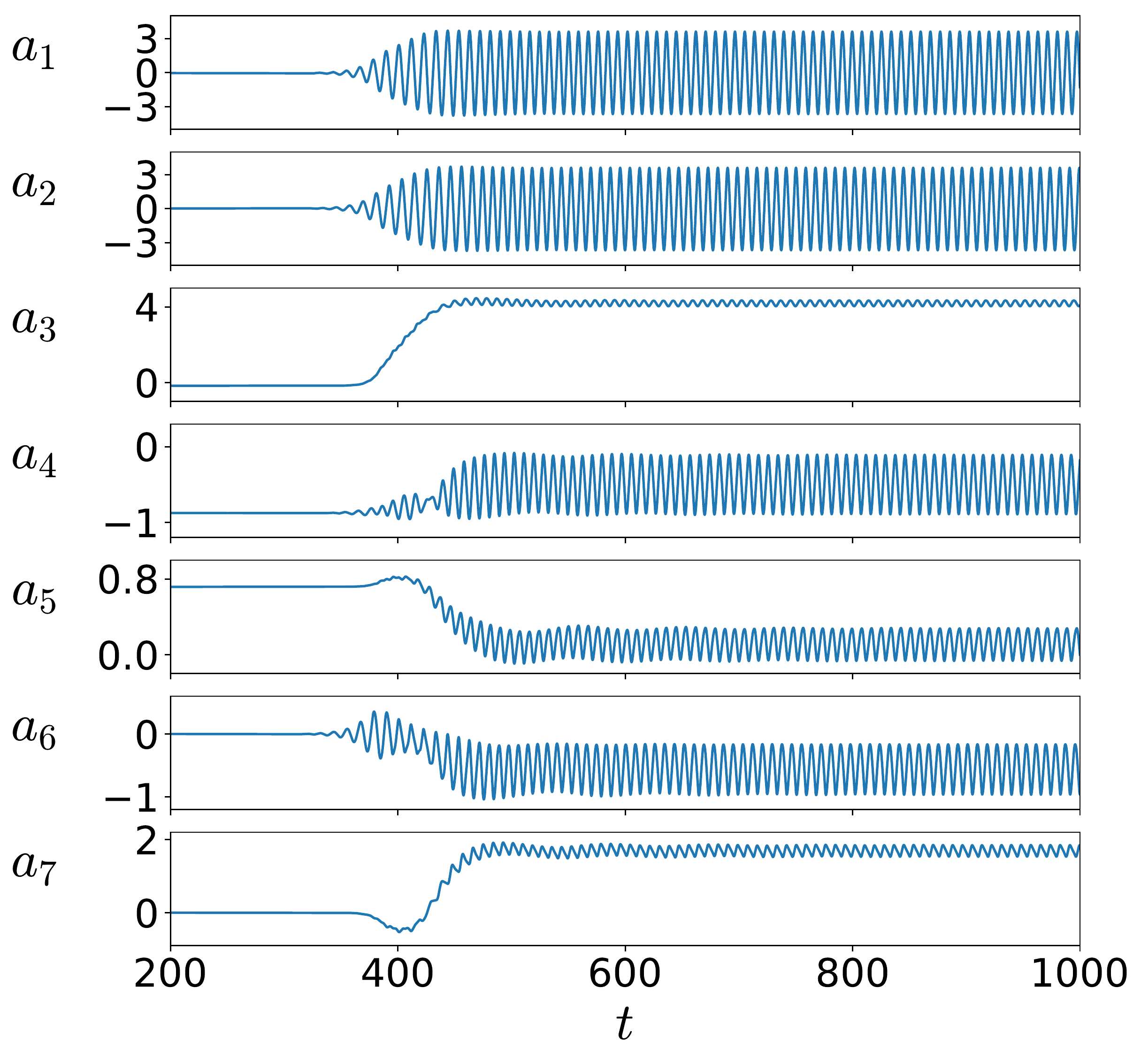} 
\end{tabular}
}
\caption{Mode amplitudes $a_{1,\dots,7}(t)$ in the full-flow dynamics starting (a) from the symmetric steady solution $\bm{u}_s$, (b) from the asymmetric steady solution $\bm{u}_s^+$, at $Re=100$.}
\label{Fig:Ampl100_7dof}
\end{figure}

The same mode decomposition strategy is proposed, resulting in a reduced-order model with 7 modes. The mode amplitudes from two DNS, starting from either the symmetric steady solution $\bm{u}_s$ (a) or the asymmetric steady solution $\bm{u}_s^+$ (b), are shown in figure \ref{Fig:Ampl100_7dof}.

\begin{figure}
\centerline{
 \begin{tabular}{ccc}
 &$C_D$ &  $C_L$\\
 \hline \hline
(a) &  & \\
 & \includegraphics[width=.45\linewidth]{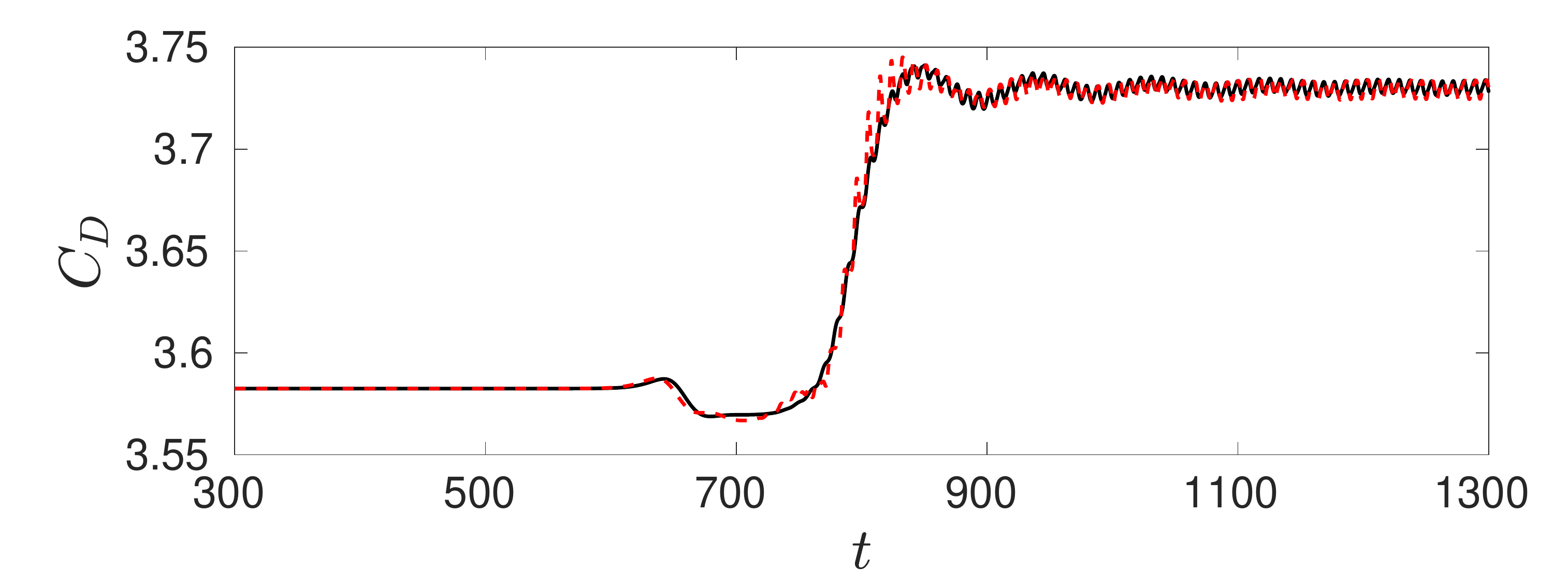} 
 & \includegraphics[width=.45\linewidth]{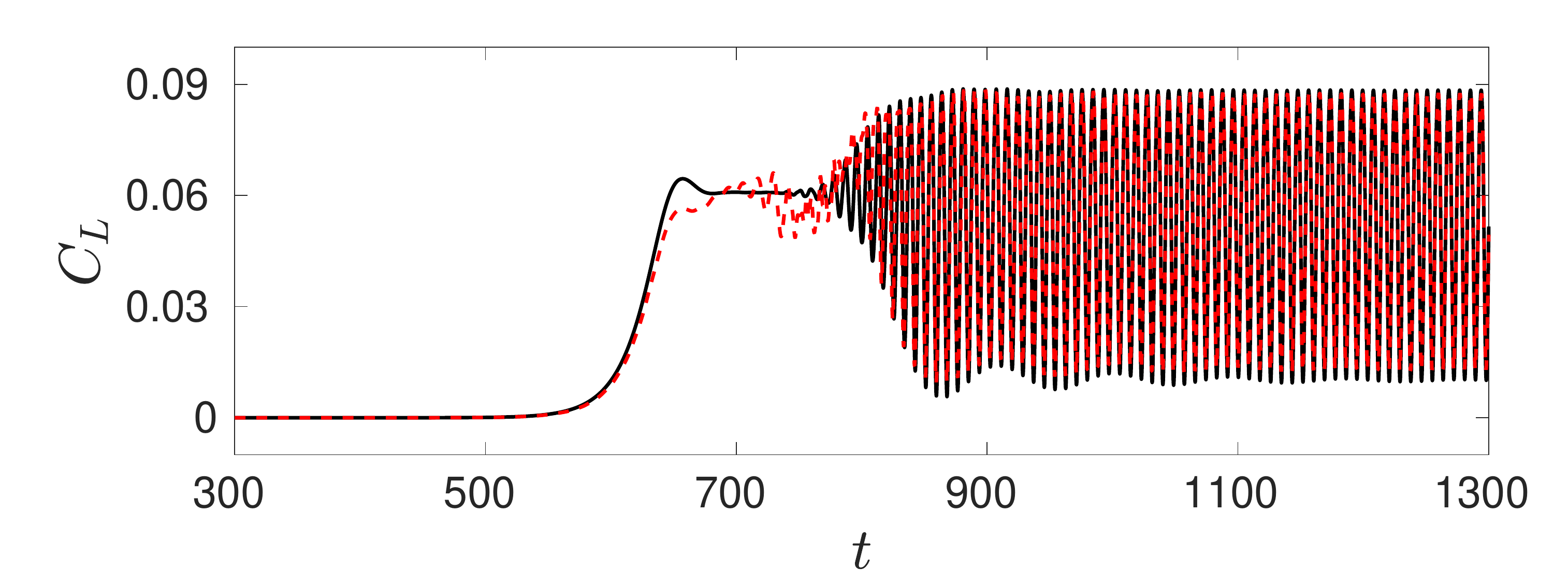} \\ 
(b) &  & \\
 & \includegraphics[width=.45\linewidth]{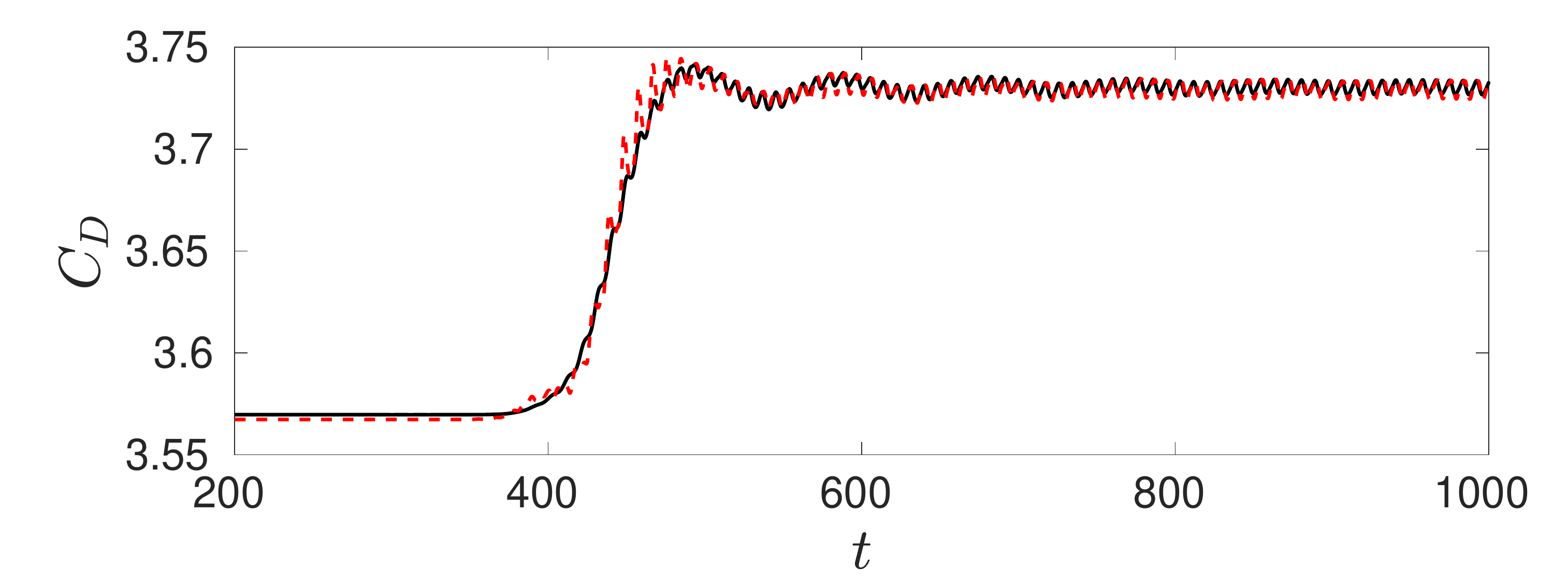} 
 & \includegraphics[width=.45\linewidth]{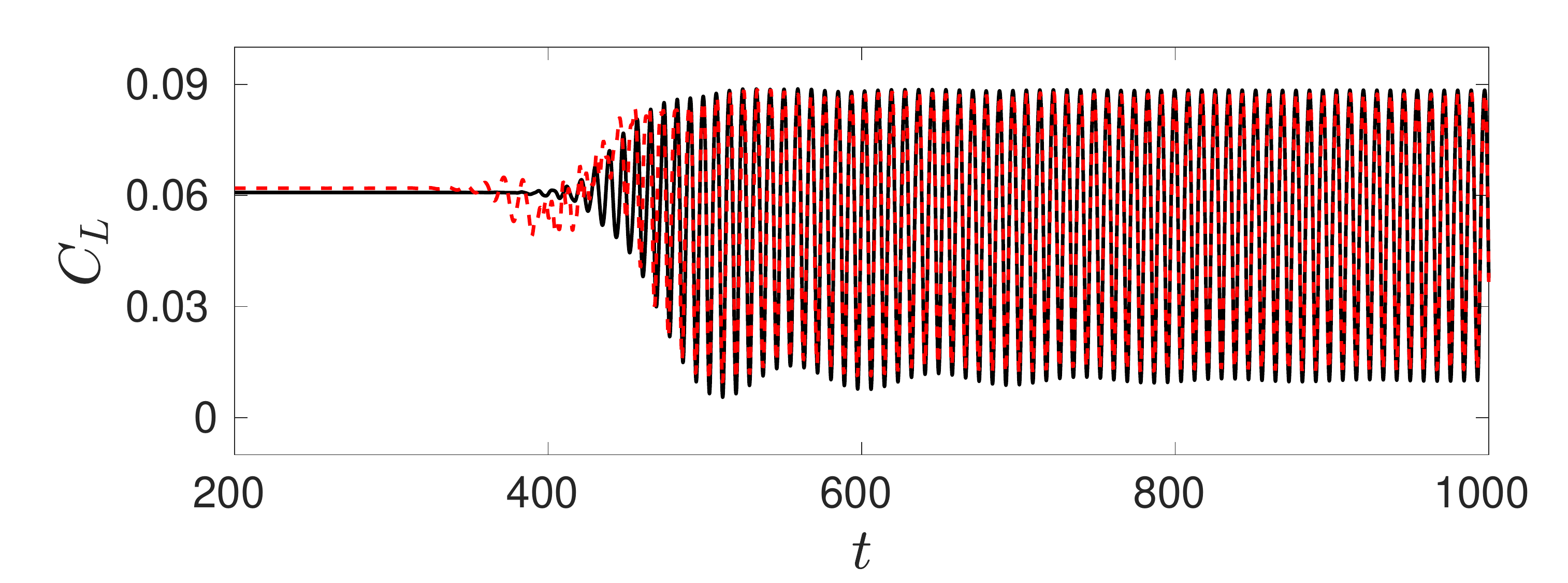} \\ 
 \end{tabular}
}
\caption{Performance of the force model with two additional slaved corrective modes. Time evolution of the drag $C_D$ (left) and the lift $C_L$ (right) coefficients in the full flow dynamics (solid black line) and for the force model (red dashed line), at $Re=100$. Initial condition: (a) symmetric steady solution $\bm{u}_s$, (b) asymmetric steady solution $\bm{u}_s^+$. }
\label{Fig:ForceTrans_Re100-7dof}
\end{figure}

As already observed in figure \ref{Fig:ForceModel-PF}, the drag coefficient (solid black line) in figure~\ref{Fig:ForceTrans_Re100-7dof}(a) exhibits a minimal value for a transient state around $t\approx 700$. This transient state is the asymmetric steady solution $\bm{u}_s^+$ (red circle of figure \ref{Fig:scheme100}). 
In the frame of our modal decomposition \eqref{Eqn:BifurcationExpansion}, $\bm{u}_s^+$ is approximated as
\begin{equation}
\label{Eqn:u+}
    \bm{u}_s^+ \approx \bm{u}_s + a_4(700)\bm{u}_4 + a_5(700)\bm{u}_5,
\end{equation}
with only $a_4$ and $a_5$ being active in the dynamics of the fluidic pinball, as can be seen in figure \ref{Fig:Ampl100_7dof}(a). From Eq.~\eqref{Eqn:forcemodel_2dof_a}, the drag coefficient only depends on $a_5$ and $a_5^2$, which actually contribute to the transitory increase and an overall decrease on the drag. This is fully consistent with the transition of the drag coefficient observed in figures \ref{Fig:ForceModel-PF}(a) and \ref{Fig:ForceTrans_Re100-7dof}(a) from $t=300$ to $700$; $a_5$ is found to contribute to the initial rising of $C_D$, around $t\approx 650$, while $a_5^2$ contributes to the subsequent decrease of the drag coefficient, around $t=700$. The degrees of freedom associated with the Hopf bifurcation become active later during the transient dynamics, when the state space orbit leaves the unstable asymmetric steady solution $\bm{u}_s^+$ toward the stable attracting limit cycle around $\bar{\bm{u}}_s^+$. 

The training data is the real force coefficients and the mode amplitudes taken from the DNS starting with the three different steady solutions to the final asymptotic regimes.
Following the same calibration procedure as for $Re=80$, we first apply the LASSO regression for the force model with the five leading degrees of freedom, and then introduce the two additional degrees of freedom $a_6$, $a_7$ into the regression for optimization.
Performing the sparse regression in this way can prevent the elimination of $a_4$, $a_5$ and ensure the corrective effect of $a_6$, $a_7$, thereby improving the robustness of the identification. 
The force model at $Re=100$ reads
\begin{subequations}
\label{Eqn:forcemodel_7dof_Re100}
\begin{eqnarray}
\notag C_D &=& 3.58248992  + 0.04367604 \>a_5 - 0.00302817 \>a_1^2 - 0.00354079 \> a_2^2 \\
\notag&&  + 0.00158873 \> a_3^2 + 0.02169661 \> a_3a_5 + 0.02223079 \> a_3a_7 \\
&&  - 0.08525184 \> a_5^2 - 0.04763643 \> a_5a_7 ,\label{Eqn:forcemodel_7dof_Re100_a} \\
\notag C_L &=& 0.00346208 \> a_1 + 0.00269236 \> a_2 - 0.13611053 
\> a_4 + 0.05962648 \> a_6\\
\notag&& + 0.00029274 \> a_1a_3 - 0.00045784\> a_2a_3 + 0.00389912 \>a_3a_4 - 0.02102284\> a_3a_6 \\
&& + 0.09194312 \> a_4a_5 + 0.02056288 \> a_4a_7 - 0.10980990\> a_5a_6. \label{Eqn:forcemodel_7dof_Re100_b}
\end{eqnarray}
\end{subequations}
with $r^2 = 0.9984$ for the drag model of Eq.~\eqref{Eqn:forcemodel_7dof_Re100_a}, and 
$r^2 = 0.9901$ for the lift model of Eq.~\eqref{Eqn:forcemodel_7dof_Re100_b}.
As shown in figure \ref{Fig:ForceTrans_Re100-7dof}, the force model fits well the time evolution of the drag and lift coefficients. Moreover, Eqs.~\eqref{Eqn:forcemodel_7dof_CD}, \eqref{Eqn:forcemodel_7dof_CL} and \eqref{Eqn:forcemodel_7dof_Re100} own the same active terms. 
Henceforth, the drag force model preserves the same structure with the same signs of the active terms as the Reynolds number is increased. 
In addition, although the transient dynamics at $Re=80$ and 100 are qualitatively very different, with the seven degrees of freedom differently activated during the transient, the force model of Eq.~\eqref{Eqn:forcemodel_7dof_CD}--\eqref{Eqn:forcemodel_7dof_CL} is still consistent at $Re=100$, with the correctly identified mean-field model. 
For the lift model \eqref{Eqn:forcemodel_7dof_Re100_b}, we notice the same structure with the sign changes for the terms $a_1a_3$ and $a_6$, compared to Eq.~\eqref{Eqn:forcemodel_7dof_CL}, which is acceptable for the oscillating dynamics. Compatible with the basic lift force model, the lift force model with seven degrees of freedom also correctly identifies the force transitions, as shown in figure \ref{Fig:ForceTrans_Re100-7dof}(right).

\section{Conclusions and outlook}
\label{Sec:Conclusions}

We proposed an aerodynamic force formulae 
complementing mean-field POD Galerkin models for the unforced fluidic pinball.
The starting point is a general Galerkin method 
for unsteady incompressible viscous flow around a stationary body.
First, the instantaneous force is derived as
a constant-linear-quadratic function of the mode amplitudes from first principles.
The viscous and pressure contributions to the force 
are directly obtained from the Galerkin expansion
and lead to a constant-linear-quadratic force in terms of the mode amplitudes.

These terms lead to corresponding changes in the flow
from which the force can also be derived.
One contribution from the convective term 
describes the momentum flux contribution. 
The additional contribution from the local acceleration
requires the Galerkin system to replace 
the time derivatives of the mode amplitudes by a state function.
In contrast to the pioneering work by \citet{noca1999comparison},
the derivation is valid for arbitrary multiply connected domains.
 
The drag and lift formula is simplified for the fluidic pinball model
exploiting the symmetry of the modes.
About half of the terms can be discarded on the grounds of symmetry.
A second simplification is performed 
with a sparse calibration 
of the remaining coefficients.
The sparsity parameter $\lambda$ penalizes any non-vanishing term 
and yields sparse human-interpretable expressions. 
The challenges of the purely projection-based approach is discussed in Appendix~\ref{Sec:RealForce}, and the challenges of using standard POD modes is elaborated in Appendix~\ref{Sec:Force-POD}. 

The sparse force model methodology is applied 
to three transient dynamics:
(1) the periodic regime of statistically symmetric vortex shedding at $Re=30$,
(2) the periodic regime of statistically asymmetric vortex shedding at $Re=100$, 
and
(3) the same regime at $Re=80$ but with metastable statistically symmetric periodicity.

The transient dynamics at $Re=30$ from the steady solution to the limit cycle
is resolved by standard third-order mean-field Galerkin model 
with two oscillatory modes for vortex shedding 
and one shift mode for the mean-field distortion \citep{Noack2003jfm}.
The drag formula includes the squares of all mode amplitudes consistent with the second harmonic fluctuations.
The drag monotonically increases during the transient.
The lift formula includes the amplitudes of the von K\'arm\'an modes 
and their products with the shift mode, consistent with expectations.
Its oscillation increases until the limit cycle is reached.

The dynamics at $Re=100$ after the Hopf and pitchfork bifurcation
has three unstable fixed points,
one symmetric steady solution and a mirror-symmetric pair of asymmetric ones.
The transients from these fixed points terminate in one of the asymmetric limit cycles 
corresponding to the asymmetric shedding states.
This dynamics is described by a fifth-order Galerkin model \citep{deng2020jfm},
where the first three modes resolve the Hopf bifurcation and the next two modes the pitchfork bifurcation.
The associated drag formula contains the terms of $Re=30$.
In addition, the drag is modified by linear and quadratic terms 
with the shift modes associated with the Hopf and pitchfork instabilities.
These additional terms vanish without pitchfork bifurcation and do not introduce harmonics of vortex shedding.
Similarly, the lift formula generalizes the expression at $Re=30$.

The intermediate Reynolds number $80$ leads to a more complex force model,
as the transients may pass through a meta-stable symmetric limit cycle.
The accuracy of the force model could significantly be increased
by two additional Galerkin modes which resolve variations between symmetric and asymmetric limit cycles.
The drag and lift formulae were correspondingly longer
and good agreement with computational data is achieved.

Summarizing, the sparse force model 
describes multi-attractor behaviour of the unforced fluidic pinball even for complex dynamics
with three steady and three periodic solutions.
For this configuration, we have the advantage of a thorough understanding of the dynamics
via a low-dimensional mean-field Galerkin model.
We envision successful applications of  sparse regression 
for aerodynamic forces for turbulent flows,
e.g., for the bi-stable behaviour of the Ahmed body wake \citep{Grandemange2013jfm,Osth2014jfm,Barros2017jfm}.

The force formula may be particularly instructive for drag reduction with active control \citep{Choi2008arfm}.
Given a Galerkin model, the force formula indicates beneficial regions of the state space.
Thus, an upfront kinematical insight is gained 
in which direction control needs to `push' the attractor.
For instance, 
the third-order mean-field model and the force formula
implies that stabilization is required for drag reduction 
consistent with earlier studies of \citet{Protas2004pf,bergmann2008jcp}.
Future generalizations may also profit
from stochasticity \citep{sapsis2009dynamically}. 

\section*{Acknowledgements}
N. Deng appreciates the support of the China Scholarships Council (No.201808070123) 
during his Ph.D. Thesis in the ENSTA Paris of Institut Polytechnique de Paris, 
and numerical supports from the laboratories LIMSI (CNRS-UPR 3251) and IMSIA (UMR EDF-ENSTA-CNRS-CEA 9219).

This work is supported by a public grant overseen by the French National Research Agency (ANR) 
by grant `FlowCon' (ANR-17-ASTR-0022),
and by Polish Ministry of Science and Higher Education (MNiSW) under the Grant No.: 0612/SBAD/3567.

We appreciate valuable discussions with Guy Cornejo Maceda, Fran\c{o}is Lusseyran, and Colin Leclercq.
We thank the anonymous referees for their insightlful suggestions 
which have inspired some of our investigations.

{\bf Declaration of Interests}. The authors report no conflict of interest.

\appendix
 \section{Forces from the momentum balance}
\label{Sec:ForceVolume}

The forces can be alternatively derived from the residual of the Navier-Stokes equations
\begin{equation}
\label{Eqn:NavierStokesResidual}
\bm{R} ( \bm{u},p)  := \partial_t \bm{u} 
   + \nabla \cdot \bm{u} \otimes \bm{u} 
   - \nu \triangle \bm{u}
   + \nabla p
\end{equation}
in the domain $\Omega$.
This domain is assumed to enclose the obstacle
and extend sufficiently far away from the obstacle such that the free-stream condition 
$\bm{u} = \bm{e}_x$ can be applied on the left, top and bottom boundaries of the fluid domain $\Omega$. 
The domain boundary $\partial \Omega$ contains 
the surface of the immersed body $\Gamma$ and the outer surface $S_{\infty}$.
It should be noted that the surface element $dS$ on the body 
points inside the body, i.e., opposite to the direction in \S\ \ref{Sec:ForceSurface}.

The force in direction $\bm{e}_{\alpha}$ 
is derived from the integrated momentum balance in that direction.
\begin{equation}
 \label{Eqn:MomentumBalance}
 \left( \bm{e}_{\alpha}, \bm{R}(\bm{u},p) 
 \right)_{\Omega} =0.
\end{equation}
Four terms are obtained. 
The first contribution is the viscous term.
This term can be converted into a skin friction integral over $\Gamma$ and $S_{\infty}$.
The contribution over the outer integral vanishes under free-stream conditions.
The remaining contribution is the viscous force applied to the immersed body:
\begin{equation}
 \label{Eqn:ViscousForceVolume}
 \left(\bm{e}_{\alpha}, \nu \triangle \bm{u} \right)_{\Omega} 
 = \nu \bm{e}_{\alpha}\cdot \oint\limits_{\Gamma + S_{\infty}}   \left( \nabla \bm{u} +  (\nabla \bm{u})^{\rm T} \right) \cdot \bm{n}\, dS \> 
= F_{\alpha}^{\nu},
\end{equation}
where $ \bm{e}_{\alpha} \cdot \left( \nabla \bm{u} +  (\nabla \bm{u})^{\rm T} \right) \cdot \bm{n} = 2 \sum_{\alpha, \beta=x,y,z} {S}_{\alpha,\beta} \> n_{\beta} \>$.

The second contribution is the pressure term
which can analogously reduce to the pressure force on the immersed body:
\begin{equation}
\label{Eqn:PressureForceVolume}
\left(\bm{e}_{\alpha},-\nabla p\right)_{\Omega} 
 = - \oint\limits_{\Gamma + S_{\infty}}   p \> n_{\alpha} \> dS
= F_{\alpha}^{p}.
\end{equation}
Not surprisingly, we arrive at the formula of \S~\ref{Sec:ForceSurface}.
The force exerted on the body is equal but opposite to the force exerted on the fluid.

The third term is the local acceleration:
\begin{equation}
 \label{Eqn:ForceLocalAcceleration}
 \left(\bm{e}_{\alpha}, 
       \partial_t \left[  \sum_{j=0}^{N} a_j(t) \bm{u}_j(\bm{x}) \right]\right)_{\Omega} 
     = \sum_{j=1}^{N} m^{t}_{\alpha; j}  \frac{{da}_j}{dt}(t) , 
\end{equation}
where $ m^{t}_{\alpha; j} = (\bm{e}_{\alpha},\bm{u}_j)_{\Omega}$.

The fourth term arises from the convective acceleration:
\begin{equation}
 \label{Eqn:ForceConvectiveAcceleration}
\left(\bm{e}_{\alpha},
 \nabla \cdot \left( \left[ \sum_{j=0}^{N} a_j \bm{u}_j \right] \otimes
                       \left[ \sum_{k=0}^{N} a_k \bm{u}_k \right] 
                 \right) \right)_{\Omega} = \sum_{j,k = 0}^N q^{c}_{\alpha; j k} a_j a_k
\end{equation}
where 
$q^{c}_{\alpha; j k} = 
\left (\bm{e}_{\alpha}, 
      \nabla \cdot \left[ \bm{u}_j \otimes \bm{u}_k\right ]
\right)_{\Omega}.$
The volume integral over $\Omega$ can be converted into a momentum flux surface integral over the boundary.

Making use of the momentum balance \eqref{Eqn:MomentumBalance},
the third and fourth contributions from the acceleration terms 
equal the total force:
\begin{equation}
 \label{Eqn:AccelerationForce}
F_{\alpha} = \sum_{j=1}^N m^{t}_{\alpha; j} \frac{{da}_j}{dt}
           + \sum_{j,k=0}^N q^{c}_{\alpha; j k} a_j a_k.
\end{equation}
This force formula contains constant, linear and quadratic terms
of the mode amplitudes as well as their time derivatives.
The state-dependent formula \eqref{Eqn:ForceExpansion}
may be obtained from \eqref{Eqn:AccelerationForce} 
by replacing the time derivatives with \eqref{Eqn:GalerkinSystem}.
The total forces on the immersed body are here again represented by a constant-linear-quadratic expression.

The above mentioned formulae dresses Newton's second law 
$\bm{F} = m \bm{a}$ 
in a Galerkin framework for fluid flow.
Eq.~\eqref{Eqn:AccelerationForce} corresponds to `$m \bm{a}$'
and is purely based on the fluid motion.
Eq.~\eqref{Eqn:ForceExpansion} corresponds to `$\bm{F}$'
and allows distinguishing between the contribution of viscous and pressure stresses.

\section{Influence of the sparsity parameter and regression methods}
\label{Sec:TwoRegressionMethods}

In the SINDy algorithms, the sparsity parameter is either the $L1$-penalty for the LASSO regression or the threshold for the sequential thresholded least squares (STLS) regression. 
We denote the $L1$-penalty and the threshold as the sparsity parameter $\lambda$ in both cases.
These two methods can however lead to different results. We can choose the one with a better performance according to the actual needs.

\begin{figure}
\centering
\begin{tabular}{cc}
(a) & \raisebox{-0.55\height}{\includegraphics[width=.9\linewidth]{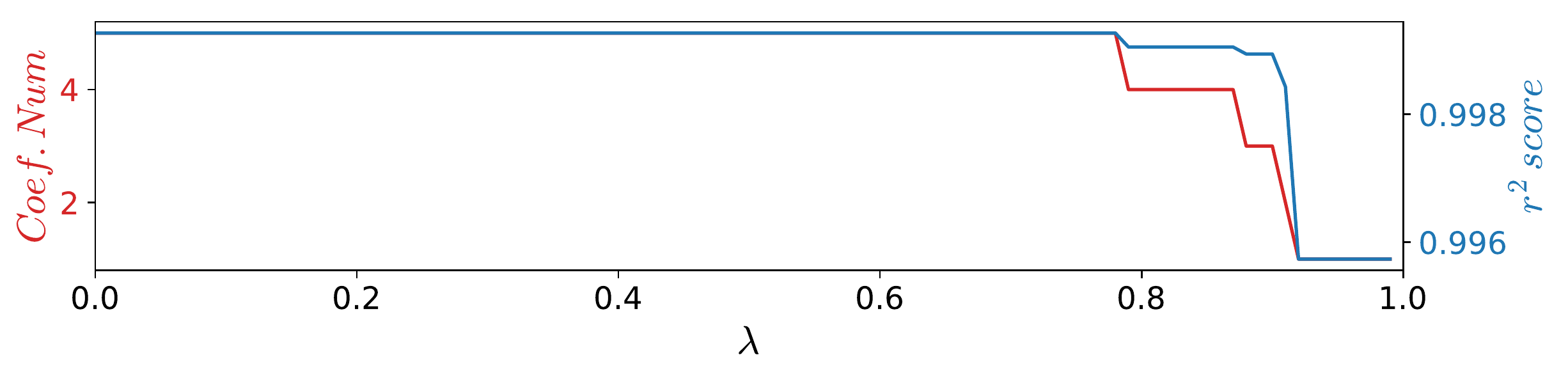} }
\end{tabular}
\begin{tabular}{ccc}
\includegraphics[width=.25\linewidth]{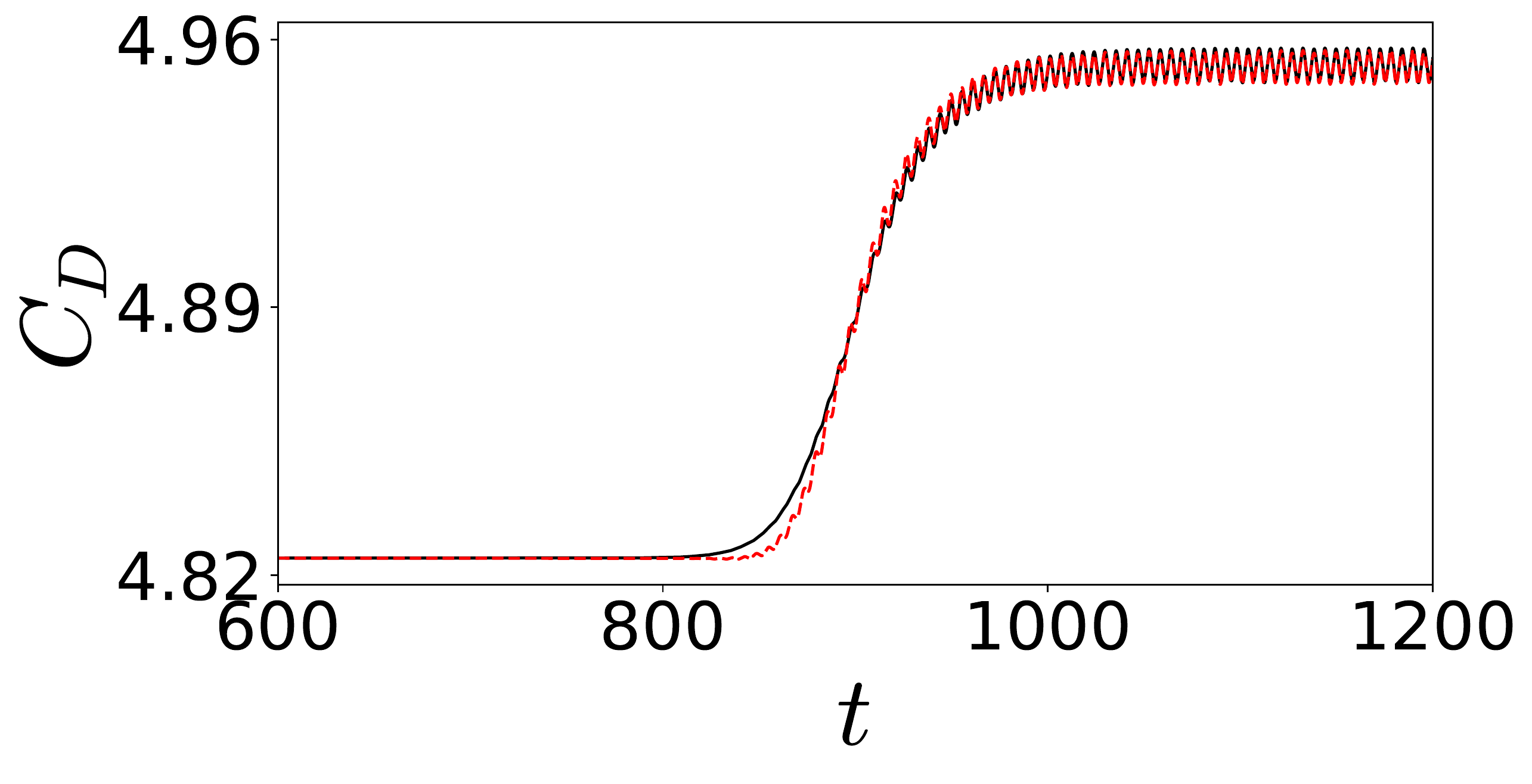} &
\includegraphics[width=.25\linewidth]{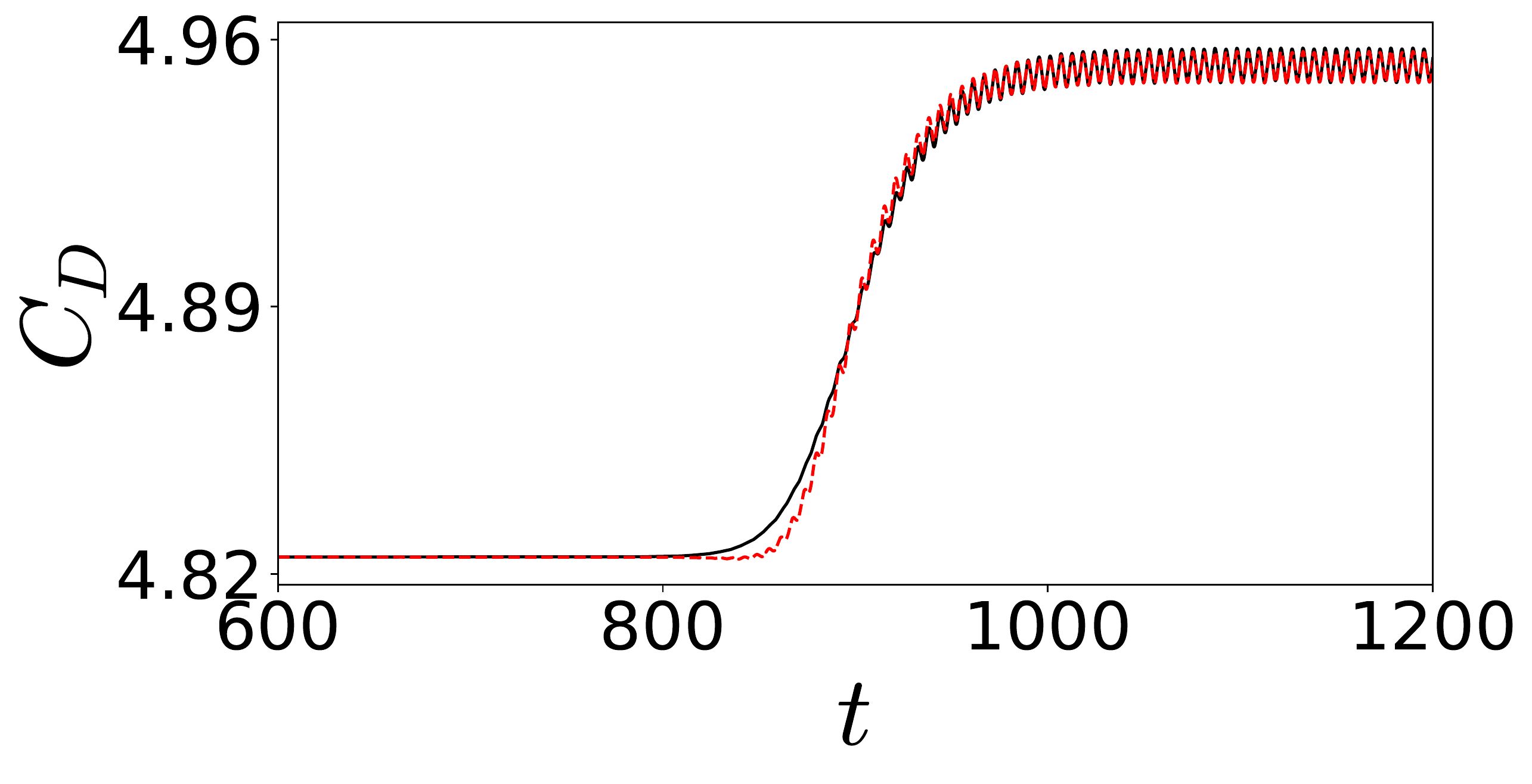} &
\includegraphics[width=.25\linewidth]{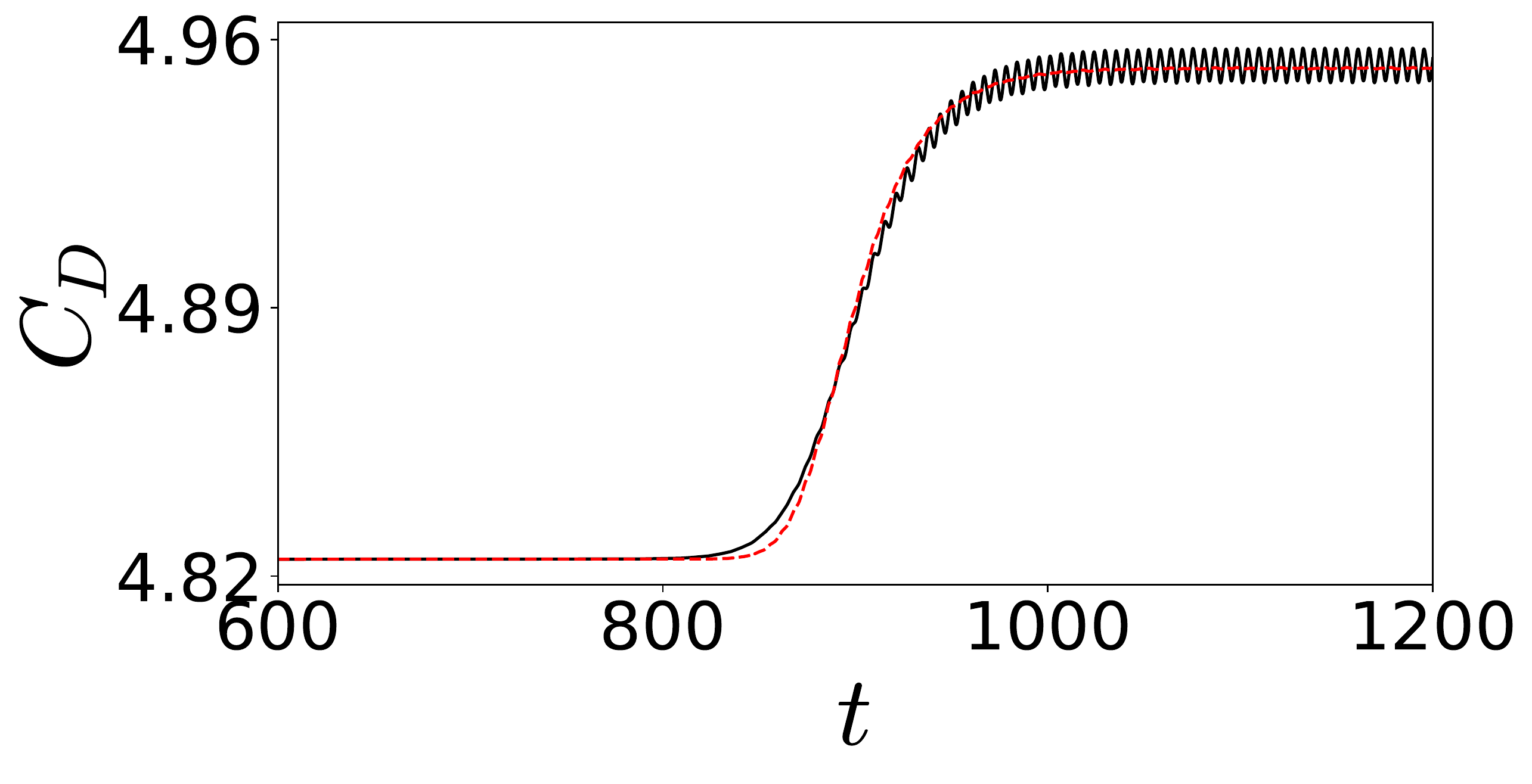} \\
(b) & (c) & (d) 
\end{tabular}
\caption{Illustration of the influence of the sparsity parameter $\lambda$ on both the complexity and accuracy of the identified drag model by the LASSO regression with three degrees of freedom at $Re = 30$.
(a) Evolution of the number of non-zero coefficients (red) and of the $r^2$ score (blue) as a function of the sparsity parameter $\lambda$. 
Performance of the identified drag model at $\lambda = 0.8$ (b), $0.9$ (c), and $0.95$ (d). Time evolution of the drag $C_D$ coefficients in the full flow dynamics (solid black line) and for the force model (red dashed line). Initial condition: symmetric steady solution.}
\label{Fig:LASSO_Drag30}
\end{figure}

\begin{figure}
\centering
\begin{tabular}{cc}
(a) & \raisebox{-0.55\height}{\includegraphics[width=.9\linewidth]{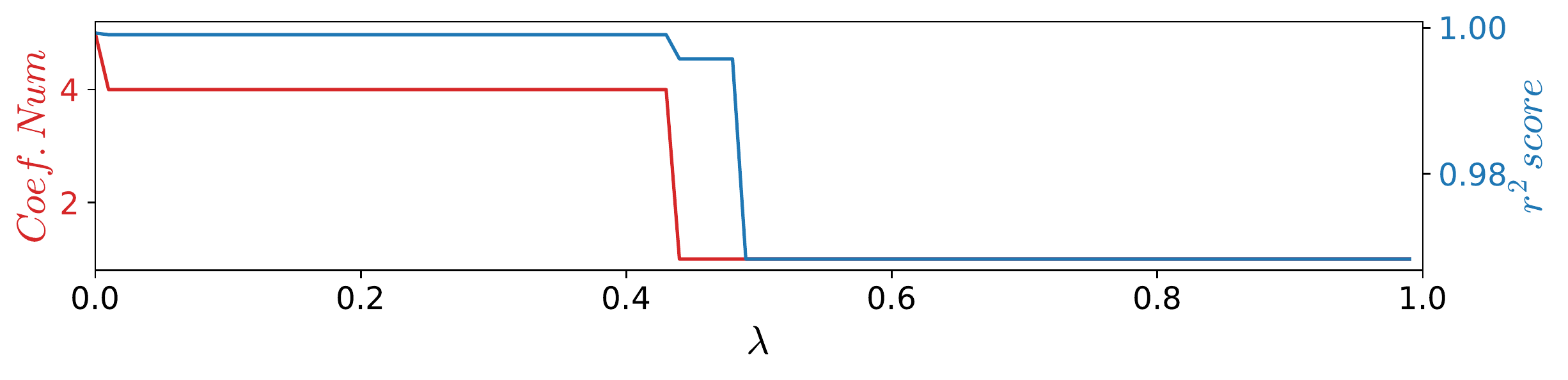} }
\end{tabular}
\begin{tabular}{ccc}
\includegraphics[width=.25\linewidth]{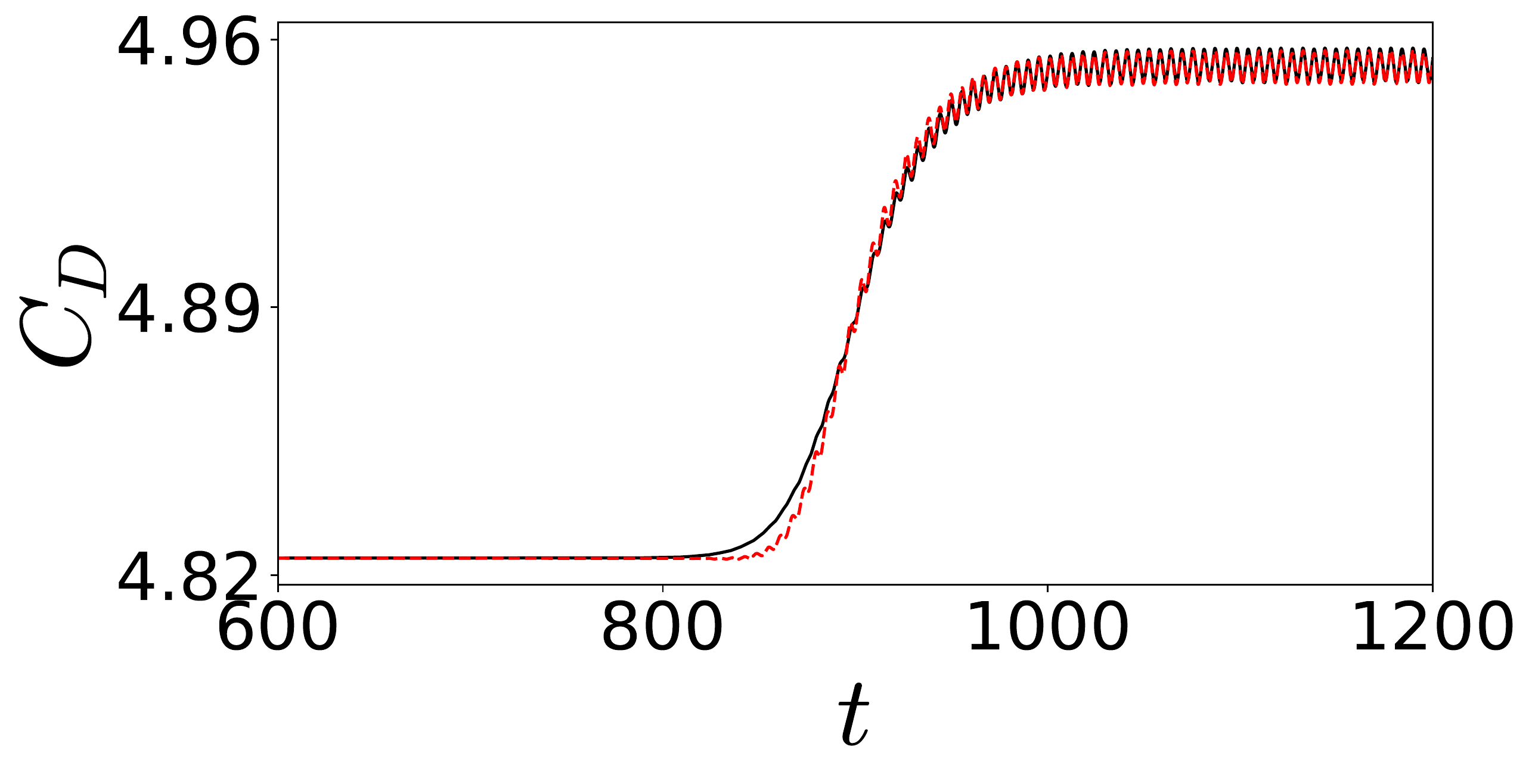} &
\includegraphics[width=.25\linewidth]{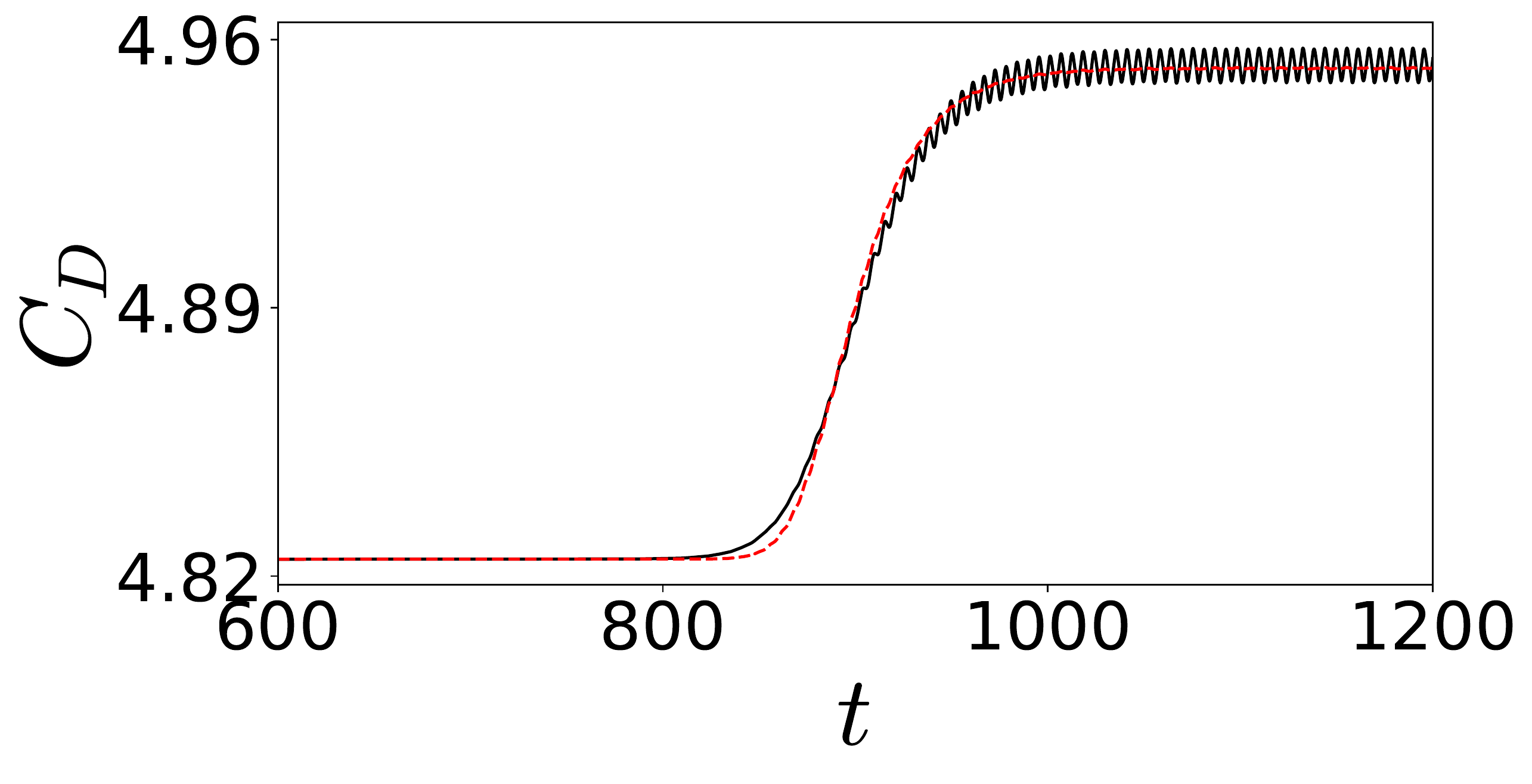} &
\includegraphics[width=.25\linewidth]{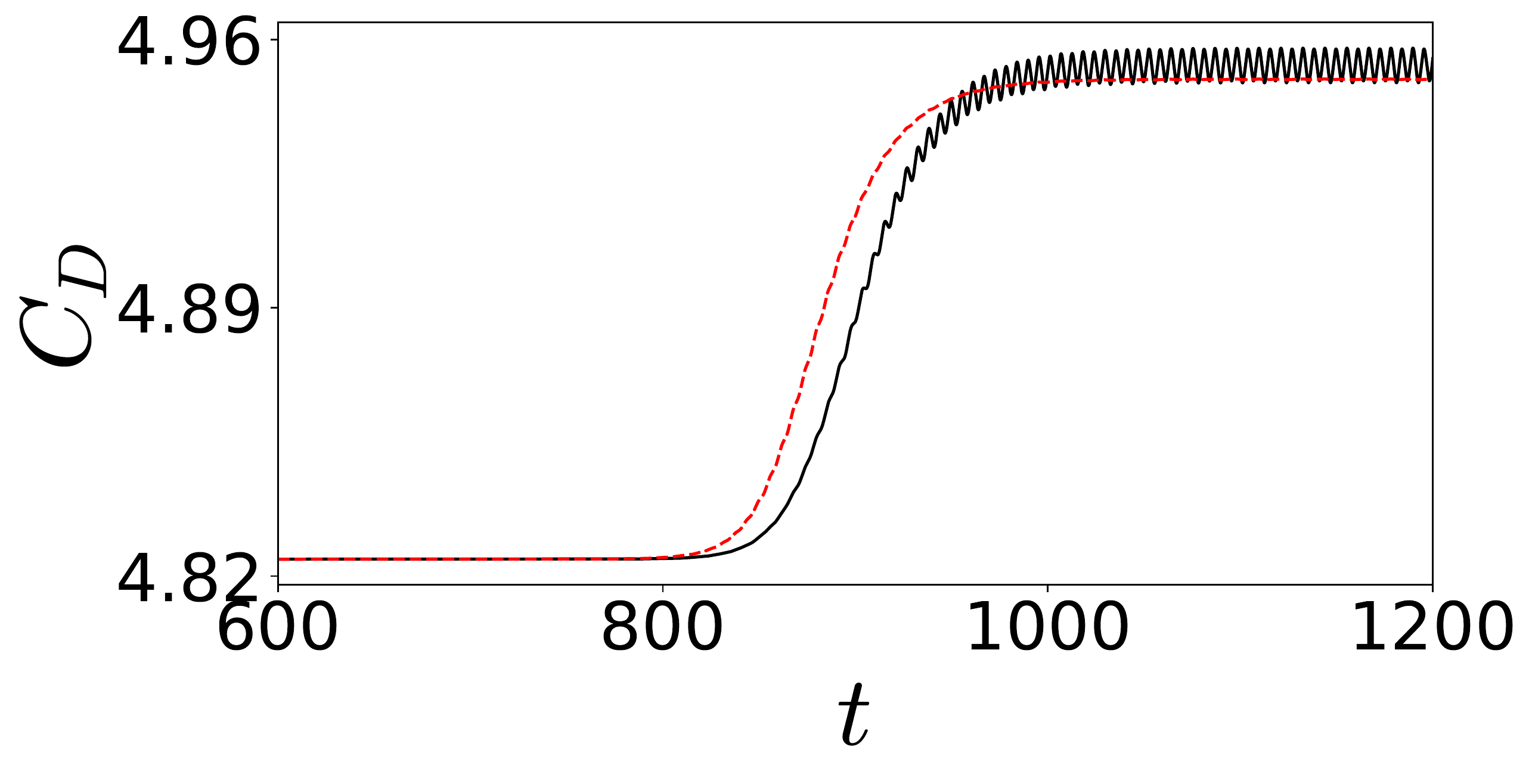} \\
(b) & (c) & (d) 
\end{tabular}
\caption{Illustration of the influence of the sparsity parameter $\lambda$ on both the complexity and accuracy of the identified drag model by the sequential thresholded least square regression with three degrees of freedom at $Re = 30$.
(a) Evolution of the number of non-zero coefficients (red) and of the $r^2$ score (blue) as a function of the sparsity parameter $\lambda$. 
Performance of the identified drag model at $\lambda = 0.3$ (b), $0.45$ (c), and $0.9$ (d). Time evolution of the drag $C_D$ coefficients in the full flow dynamics (solid black line) and for the force model (red dashed line). Initial condition: symmetric steady solution.}
\label{Fig:LSTSQ_Drag30}
\end{figure}

In \S~\ref{Sec:Hopf}, we derived the sparse drag model with three degrees of freedom at $Re = 30$.
Benefit from the low cost of computation for the regression test, we can iteratively run the algorithm with changing the sparsity parameter $\lambda$ and investigate the performance changes of the identified model.
The performance of the identified drag model by these two regression methods when varying $\lambda$ is illustrated in figure~\ref{Fig:LASSO_Drag30}(a) and figure~\ref{Fig:LSTSQ_Drag30}(a), together with a comparison with the real force dynamics for three typical values of $\lambda$.  

The sparsity parameter starts with $0$ (pure least square regression) and increases up to nearly $1$.
The structures of the resulting models at $\lambda=0.95$ for the LASSO regression, see figure~\ref{Fig:LASSO_Drag30}(d), and $\lambda=0.45$ for the STLS regression, see figure~\ref{Fig:LSTSQ_Drag30}(c), are identical, where only $a_3^2$ remains. 
However, the STLS regression is more sensitive to the sparsity parameter, as shown in figure~\ref{Fig:LSTSQ_Drag30}(a). The terms $a_{3}$, $a_1^2$ and $a_2^2$ are eliminated at the same time. The remaining $a_3^2$ is replaced by $a_3$ with $\lambda > 0.48$, and the identified models are obviously under-fitted, as shown in figure~\ref{Fig:LSTSQ_Drag30}(d).
In contrast, the LASSO regression eliminates the terms gradually, first $a_1a_2$, then $a_3$, and eventually $a_1^2$ together with $a_2^2$. 
In figure~\ref{Fig:LASSO_Drag30}(a), the elimination of $a_1^2$ and $a_2^2$ only reduces the $r^2$ score by $0.003$. 
But the loss of the fluctuating drag dynamics indicates an under-fitting.
Hence, the optimal $\lambda$ is found for $0.85$.

\begin{figure}
\centering
\begin{tabular}{cc}
\includegraphics[width=.5\linewidth]{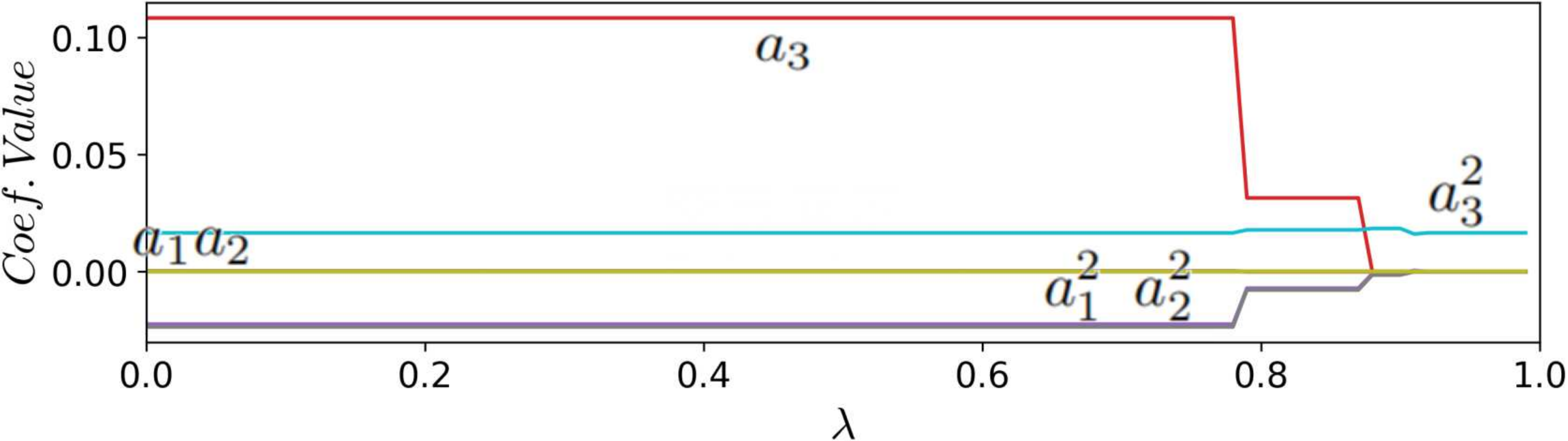} &
\includegraphics[width=.5\linewidth]{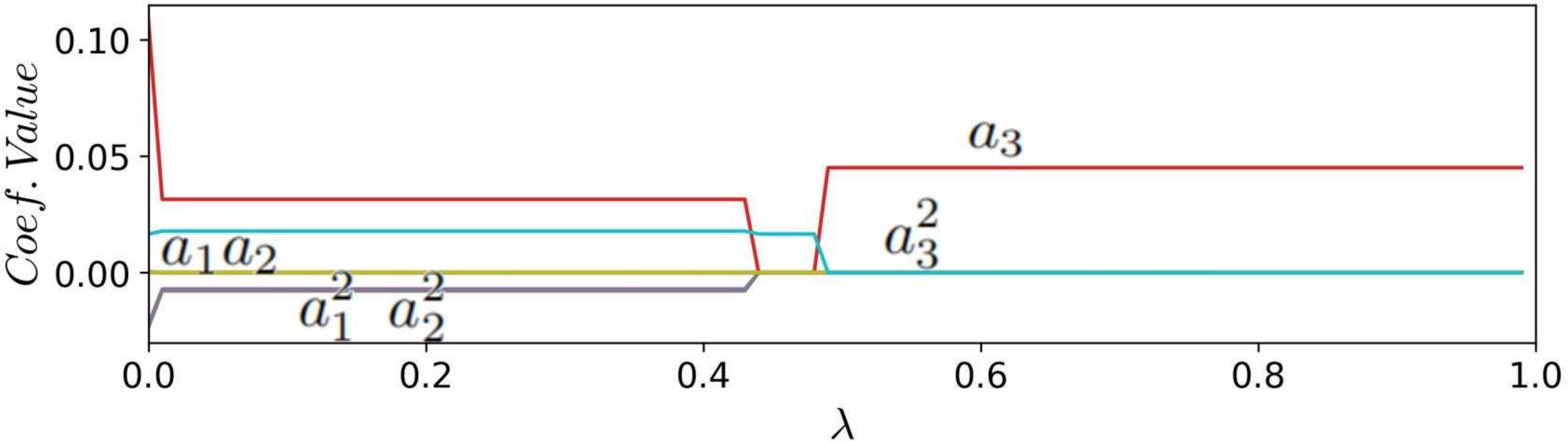} \\
(a) & (b) 
\end{tabular}
\caption{Evolution of the coefficients of the terms $a_1a_2$ (green), $a_3$ (red), $a_3^2$ (light blue), $a_1^2$ $a_2^2$ (purple), in the identified drag model as a function of the sparsity parameter $\lambda$ for (a) the LASSO regression and (b) the sequential thresholded least square regression.}
\label{Fig:Coef_Drag30}
\end{figure}

To figure out the reason for the failure of the identification with the STLS regression when $\lambda > 0.48$, we compare the evolution of the coefficients with increasing $\lambda$ in figure~\ref{Fig:Coef_Drag30}.  
$a_1a_2$ is the first eliminated term in both cases. 
The coefficients in the initial stage before the elimination of $a_3$ are almost the same. 
After the elimination of $a_3$ with the LASSO regression, as shown in figure~\ref{Fig:Coef_Drag30}(a), the coefficients of $a_1^2$ and $a_2^2$ become of order $O(10^{-3})$.
Since the STLS regression algorithm thresholds the terms with smaller coefficients, the tiny coefficients of $a_1^2$ and $a_2^2$ will be set to zero simultaneously. 
When the STLS regression is used with a too large sparsity parameter $\lambda$, the term with larger coefficient can survive.
As illustrated in figure~\ref{Fig:Coef_Drag30}(b), the remaining term $a_3^2$ is replaced by $a_3$ with a larger coefficient.
This explains the reason why the STLS regression final converges to $a_3$, which is obviously the wrong term for the real drag force dynamics in figure~\ref{Fig:LSTSQ_Drag30}(d).

\begin{figure}
\centering
\begin{tabular}{cc}
(a) & \raisebox{-0.55\height}{\includegraphics[width=.9\linewidth]{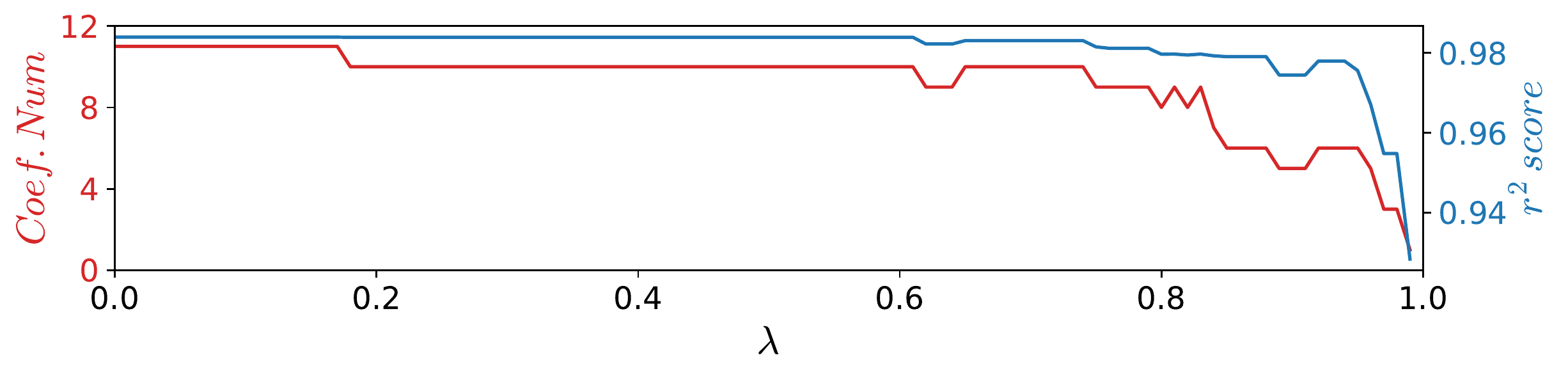} } \\
(b) & \raisebox{-0.55\height}{\includegraphics[width=.9\linewidth]{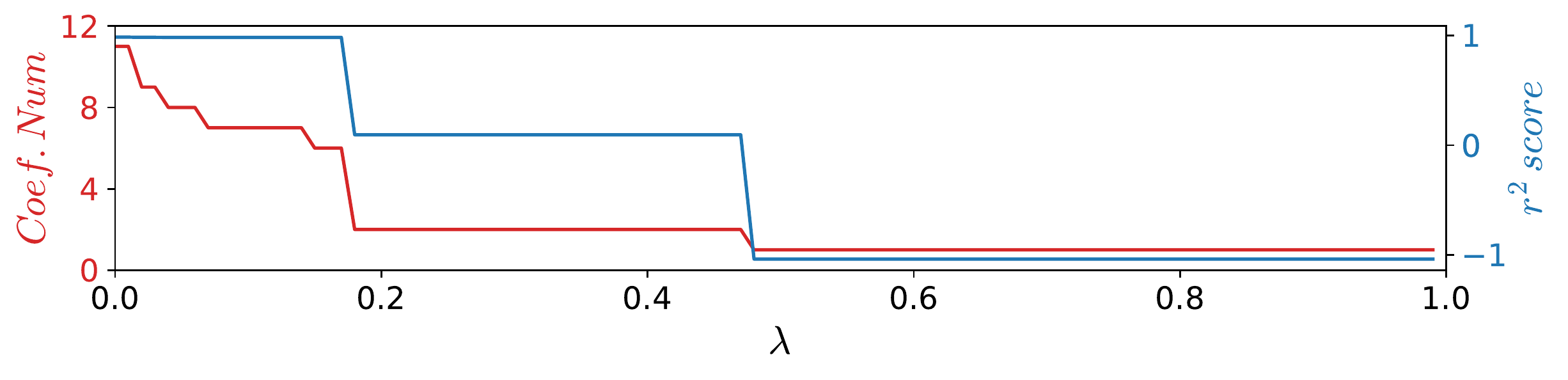} }
\end{tabular}
\caption{Illustration of the influence of the sparsity parameter $\lambda$ on both the complexity and accuracy of the identified drag model by (a) the LASSO regression, 
(b) the sequential thresholded least square regression,with three degrees of freedom at $Re = 80$.
Evolution of the number of non-zero coefficients (red) and of the $r^2$ score (blue) as a function of the sparsity parameter $\lambda$.}
\label{Fig:Nscore_Drag80}
\end{figure}

We apply the same analysis for the sparse drag model with five degrees of freedom at $Re = 80$, as described in \S~\ref{Sec:ForceModel5DOF}. 
The evolution of the performances under the two regression methods are shown in figure~\ref{Fig:Nscore_Drag80}.
The STLS regression goes in the wrong direction as $\lambda > 0.17$.
After checking the list of coefficients, the key term $a_3^2$ is deleted irretrievably, resulting in the inability of the model to fit correctly.
However, the regression result right before the critical value provides the most simplified and relevant drag model of Eq.~\eqref{Eqn:forcemodel_5dof_a} with $r^2 = 0.9816$. 

The LASSO regression is much safer on the elimination of terms. $a_3^2$ can survive during the regression in all the range of $\lambda$ from $0$ to almost $1$.
This further indicates that the key terms can own better robustness in the LASSO regression. 
From figure~\ref{Fig:Nscore_Drag80}(a), the optimal $\lambda$ is chosen at $0.85$, involving six terms and $r^2 = 0.9791$. 
Although there are only five terms remaining when $\lambda = 0.9$, the resulting model is not stable.
It returns to six terms and $r^2 = 0.9755$ at $\lambda = 0.95$, with different active terms compared to the model at $\lambda = 0.85$.
At the optimal value, the identified drag model consists in terms $a_5$, $a_1^2$, $a_1a_2$, $a_2^2$, $a_3^2$, $a_3a_5$, where $a_5^2$ is missing. 
Since $a_1a_2$ is of order $O(10^{-4})$, we can directly apply the least square regression on the updated library with deleting $a_1a_2$ and adding $a_5^2$. The regression result is the same as for the STLS regression.

\section{Limitations of the purely projection-based approach}
\label{Sec:RealForce}

From the expression of pressure and viscous force on the body in \S~\ref{Sec:ForceSurface}, 
the force contribution of each velocity mode in the Galerkin expansion can be numerically determined, as in \citet{liang2014virtual}.

The viscous force associated with mode $\bm{u}_j$ can be explicitly calculated through $q^{\nu}_{\alpha; j}$ in Eq.~\eqref{Eqn:ViscousForce-real}.
However, solving $q^{p}_{\alpha; jk}$ in Eq.~\eqref{Eqn:PressureForce-real} needs a homogeneous Neumann boundary condition for the pressure,
i.e. the normal derivative of $p$ in the outward direction $\bm{n}$ must vanish on the whole domain boundary $\partial \Omega$,
\begin{eqnarray} 
\label{Eqn:Neumann-P}
\partial_n p = \bm{n} \cdot \nabla p = 0 .
\end{eqnarray}
In this study, we apply a no-slip condition on velocity without the above-mentioned Neumann boundary condition on pressure. Hence, the partial pressure fields $p_{jk}$ can not be determined to a constant pressure field. Analogously, $q^{p}_{\alpha; jk}$ can not be solved with an exact value. 
Even if we assume Neumann boundary conditions for the pressure field $p$, it is still a numerically challenging work since the pressure field are expanded to numerous partial pressure fields $p_{jk}$, see Eq.~\eqref{Eqn:PressureExpansion}.
 
\begin{figure}
 \centering
 \begin{tabular}{cc}
 \includegraphics[width=.45\linewidth]{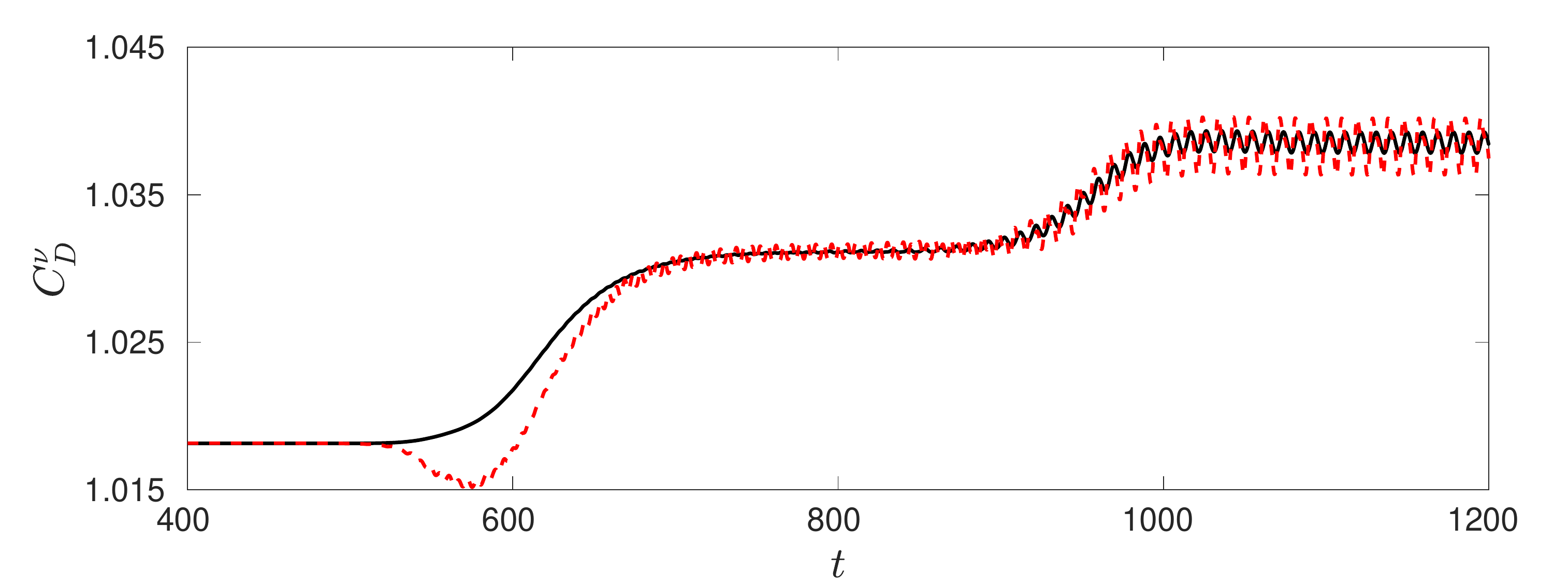} &
 \includegraphics[width=.45\linewidth]{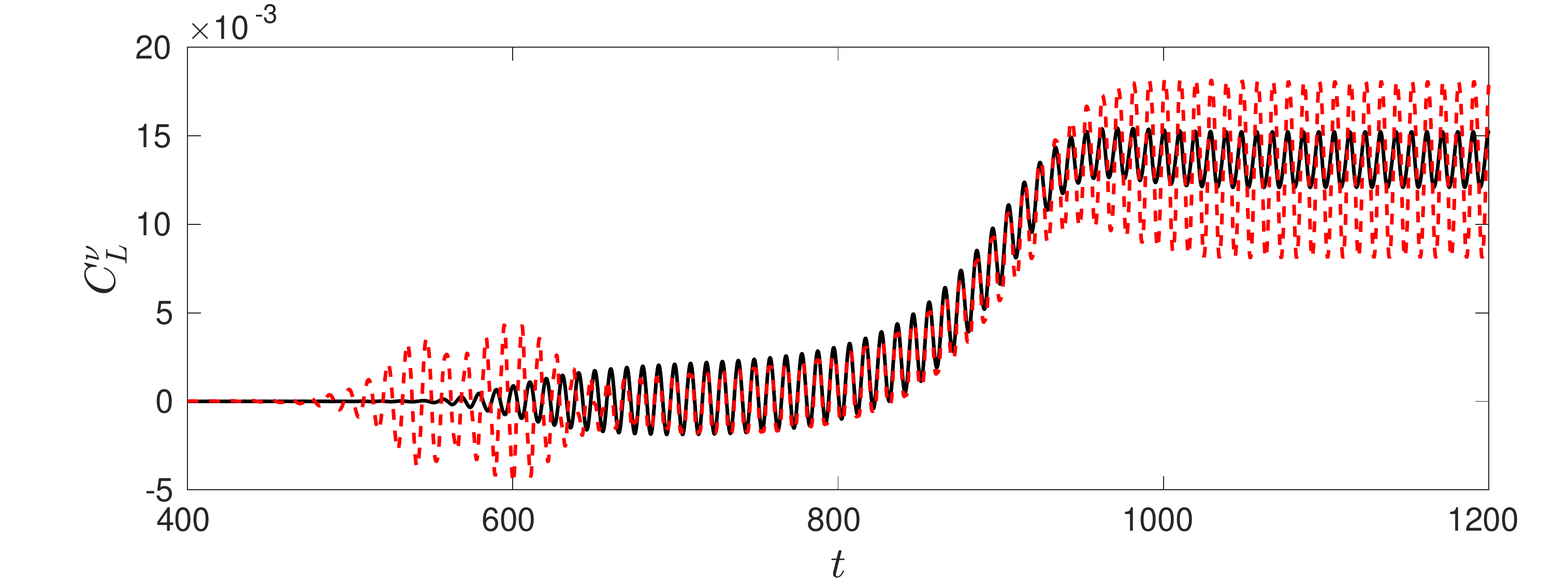} \\ 
 \end{tabular}
\caption{Performance of the force model with the real forces contribution of seven bifurcation modes. Time evolution of the viscous drag $C_D^{\nu}$ (left) and the viscous lift $C_L^{\nu}$ (right) coefficients in the full flow dynamics (solid black line) and for the force model (red dashed line) for DNS starting from the symmetric steady solution $\bm{u}_s$ at $Re=80$.}
\label{Fig:Realforcemodel_Re80}
\end{figure}

Without considering the pressure force contribution, we can reconstruct the viscous force from the viscous force contribution of the bifurcation modes. 
The resulting viscous force model only contains linear terms and reads
\begin{subequations}
\label{Eqn:Realforcemodel_Re80}
\begin{eqnarray}
\notag C_D^{\nu} &=& 1.01814664 + 0.00159948\> a_3 - 0.0023798\> a_5 + 0.00601715 \> a_7,\\
\notag C_L^{\nu} &=& 0.000267167\> a_1 + 0.00004522\> a_2 - 0.01409768\> a_4 - 0.0055717\> a_6.
\end{eqnarray}
\end{subequations}
The viscous force contributions of each bifurcation mode is explicitly computed without any symmetry assumption, no sparsity can be expected in this model.
Yet, after eliminating terms with a coefficient less than $O(10^{-5})$, the resulting force models \eqref{Eqn:Realforcemodel_Re80} only involve the terms associated with the bifurcations modes with the appropriate symmetry, indicated as the symmertic modes $\bm{u}_3$, $\bm{u}_5$, $\bm{u}_7$ in $C_D^{\nu}$ and the symmertic modes $\bm{u}_1$, $\bm{u}_2$, $\bm{u}_4$, $\bm{u}_6$ in $C_L^{\nu}$.
The performance of the force model using the real viscous force contribution of the seven bifurcation modes is illustrated in figure~\ref{Fig:Realforcemodel_Re80}.
The $r^2$ score for the viscous drag model is $0.9786$ and $0.9183$ for the viscous lift model. The accuracy and the predictive ability of the force model are acceptable for the drag model with only three items and the lift model with only four terms.

\section{Limitation of the POD-based force model}
\label{Sec:Force-POD}

We apply POD on the fluctuating flow field $\bm{u}(\bm{x},t)- \bm{u}_s(\bm{x})$, where $\bm{u}_s$ is the symmetric steady Navier-Stokes solution described in \S~\ref{Sec:Galerkin}. 
The snapshots used for the POD come from the two mirror-conjugated DNS trajectories started close by the symmetric steady solution.
The POD mode expansion of the flow field reads:
\begin{eqnarray} 
\label{Eqn:PODExpansion}
\bm{u} (\bm{x},t ) & = & \bm{u}_s + \sum\limits_{j=1}^N a_j(t) \bm{u}_j ( \bm{x} ),
\end{eqnarray}

\begin{figure}
 \centering
 \begin{tabular}{cc}
 (a) & (b) \\
 \includegraphics[width=.45\linewidth]{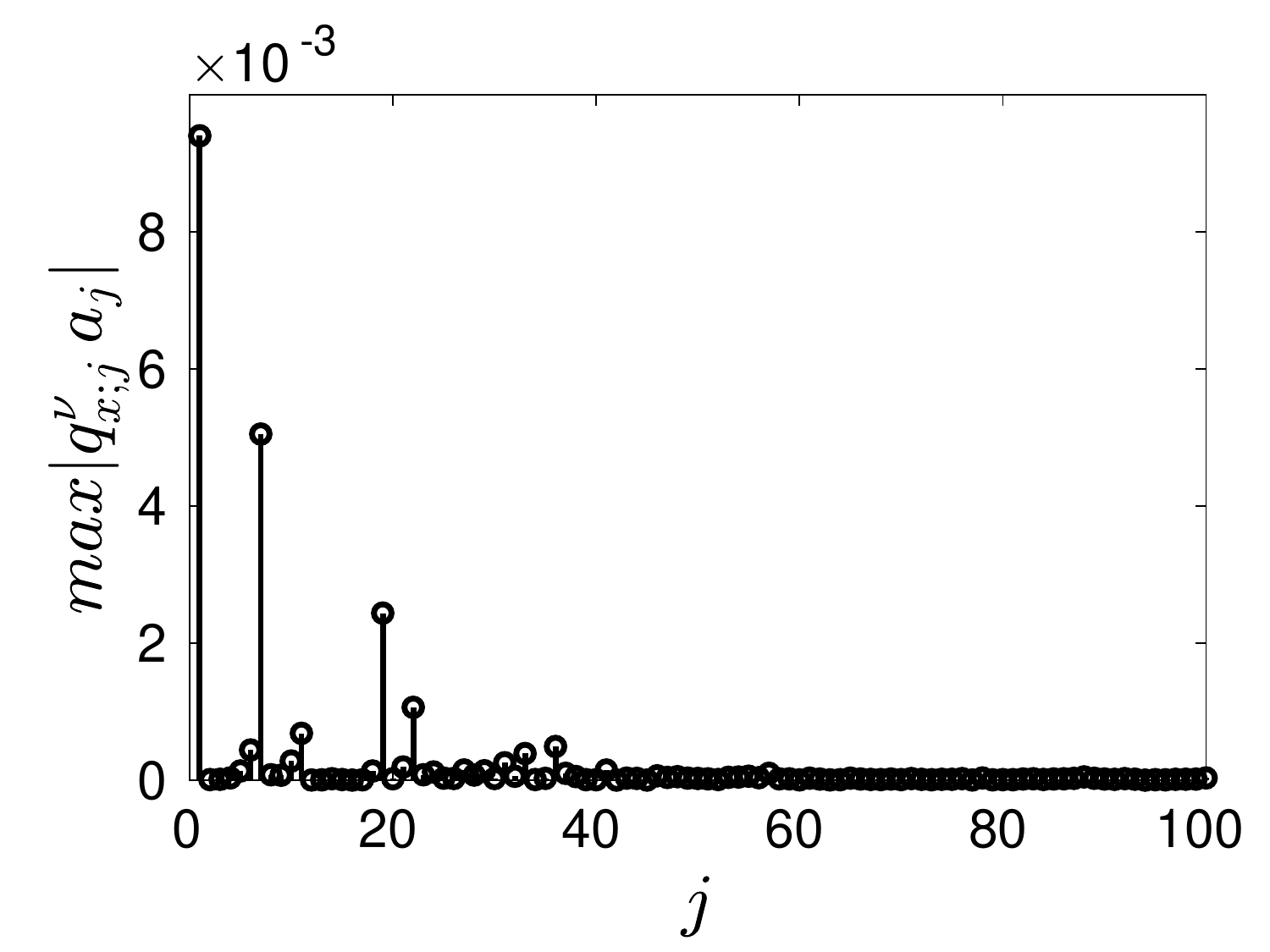} 
 & \includegraphics[width=.45\linewidth]{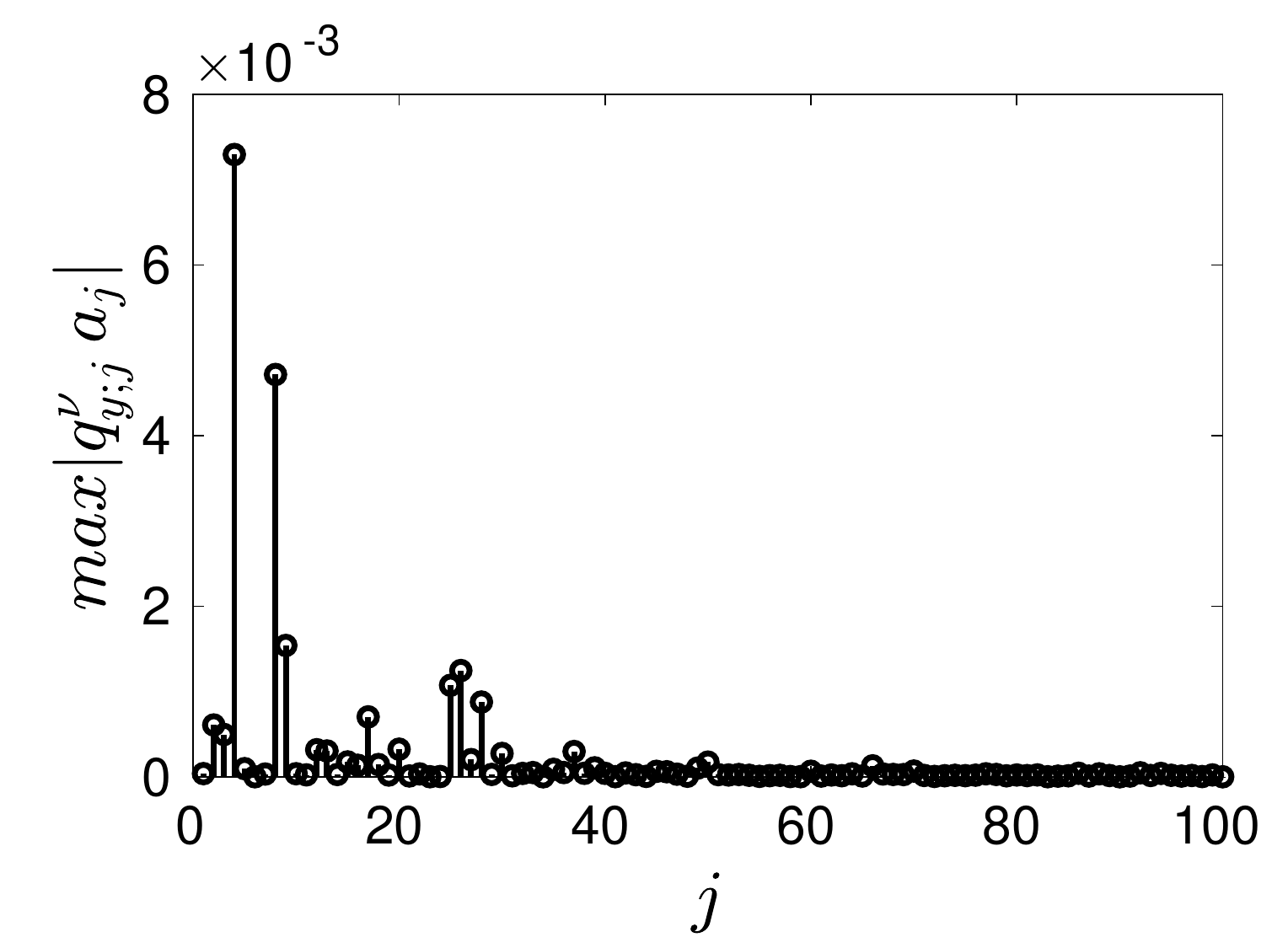} \\ 
 \end{tabular}
\caption{Contribution of the POD modes $\bm{u}_j$ to the viscous (a) drag and (b) lift forces for DNS starting from the symmetric steady solution $\bm{u}_s$ at $Re=80$.}
\label{Fig:POD80}
\end{figure}

\begin{figure}
 \centerline{
 \begin{tabular}{ccc}
 &$C_D^{\nu} $ &  $C_L^{\nu} $\\
 \hline \hline
(a) &  & \\
 & \includegraphics[width=.45\linewidth]{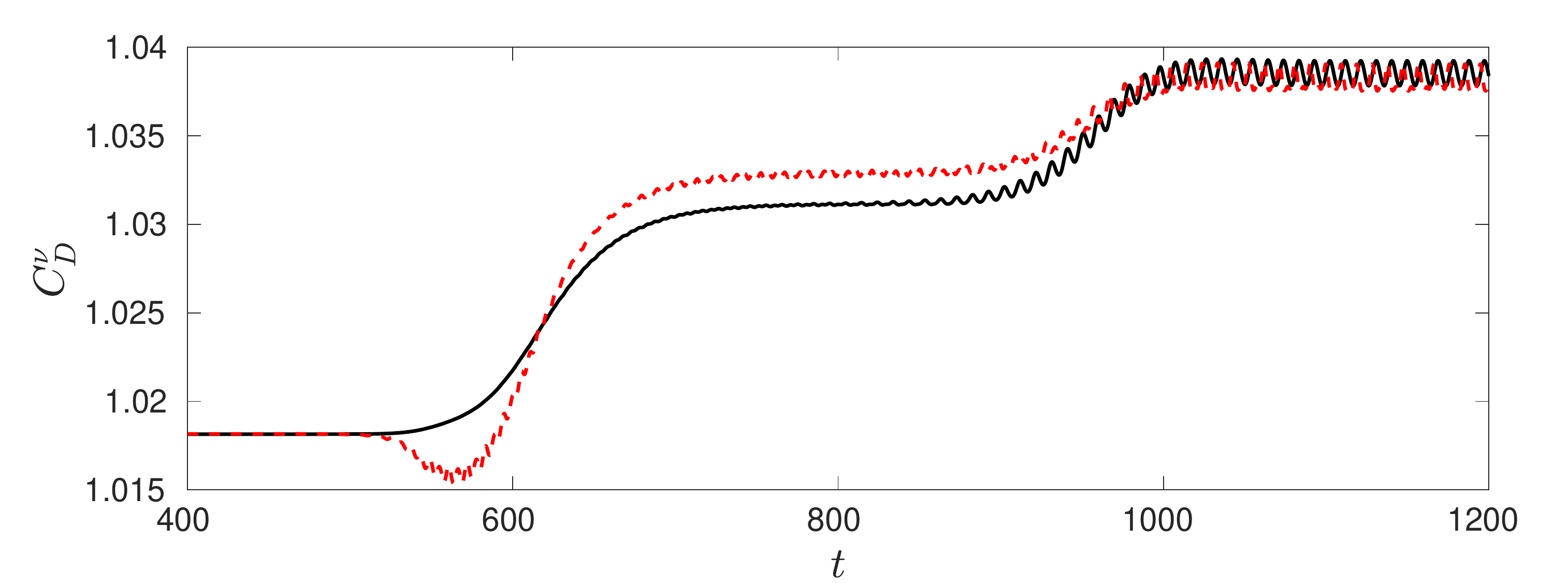} 
 & \includegraphics[width=.45\linewidth]{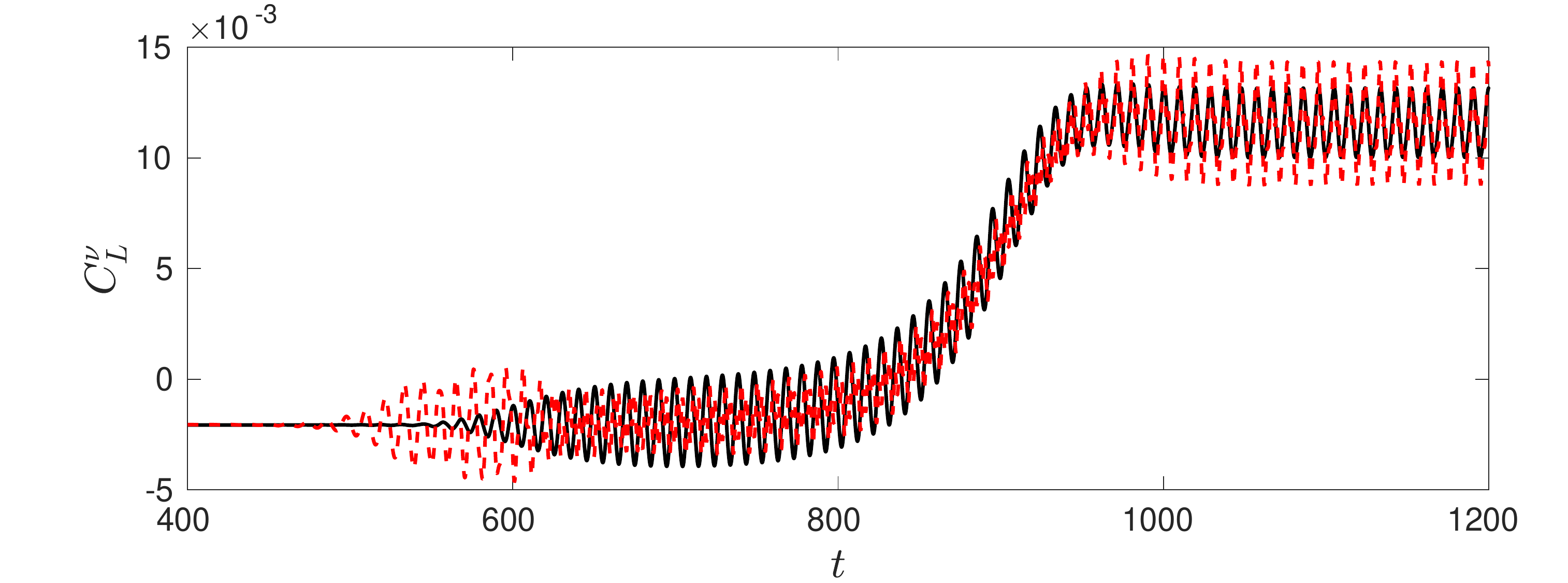}  \\ 
(b) &  & \\
 & \includegraphics[width=.45\linewidth]{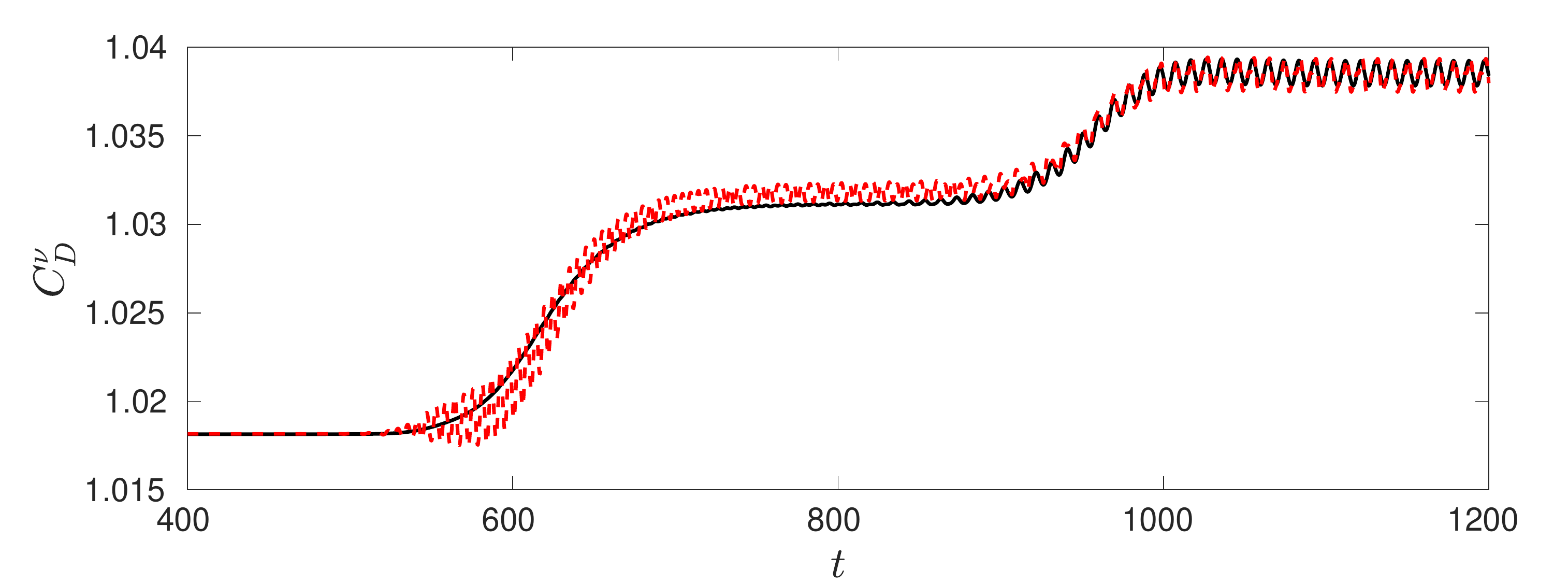} 
 & \includegraphics[width=.45\linewidth]{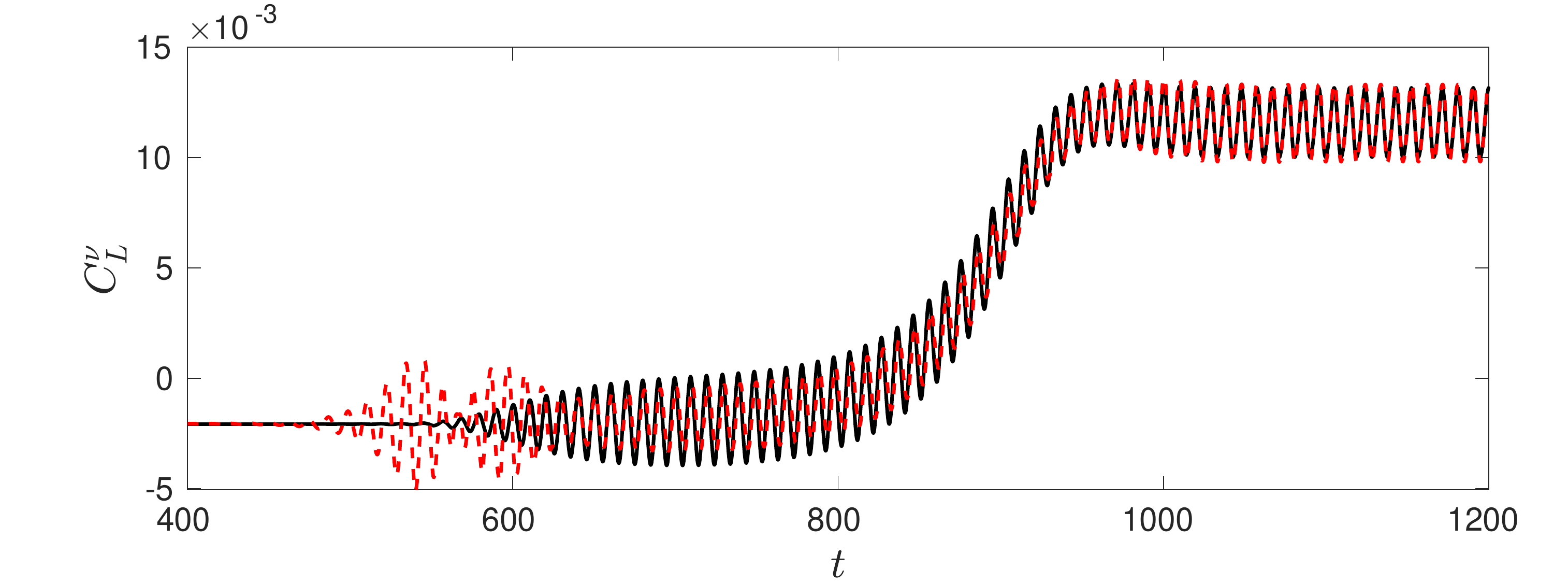}  \\
(c) &  & \\
 & \includegraphics[width=.45\linewidth]{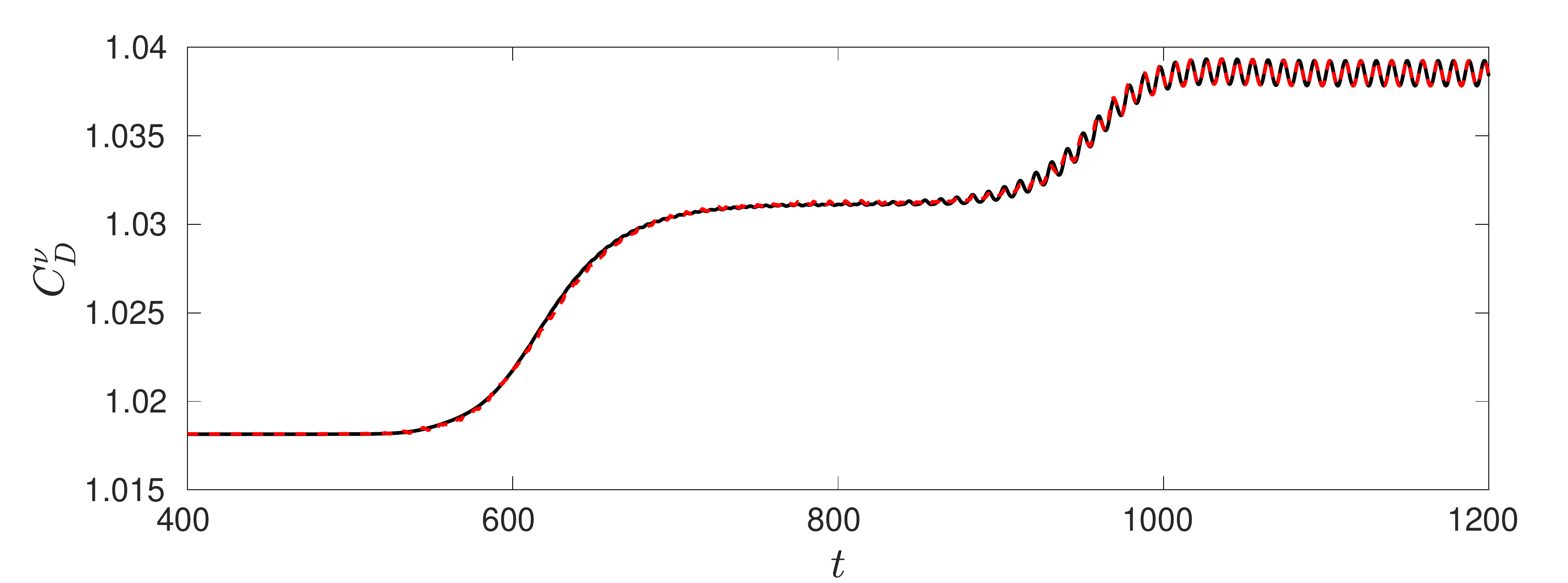}  
 & \includegraphics[width=.45\linewidth]{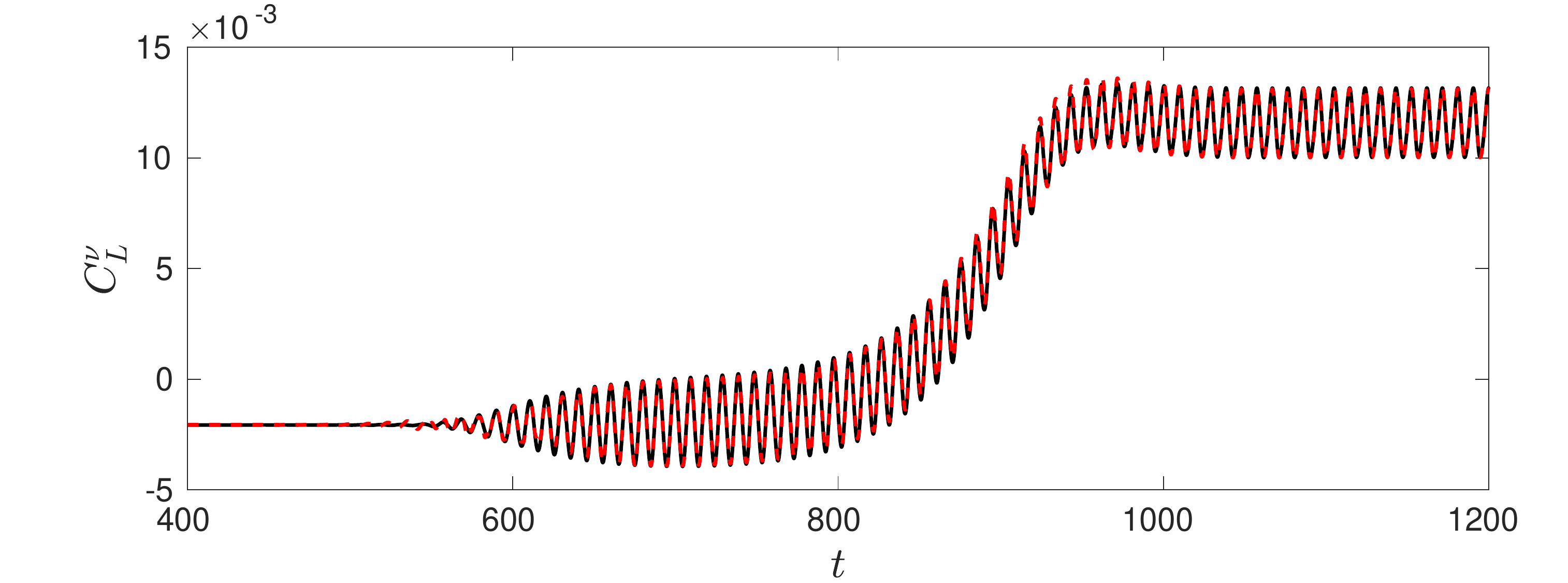} \\
 \end{tabular}
 }
\caption{Viscous drag (left) and lift (right) force reconstruction with the (a) $N=10$, (b) $N=20$, (c) $N=50$ leading POD modes starting from the symmetric steady solution $\bm{u}_s$ at $Re=80$. Real force dynamics computed from the DNS (black curve), reconstructed forces from the $N$ leading POD modes (dashed red line).}
\label{Fig:ForceReconstruction_POD80}
\end{figure}

Due to the lack of boundary conditions for the pressure field contribution, we only focus on the reconstruction of the viscous force with the purely projection-based approach. 
The contribution to the viscous drag and lift forces, given by $\max{|q^{\nu}_{\alpha; j} a_j|}$, with $\alpha=x, y$, are shown in figure~\ref{Fig:POD80}.
The main force contribution comes from the leading $50$ POD modes. 
The viscous force reconstructed with the $N$ leading POD mode amplitudes reads
\begin{equation}
\label{Eqn:ViscousForceExpansion}
F_{\alpha}^{\nu} 
= c_{\alpha}^{\nu} 
+ \sum\limits_{j=1}^N q_{\alpha; j}^{\nu} a_j.
\end{equation}
The viscous drag $C_D^{\nu}$ and lift $C_D^{\nu}$ coefficients reconstructed with different numbers of POD modes are compared to the real force dynamics in figure~\ref{Fig:ForceReconstruction_POD80}.

For a sequential $N$, the error of the reconstructed force coefficients with $N$ leading POD modes can be also evaluated with the $r^2$ score. 
A higher $r^2$ score indicates less error in the reconstructed force.
As expected, the error tends to decrease when the number of POD modes is increased. 
To achieve $r^2 > 0.999$, $N = 36$ leading POD modes are required for the drag force, and $N = 51$ modes for $r^2 > 0.9999$. For the lift force, these two critical numbers are respectively $N = 30$ and $N = 86$. In actual situations, the model with $r^2 > 0.999$ has enough accuracy.
\begin{figure}
 \centering
 \begin{tabular}{cc}
  (a) & (b) \\
 \includegraphics[width=.45\linewidth]{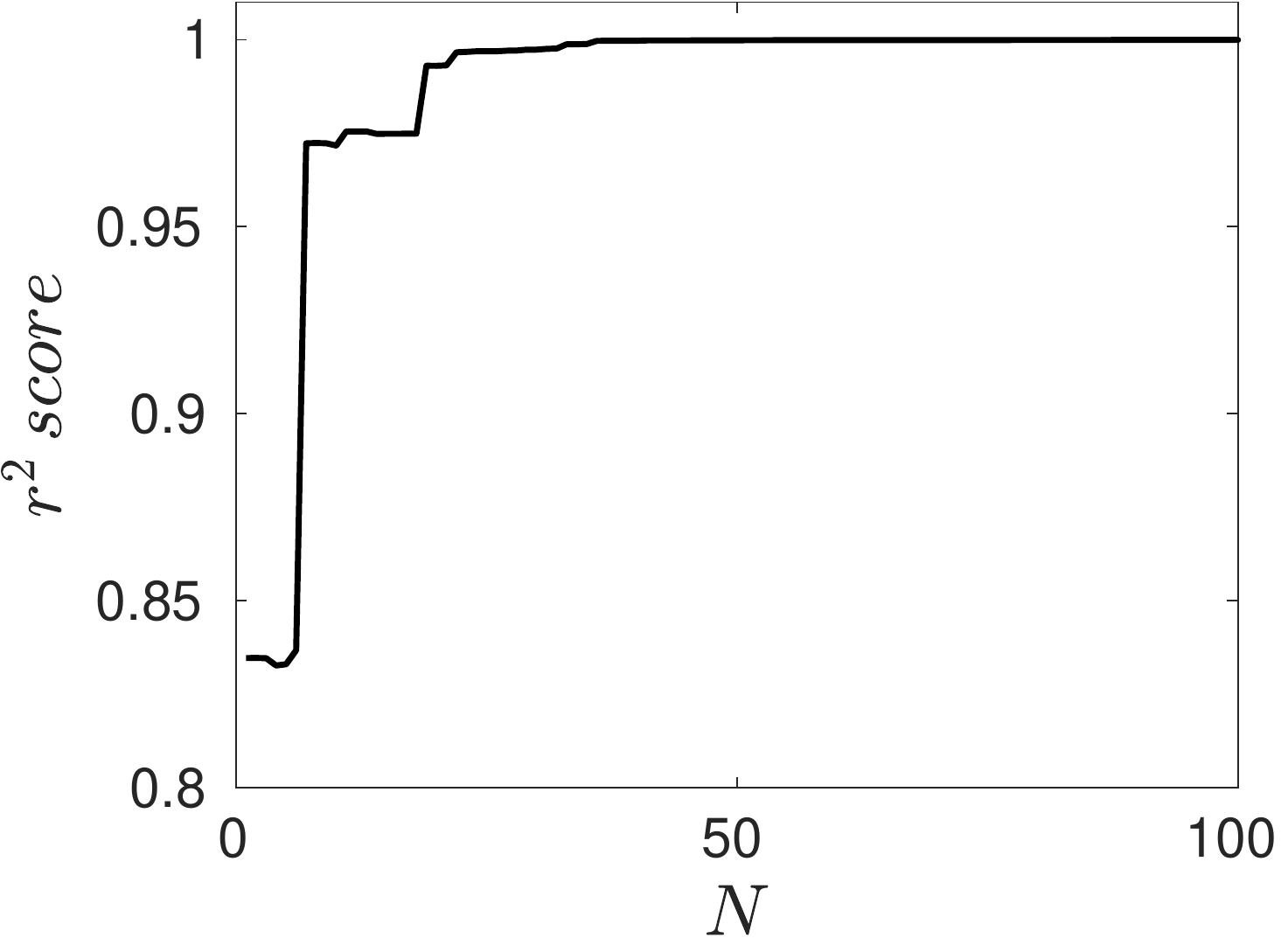} &
 \includegraphics[width=.45\linewidth]{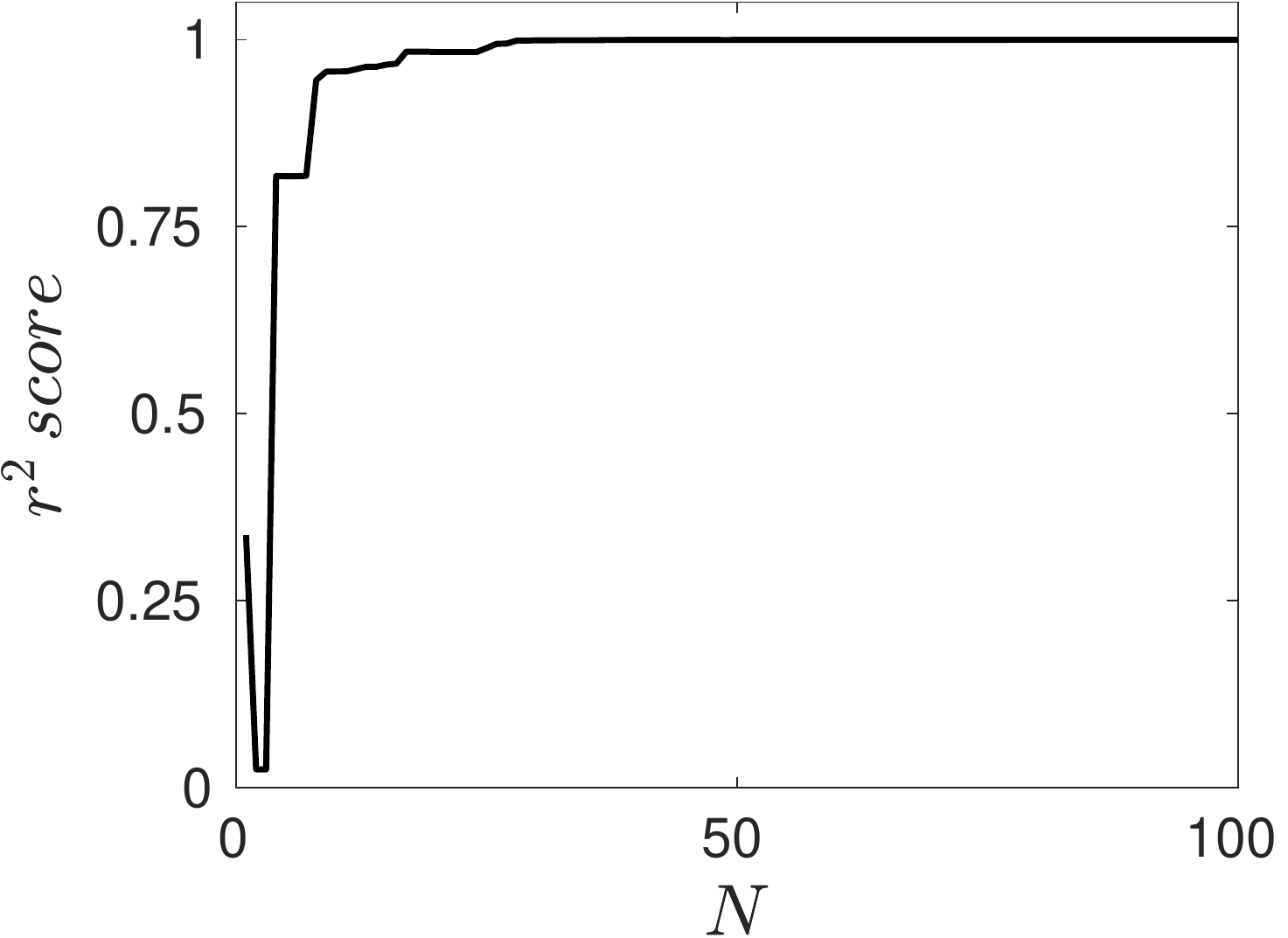} \\ 
 \end{tabular}
\caption{Error on the viscous (a) drag and (b) lift force reconstruction with the $N$ leading POD modes starting the DNS starting from the symmetric steady solution $\bm{u}_s$ at $Re=80$.}
\label{Fig:ErrorPOD80}
\end{figure}
Note that no sparsity is involved in the model because the force contribution of each POD mode is computed explicitly.

We now focus on the regression-based approach, we set a truncation of the model with $N=10, 20, 50$ leading POD modes, and try to use the sparse regression to find a drag model with a balance between accuracy and complexity. 
To be noted, the drag force considered here involves both the pressure and viscous contributions to the force. 
To reach the same $r^2$ score, it requires more POD modes due to the additional quadratic complexity of the pressure force contribution.
\begin{figure}
\centering
\begin{tabular}{cc}
(a) & \raisebox{-0.55\height}{\includegraphics[width=.9\linewidth]{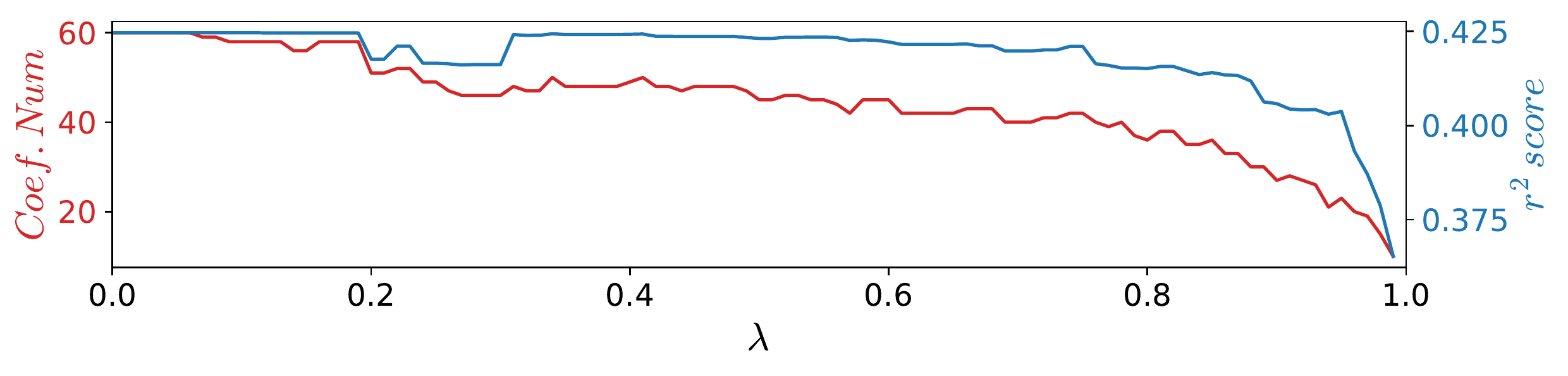} } \\
(b) & \raisebox{-0.55\height}{\includegraphics[width=.9\linewidth]{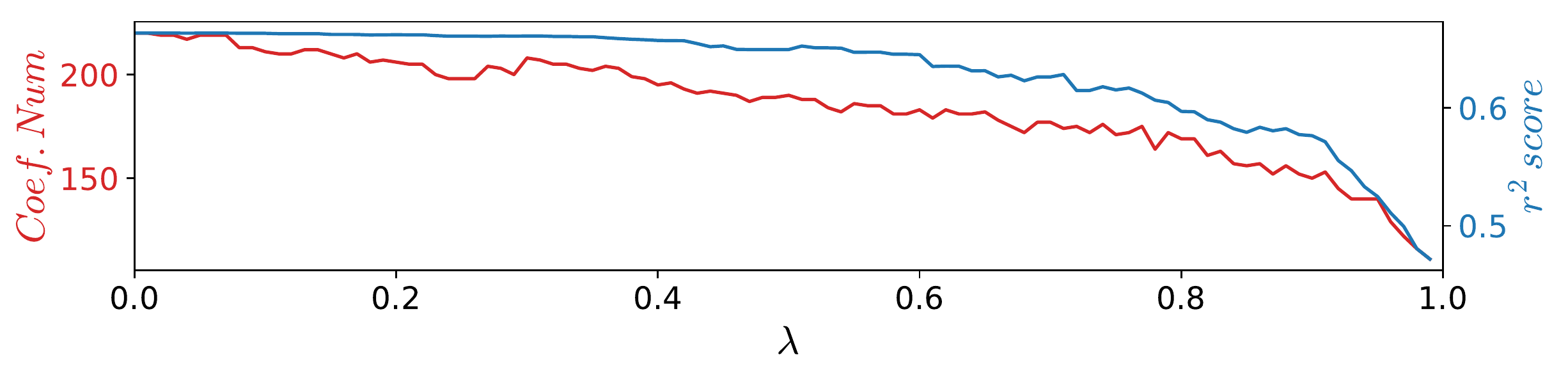} } \\
(c) & \raisebox{-0.55\height}{\includegraphics[width=.9\linewidth]{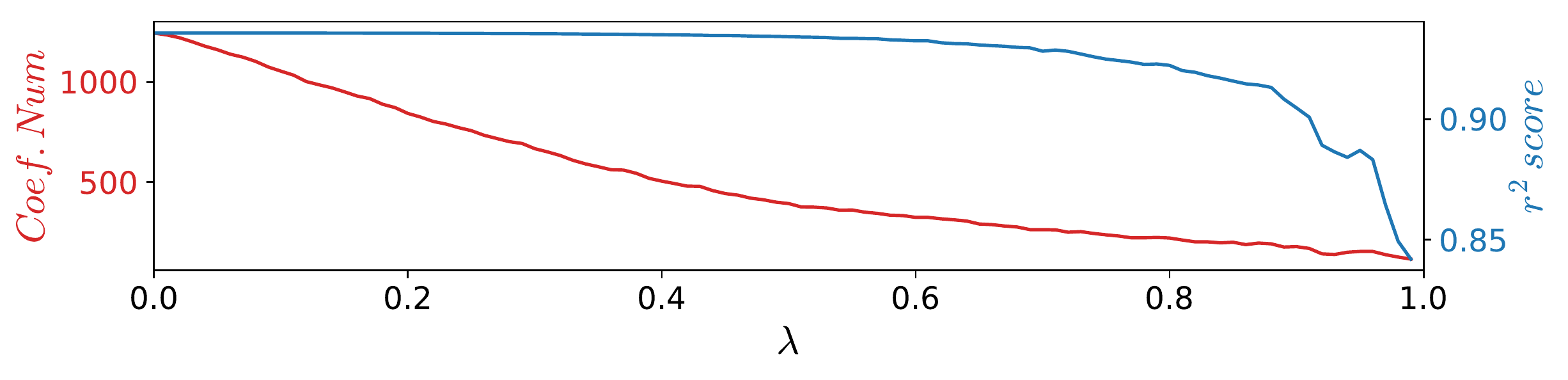} }
\end{tabular}
\caption{Illustration of the influence of the sparsity parameter $\lambda$ on both the complexity and accuracy of the identified drag model by the LASSO regression with the (a) $N=10$, (b) $N=20$, (c) $N=50$ leading POD modes at $Re = 80$.
Evolution of the number of non-zero coefficients (red) and of the $r^2$ score (blue) as a function of the sparsity parameter $\lambda$.}
\label{Fig:Nscore_POD80}
\end{figure}

The library of mode amplitudes contains $66$, $231$ and $1326$ candidate terms for the $N=10,20,50$ leading POD modes.
However, as shown in figure~\ref{Fig:Nscore_POD80}, the least square regression result for $N=10,20$ cannot reach a $r^2$ score higher than $0.7$. 
Only the situation with $N=50$ can start with $r^2=1$, but hundreds of terms are still required for an acceptable accuracy. The interpretability of the identified model is hopeless.

In summary, POD modes are decomposed and sorted according to energy criteria.
The constant-linear-quadratic expression for the drag and lift forces can still be derived, but it requires a large number of POD modes. 
From a purely numerical approach, no sparsity is imposed in the model.
For the regression-based approach, the library of sparse regression is polluted with harmonic modes and noise. 
Too many degrees of freedom and the harmonic relationships between them make it hard to derive a simple model from sparse regression.
The most feasible solution is to find the dynamically related degrees of freedom between these modes --- as we actually did in our approach. 
Another possible direction is to optimize the sparse regression process, in order to select the key degrees of freedom out of a polluted library of too many degrees of freedom.

\section{Reduced-order model with seven degrees of freedom}
\label{Sec:MFM-7dof}
In \cite{deng2020jfm}, the reduced-order model of the fluidic pinball dynamics was derived for five degrees of freedom at $Re=80$, namely $a_1$ to $a_5$. Here we generalize the reduced-order model for seven modes, by adding $a_6$ and $a_7$ to the model. The new system reads
\begin{subequations}
\label{Eqn:Hopf+Pitchfork}
\begin{eqnarray}
	da_1/dt & = & a_1(\sigma_1 - \beta\ a_3 - \beta_{15}\ a_5) - a_2(\omega_1 + \gamma\ a_3 + \gamma_{15}\ a_5) + l_{14}\ a_4 + q_{134}\ a_3a_4,\\
	da_2/dt & = & a_2(\sigma_1 - \beta\ a_3  - \beta_{15}\ a_5) + a_1(\omega_1 + \gamma\ a_3 + \gamma_{15}\ a_5) + l_{24}\ a_4 + q_{234}\ a_3a_4,\\
	da_3/dt & = & \sigma_3\ a_3 +\beta_3\ r + l_{35}\ a_5 + q_{314}\ a_1a_4 + q_{335}\ a_3a_5 + q_{355}\  a_5^2, \\
	da_4/dt & = & \sigma_4\ a_4 - \beta_4\ a_4a_5 + a_1(l_{41} + q_{413}\ a_3 + q_{415}\ a_5) + a_2(l_{42} + q_{423}\ a_3 + q_{425}\ a_5), \qquad \qquad   \\
	da_5/dt & = & \sigma_5\ a_5 + \beta_{5}\ a_4^2 + l_{53}\ a_3 + q_{514}\ a_1a_4 + q_{533}\ a_3^2 + q_{535}\ a_3a_5,\\	da_6/dt & = & \sigma_6\ a_6 - \beta_6\ a_6a_7 + a_1(l_{61} + q_{613}\ a_3 + q_{617}\ a_7) + a_2(l_{62} + q_{623}\ a_3 + q_{627}\ a_7), \qquad \qquad   \\
	da_7/dt & = & \sigma_7\ a_7 + \beta_{7}\ a_6^2 + l_{73}\ a_3 + q_{716}\ a_1a_6 + q_{726}\ a_2a_6 + q_{733}\ a_3^2 + q_{737}\ a_3a_7.
\end{eqnarray}
\end{subequations}

\begin{figure}
\begin{center}
 \includegraphics[width=.9\textwidth]{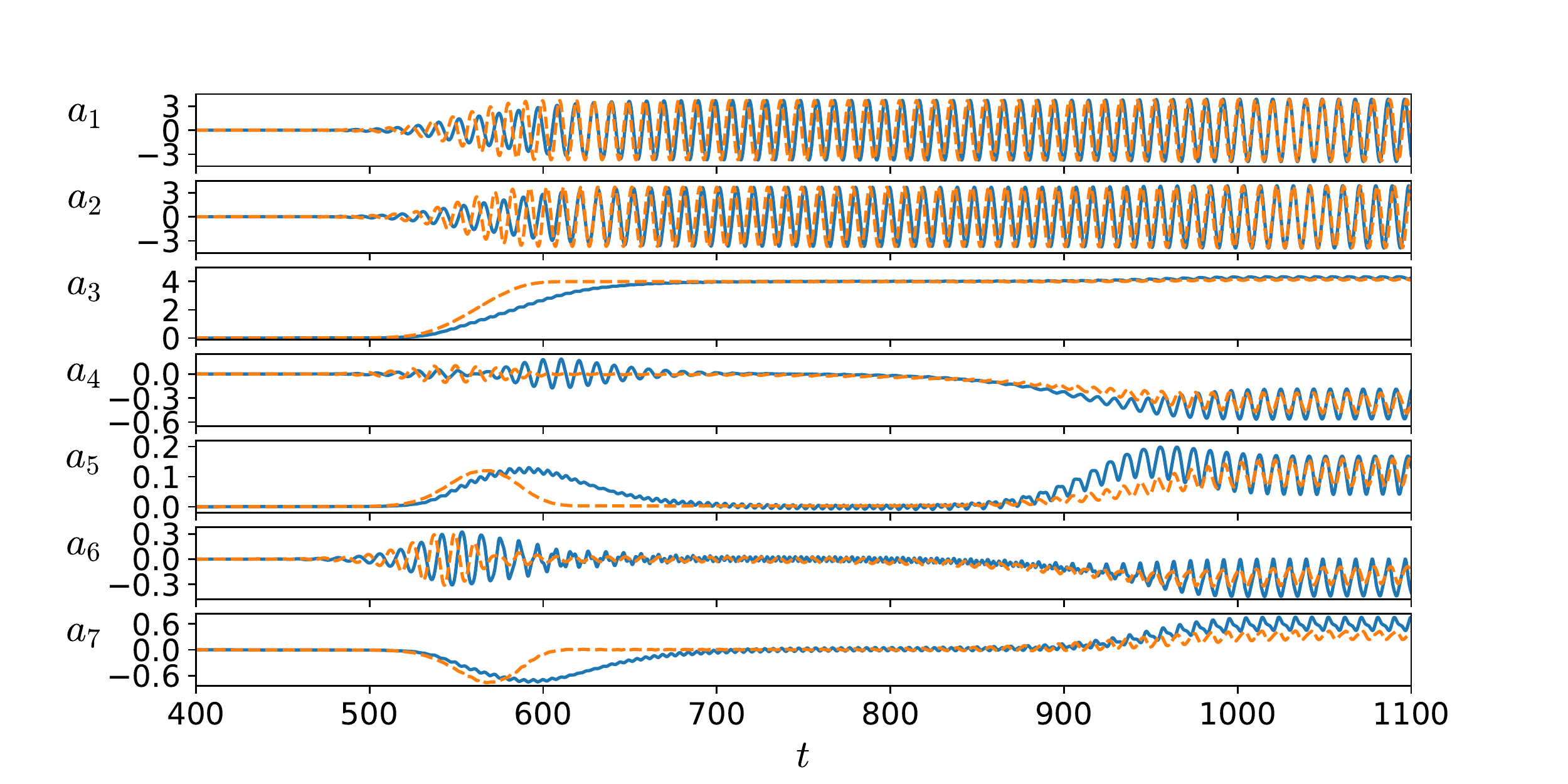}  
\caption{Performance of the reduced-order model with cross-terms. Time evolution of coefficients $a_1$ to $a_7$ in the full flow dynamics (solid blue line) and for the reduced-order model (red dashed line). The initial condition is the same for the reduced-order model and the full flow dynamics. }
\label{Fig:TS7modes-Re80} 
\end{center}
\end{figure}

The identified system coefficients are recorded in table~\ref{Tab:Pitchfork}, and the model performance is exemplified in figure~\ref{Fig:TS7modes-Re80}.
\begin{table}
 \begin{center}
 \begin{tabular}{lrlrlrlr}
  $\sigma_1$ & $5.22 \times 10^{-2}$ \quad & \quad $\beta$ & $1.31 \times 10^{-2}$ \quad & \quad $l_{14}$ & $2.93 \times 10^{-1}$ \quad & \quad $l_{24}$ & $-4.87 \times 10^{-1}$ \\
 $\omega _1$ & $5.24 \times 10^{-1}$ \quad & \quad $\gamma$ & $2.95 \times 10^{-2}$ \quad & \quad $q_{134}$ & $-5.87 \times 10^{-2}$ \quad & \quad $q_{234}$ & $1.18 \times 10^{-1}$ \\
 $\sigma_3$ & $-5.22 \times 10^{-1}$ \quad & \quad $\beta_{3}$ & $1.53 \times 10^{-1}$ \quad & \quad $l_{41}$ & $3.14 \times 10^{-2}$ \quad & \quad $l_{42}$ & $-5.14 \times 10^{-2}$ \\
 $\sigma_4$ & $2.72 \times 10^{-2}$ \quad & \quad $\beta_{4}$ & $5.78 \times 10^{-2}$ \quad & \quad $q_{413}$ & $-7.56 \times 10^{-3}$ \quad & \quad $q_{423}$ & $1.28 \times 10^{-2}$ \\
 $\sigma_5$ & $-2.72 \times 10^{-1}$ \quad & \quad $\beta_{5}$ & $1.91 \times 10^{-1}$ \quad & \quad $q_{415}$ & $2.99 \times 10^{-2}$ \quad & \quad $q_{425}$ & $1.71 \times 10^{-1}$ \\
 & & \quad $\beta_{15}$ & $-2.42 \times 10^{-2}$ \quad & \quad $l_{35}$ & \quad $4.28$ \quad & \quad $l_{53}$ & $2.89 \times 10^{-2}$ \\
 & & \quad $\gamma_{15}$ & $1.70 \times 10^{-2}$ \quad & \quad $q_{335}$ & $-1.11$ \quad & \quad $q_{533}$ & $-7.22 \times 10^{-3}$\\
 & & & & \quad $q_{355}$ & $-5.13 \times 10^{-1}$ \quad & \quad $q_{535}$ & $1.48 \times 10^{-2}$ \\
 & & & & \quad $q_{314}$ & $1.57 \times 10^{-2}$ \quad & \quad $q_{514}$ & $-9.44 \times 10^{-3}$ \\ 
 \hline
 $\sigma_6$ & $-7.6 \times 10^{-2}$ \quad & \quad $\beta_{6}$ & $2.8 \times 10^{-2}$ \quad & \quad $q_{613}$ & $-3.18 \times 10^{-2}$ \quad & \quad $q_{623}$ & $3.23 \times 10^{-2}$ \\
 $\sigma_7$ & $-7.6 \times 10^{-1}$ \quad & \quad $\beta_{7}$ & $6.27 \times 10^{-1}$ \quad & \quad $q_{617}$ & $3.82 \times 10^{-2}$ \quad & \quad $q_{627}$ & $-5.62 \times 10^{-2}$ \\
 & & \quad $l_{61}$ & $1.23 \times 10^{-2}$ \quad & \quad $q_{716}$ & \quad $-9.18 \times 10^{-2}$ \quad & \quad $q_{726}$ & $-1.01 \times 10^{-1}$ \\
 & & \quad $l_{62}$ & $-1.33 \times 10^{-2}$ \quad & \quad $q_{733}$ & $8.21 \times 10^{-2}$ \quad & \quad $q_{737}$ & $1.37 \times 10^{-1}$\\
 & & \quad $l_{73}$ & $-3.27 \times 10^{-1}$ \quad & & & & 
 \end{tabular}
 \end{center}
 \caption{Coefficients of the reduced-order model at $Re=80$. See text for details.}
 \label{Tab:Pitchfork}
\end{table}

\bibliographystyle{jfm}
\bibliography{Main}

\end{document}